\documentclass[twocolumn,aps,prc,tightenlines,floats,floatfix,nofootinbib]{revtex4}

\def\st#1{{\kern-4pt} \not\!#1}

\def\sp{\kern +3pt}
\def\sm{\kern -3pt}

\def\spQ{\kern +6pt}

\def\etal{{\it et.\ al.\/},$\,$}

\def\bea{\begin{eqnarray}}
\def\eea{\end{eqnarray}}
\def\etal{{\it et al.\/}}

\def\etal{{\it et al.\/}}
\def\sfrac#1#2{{\textstyle \frac{#1}{#2}}}

\newcommand{\bra}[1]{\langle #1|}
\newcommand{\ket}[1]{|#1\rangle}

\def\be{\begin{equation}}
\def\ee{\end{equation}}
\def\ba{\begin{eqnarray}}
\def\ea{\end{eqnarray}}

\usepackage{graphics}
\usepackage{graphicx}
\usepackage{epsf}
\usepackage{amsmath}
\usepackage{amssymb}

\begin{document}

\phantom{0}
\vspace{-0.2in}
\hspace{5.5in}
\parbox{1.5in}{\leftline{ADP-11-24/T746}}

\vspace{-1in}

\title
{\bf Octet baryon electromagnetic form factors
in a relativistic quark model}
\author{G. Ramalho$^{1}$ and K. Tsushima$^2$
\vspace{-0.1in}  }

\affiliation{
$^1$CFTP, Instituto Superior T\'ecnico,
Universidade T\'ecnica de Lisboa,
Av. Rovisco Pais, 1049-001 Lisboa, Portugal
\vspace{-0.15in}}
\affiliation{$^2$CSSM, School of Chemistry and Physics,
University of Adelaide, Adelaide SA 5005, Australia}

\vspace{0.2in}
\date{\today}

\phantom{0}

\begin{abstract}
We study the octet baryon electromagnetic properties
by applying the covariant spectator quark model,
and provide covariant parametrization that can be
used to study baryon electromagnetic reactions.
While we use the lattice QCD data in the large pion mass regime
(small pion cloud effects)
to determine the parameters of the model in the valence quark sector,
we use the nucleon physical and octet baryon magnetic
moment data to parametrize the pion cloud contributions.
The valence quark contributions
for the octet baryon electromagnetic form factors are estimated
by extrapolating the lattice parametrization in the large pion
mass regime to the physical regime.
As for the pion cloud contributions,
we parametrize them in a covariant, phenomenological manner,
combined with SU(3) symmetry.
We also discuss the impact of the
pion cloud effects on the octet baryon electromagnetic
form factors and their radii.
\end{abstract}

\vspace*{0.9in}  
\maketitle

\section{Introduction}

Low-lying baryons are classified into the octet \mbox{(spin 1/2)}
and decuplet (spin 3/2) baryon members.
This is based on SU(3) symmetry, which is presently understood
in terms of quantum chromodynamincs (QCD).
To study the electromagnetic structure of the light baryons,
they are probed by electron beams, and
the form factors are measured
as functions of momentum transfer
squared $q^2=-Q^2$, and they reveal the
nonpointlike structure of the baryons.
In the octet and decuplet baryons,
only the weak and electromagnetic structure of the nucleon
has been well measured in experiments at finite $Q^2$.
Our present knowledge of the electromagnetic structure
for the octet and decuplet baryons is restricted
to magnetic moments of some octet~\cite{Nucleon,Octet}
and decuplet~\cite{Omega} baryons.
Thus, except for the nucleon system~\cite{Nucleon},
studies of the weak and electromagnetic structure
for the other octet baryon members are
very scarce~\cite{KuboderaCBM,OctetWeak,TsushimaCBM,Jakob93,Kim96,Williams96,Kubis99,Kubis01,Cheedket04,Silva04,Wang08,Liu09a,Liu09b}.
Therefore, more data and studies of the octet
baryon structure are desired.

In this work we study the electromagnetic structure
of the octet baryons using the covariant
spectator quark model~\cite{Nucleon,Octet,Omega}.
In particular, one of our goals is to obtain covariant parametrization
associated with the valence quark degrees of freedom,
that provides insight into the octet baryon electromagnetic internal structure.
This will open tremendous possibilities for future applications.
A study of the decuplet baryons with the same formalism
was already performed successfully in Ref.~\cite{Omega}.

Generally, phenomenological treatment of baryon structure
based solely on the valence quark degrees of freedom,
can be improved by the inclusion of the meson cloud effects,
guided by chiral effective field theory~\cite{Perdrisat07}.
The effects of the meson cloud are particularly important
for the neutron electric form factor~\cite{Nucleon},
and the electromagnetic transition,
$\gamma N \to \Delta(1232)$~\cite{NDelta,NDeltaD,LatticeD}.
Although chiral perturbation theory is very useful
to infer the $Q^2$ dependence of the
nucleon form factors,
it is usually applicable in the very low $Q^2$ region
($\lesssim  0.4$ GeV$^2$), and cannot be used
to define the scale of separation between
the valence quark dominant region from
that of the meson-baryon excitations (meson cloud effects)~\cite{ChPT1}.
The separation is possible only within a specific model.
QCD in the limit of high $Q^2$, and many phenomenological calculations suggests
that the meson cloud effects should falloff with increasing $Q^2$, leading
to the valence quark dominance in the high $Q^2$ region.
However, the magnitude, sign, and the rate of the falloff
of the meson cloud excitations depend on
models~\cite{ChPT1,ChPT2,Leinweber99,Thomas83,Thomas84,TsushimaCBM,Liu98,Miller02,Friedrich03,Faessler06b,Wang07,Wang09,Cloet09}.
Among the mesons, the pion is expected to give
the most important contributions according to chiral
perturbation theory~\cite{ChPT1,ChPT2,Leinweber99}.
For example, the importance of the effects
in nucleon magnetic moments in different models can
be found in Ref.~\cite{Octet}.
However, for the nucleon, an accurate description
of the form factors can also be achieved by a model
without the pion cloud in the physical and
the lattice QCD regimes~\cite{Nucleon,Lattice},
e.g., model II in Ref.~\cite{Nucleon}.
This suggests that the effects of the pion cloud can
possibly be small for the nucleon form factors
in the low $Q^2$ region.
Although the pion cloud effects cannot be extracted
directly from the experimental data,
we assume that the octet baryon systems
can be described by a mixture of the valence quarks and
the pion cloud, and use lattice QCD data to constrain
the valence quark contributions, and then
to extract the pion cloud effects.

In this study we use the covariant spectator quark model
which is inspired by the covariant spectator theory~\cite{Gross}.
The covariant spectator quark model
has been successfully applied for studying
the electromagnetic properties of several
baryons~\cite{Octet,Omega,GE2Omega,Nucleon,NDelta,NDeltaD,LatticeD,Lattice,FixedAxis,DeltaFF,DeltaDFF,DeltaDFF2,Roper,ExclusiveR,Delta1600,S11,S11scaling}.
A baryon in the model is described by a wave function
for the quark-diquark system
parametrized in a covariant manner.
The photon-quark coupling is described
based on vector meson dominance (VMD)
for a constituent quark.
Because of this feature, the model can be extended easily
to the lattice QCD regime~\cite{Omega,Lattice,LatticeD},
as will be explained later.
Starting by a simplified model for the nucleon
based on an S-state configuration for
the quark-diquark system, we extend the model
for the octet baryons first, and next
to the lattice QCD regime.
That procedure allows us
to use the lattice QCD simulation data with
large pion masses to constrain better the parameters of the model,
instead of using only the physical data.
The final parametrization for the wave functions
obtained in the lattice regime, can then be extrapolated
to obtain the physical octet baryon wave functions.
This is particularly useful, since
there are no experimental data for
$\Sigma$, $\Lambda$ and $\Xi$ baryons for $Q^2 >0$.
However, this procedure can only provide
the contributions from the valence quarks,
since still lattice QCD simulations are presently
performed with large pion masses, and thus the effects
of the pion cloud are suppressed~\cite{Detmold}.
To take into account the effects of the pion cloud
which are very important for the physical regime,
we first use a simple, covariant parametrization
for the pion cloud contributions for
the nucleon electromagnetic form factors.
Then, the parametrization for the nucleon is extended
to the other octet baryon members using SU(3) symmetry.

This article is organized as follows.
In Sec.~\ref{secValPion} we describe the electromagnetic
current, and discuss the separation
of the photon couplings with
the quarks, and those with the pion cloud.
In Sec.~\ref{secValence} we give detailed explanations
of the covariant spectator quark model extended for the octet
baryons, both for the physical and lattice QCD regimes.
In Sec.~\ref{secPionCloud} we discuss
the pion cloud contributions, and give
phenomenological parametrization for
the pion cloud contributions.
In Sec.~\ref{secParam} explicit
parametrization for the octet baryon wave functions
and pion cloud contributions for the
electromagnetic form factors are presented.
The results for the bare and dressed
electromagnetic form factors of the
octet baryons are presented in Sec.~\ref{secResults}.
Summary and conclusions are
given in Sec.~\ref{secConclusions}.

\section{Valence quark  and pion cloud contributions for the current}
\label{secValPion}

The current associated with the elastic electromagnetic
interaction with a baryon $B$, with spin 1/2
positive parity and mass $M_B$,
can be represented in general as
\be
J_B^\mu=
F_{1B}(Q^2) \gamma^\mu +
F_{2B}(Q^2)
\frac{i \sigma^{\mu \nu}q_\nu}{2M_B},
\label{eqJgen}
\ee
where $F_{1B}$ and $F_{2B}$ are respectively the Dirac and
Pauli form factors which are the functions of $Q^2=-q^2$,
where \mbox{$q=P_+ -P_-$},
with $P_+$ ($P_-$) being the final (initial) momentum.
Omitted in Eq.~(\ref{eqJgen}) are the initial
and final state Dirac spinors function of $P_\pm$
and the spin projections.
For simplicity we represent the current
in units $e=\sqrt{4\pi\alpha}$, with
$\alpha \simeq 1/137$, the electromagnetic
fine structure constant.
At $Q^2=0$, these form factors are defined as
\be
F_{1B}(0)= e_B, \hspace{.9cm}
F_{2B}(0)=\kappa_B,
\label{eqCharge}
\ee
where $e_B$ is the baryon charge in units of $e$
and $\kappa_B$ is the baryon anomalous magnetic moment
in natural units $\sfrac{e}{2M_B}$.
An alternative representation of the
electromagnetic form factors of the baryon $B$
is the Sachs parametrization in terms of the
electric charge
\mbox{$G_{EB}= F_{1B} -\sfrac{Q^2}{4M_B^2}F_{2B}$}
and magnetic dipole
$G_{MB}= F_{1B}+ F_{2B}$
form factors.

\begin{figure}[t]
\includegraphics[width=3.0in]{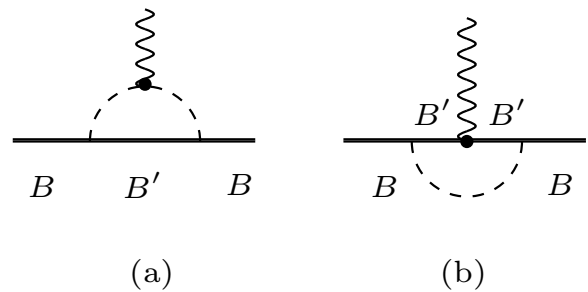}
\caption{\footnotesize
Electromagnetic interaction with the baryon $B$
within the one-pion loop level (pion cloud) through
the intermediate baryon states $B^\prime$.
A diagram including a contact vertex
$\gamma \pi B B^\prime$, as described in Ref.~\cite{Octet},
is not represented explicitly, since
the isospin structure is the same as diagram (a).
See Ref.~\cite{Octet} for details.
}
\label{figPionCloud}
\end{figure}

In a quark model
the electromagnetic interaction with a baryon $B$ may
be decomposed into the photon interaction
with valence quarks, and with sea quarks
(polarized quark-antiquark pairs).
The latter can be interpreted
as virtual mesons which dress the
baryon valence quark core.
The photon couplings with
the intermediate meson-baryon states
can be described by effective field theories
that are perturbative for the low meson energies and momenta.
According to chiral perturbation
theory, the most important meson for a given
reaction is the lightest one, the pion.
This is also expected to be the case for the octet baryons.
Thus, we can describe the
electromagnetic interaction for a member $B$
of the octet baryons using a current,
\be
J_B^\mu =
Z_B \left[
J_{0B}^\mu + J_\pi^\mu + J_{\gamma B}^\mu \right],
\label{eqJdecomp}
\ee
where $J_{0B}^\mu$ stands for the electromagnetic
interaction with the quark core without the pion cloud,
and the remaining terms are the interaction
with the intermediate pion-baryon ($\pi B$) states.
(See Fig.~\ref{figPionCloud}.)
In particular, $J_\pi^\mu$ represents the direct interaction
with the pion, and $J_{\gamma B}^\mu$ the interaction
with the baryon while one pion is in the air.
The factor $Z_B$  is a renormalization constant,
which is common to each isomultiplet:
nucleon ($N$), $\Sigma$, $\Lambda$, and $\Xi$.
$Z_B$ is related with the derivative of the
baryon self-energy~\cite{Octet}.

\begin{figure}[t]
\centerline{
\mbox{
\includegraphics[width=7.5cm]{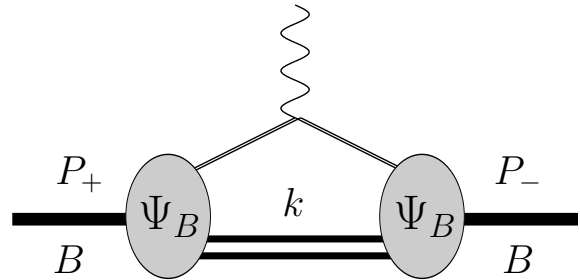} }}
\caption{
Electromagnetic interaction
with the baryon $B$ in a impulse approximation.
$P_+$ ($P_-$) represents the final (initial)
baryon momentum and $k$ the momentum of the on-shell diquark.
The baryon wave function
is represented by $\Psi_B$.}
\label{figImp}
\end{figure}

In an additive constituent quark model, the current $J_{0B}^\mu$
is given by the sum of the individual quark current.
The electromagnetic interaction processes
for the nucleon in the covariant spectator quark model~\cite{Nucleon}
is presented in Fig.~\ref{figImp}.
The decomposition given by Eq.~(\ref{eqJdecomp}) is justified 
when the pion is created by the {\it overall baryon},
but not by a single quark.
The processes where pions are created and absorbed
by the same quark are included in the constituent quark
internal structure,
and thus included in the current $J_{0B}^\mu$.

\section{Spectator quark model for the octet baryons}
\label{secValence}

In the covariant spectator quark model
a baryon $B$ is described as a system
with an off-mass-shell quark, free to
interact with photons, and two on-mass-shell quarks.
Integrating over the two on-mass-shell
quark momenta, we represent the quark pair as
an on-mass-shell diquark with an effective mass $m_D$,
and the baryon as a quark-diquark system~\cite{Nucleon}.
This quark-diquark system is then described
by a transition vertex between the three-quark bound state
and the quark-diquark state,
that describes effectively
the confinement~\cite{Nucleon,Omega}.

\subsection{Octet baryon wave functions}

The simplest representation for a
quark-diquark system with spin 1/2 and positive parity
is the S-wave configuration.
The wave function for an octet baryon $B$
with momentum $P$ and the internal diquark momentum $k$,
can be represented in the S-state
approximation~\cite{Nucleon,Octet},
\be
\Psi_B(P,k) =
\frac{1}{\sqrt{2}} \left\{
\phi_S^0 \ket{M_A} +
\phi_S^1 \ket{M_S}
\right\} \psi_B(P,k),
\label{eqPsiB}
\ee
where $\ket{M_A}$ and $\ket{M_S}$ are the flavor
antisymmetric and mixed symmetric states, respectively,
and $\phi_S^{0,1}$ are the spin (0 and 1) wave functions.
$\psi_B(P,k)$ is a scalar
function of $P$ and $k$, and it reflects the momentum
distribution of the quark-diquark system.

Explicit baryon flavor wave functions
are presented in Table~\ref{tablePHI}.
The spin wave functions are given by~\cite{Nucleon}
\ba
& &
\phi_S^0 \left(+\sfrac{1}{2}\right)=
\frac{1}{\sqrt{2}}
\left( \uparrow \downarrow - \downarrow  \uparrow  \right) \uparrow, \\
& &
\phi_S^1
\left(+\sfrac{1}{2}\right)
= -
\frac{1}{\sqrt{6}}
\left[
\left( \uparrow \downarrow + \downarrow  \uparrow \right)
\uparrow - 2 \uparrow \uparrow \downarrow \right],
\ea
and
\ba
& &
\phi_S^0
\left(-\sfrac{1}{2}\right)
=
\frac{1}{\sqrt{2}}
\left( \uparrow \downarrow - \downarrow  \uparrow  \right) \downarrow, \\
& &
\phi_S^1
\left(-\sfrac{1}{2}\right)=
\frac{1}{\sqrt{6}}
\left[
\left( \uparrow \downarrow + \downarrow  \uparrow \right)
\downarrow - 2 \downarrow \downarrow \uparrow \right].
\ea

This nonrelativistic structure is generalized
in the covariant spectator quark model~\cite{Nucleon} as
\ba
\phi_S^0&=& u_B(P,s), \nonumber \\
\phi_S^1&=& -\varepsilon_\lambda^{\alpha \ast} (P)
U_B^\alpha(P,s),
\label{eqPhiS}
\ea
where
\be
U_B^\alpha(P,s)=
\frac{1}{\sqrt{3}}
\gamma_5 \left(
\gamma^\alpha-\frac{P^\alpha}{M}
\right) u_B(P,s).
\ee
In the above $u_B(P,s)$
is the Dirac spinor of the octet baryon $B$
with momentum $P$, spin $s$,
and $\varepsilon_\lambda(P)$ the
diquark polarization in the fixed-axis
representation~\cite{Nucleon,FixedAxis}.
In Ref.~\cite{Nucleon} it is shown how
Eq.~(\ref{eqPhiS}) generalizes the nonrelativistic
spin wave functions.

\begin{table*}[t]
\begin{center}
\begin{tabular}{l c c c}
\hline
\hline
$B$   & $\ket{M_S}$  & &  $\ket{M_A}$  \\
\hline
$p$     &   $\sfrac{1}{\sqrt{6}} \left[
        (ud + du) u - 2 uu d \right]$ & &
        $\sfrac{1}{\sqrt{2}} (ud -du) u$  \\
$n$     &  $-\sfrac{1}{\sqrt{6}} \left[
         (ud + du) d - 2 ddu \right]$ & &
         $\sfrac{1}{\sqrt{2}} (ud -du) d$ \\
\hline
$\Lambda^0$ &
$\sfrac{1}{2}
\left[ (dsu-usd) + s (du-ud)
\right]$
& &
$\sfrac{1}{\sqrt{12}}
\left[
s (du-ud) - (dsu-usd) -2(du-du)s
\right]$
\\
\hline
$\Sigma^+$  & $\sfrac{1}{\sqrt{6}} \left[(us + su) u - 2 uu s \right]$ &
            &  $\sfrac{1}{\sqrt{2}} (us -su) u $ \\
$\Sigma^0$ &
$\sfrac{1}{\sqrt{12}}
\left[
s (du+ud) +(dsu+usd) -2(ud+du)s
\right]$
& &
$\sfrac{1}{2}
\left[ (dsu+usd) -s (ud+du)
\right]$ \\
$\Sigma^-$ & $\sfrac{1}{\sqrt{6}}\left[ (sd + ds) d - 2 dd s \right]$ & &
              $\sfrac{1}{\sqrt{2}} (ds -sd) d$ \\
\hline
$\Xi^0$ & $-\sfrac{1}{\sqrt{6}} \left[(ud + du) s - 2 ss u\right]$ & &
          $\sfrac{1}{\sqrt{2}} (us -su) s$ \\
$\Xi^-$ & $-\sfrac{1}{\sqrt{6}} \left[(ds + sd) s - 2 ss d\right]$ & &
          $\sfrac{1}{\sqrt{2}} (ds -sd) s$  \\
\hline
\hline
\end{tabular}
\end{center}
\caption{Flavor wave functions of the octet baryons.}
\label{tablePHI}
\end{table*}

\subsection{Electromagnetic current}

Taking into account that the wave function $\Psi_B$ is written
in terms of the wave functions
of a quark pair (12) and a single quark (3),
one can write the  electromagnetic
current associated with the baryon $B$
in a impulse approximation \cite{Nucleon,Omega},
\ba
J_{0B}^\mu&=& 3 \sum_{\Gamma}
\int_k \overline \Psi_B (P_+,k)j_q^\mu \Psi_B (P_-,k),
\label{eqJB0}
\ea
where $j_q^\mu$ is the quark current operator,
$P_+$ ($P_-$) is the final (initial)
baryon momentum and $k$ the
momentum of the on-shell diquark,
and $\Gamma= \left\{s, \lambda \right\}$ labels the
scalar diquark and the vectorial diquark
polarization $\lambda=0,\pm$.
The factor 3 in Eq.~(\ref{eqJB0}) takes into account
the contributions for the current from the pairs $(13)$ and $(23)$,
where each pair has the identical contribution with that of the pair $(12)$.
The polarization indices are suppressed for simplicity.
The integral symbol represents
\be
\int_k= \int \frac{d^3 {\bf k}}{2 E_D (2\pi)^3},
\ee
where $E_D=\sqrt{m_D^2+ {\bf k}^2}$.

Generally, the baryon electromagnetic current (\ref{eqJB0})
can be expressed as
\be
J_{0B}^\mu=
\tilde e_{0B} \, \gamma^\mu + \tilde \kappa_{0B} \,
\frac{i \sigma^{\mu \nu}q_\nu}{2M_B},
\label{eqJB1}
\ee
where $\tilde e_{0B}$ and $\tilde \kappa_{0B}$
are the functions of $Q^2$, and respectively correspond to
the valence quark contributions for the
$F_{1B}(Q^2)$ and $F_{2B}(Q^2)$ form factors.
To represent these quantities for $Q^2=0$, we suppress the tildes.
Note that in Eq.~(\ref{eqJB1}) we
omit the baryon spinors as in Eq.~(\ref{eqJgen}).

\subsection{Quark current}

The quark current operator $j_q^\mu$ has a
generic structure,
\be
j_q^\mu = j_1
\left(
\gamma^\mu - \frac{\not \! q q^\mu}{q^2}\right)
+ j_2 \frac{i \sigma^{\mu \nu} q_\nu}{2 M_N},
\label{eqJi}
\ee
where $M_N$ is the nucleon mass and
$j_i$ ($i=1,2$) are SU(3) flavor operators acting on the
third quark of the $\ket{M_A}$ or $\ket{M_S}$ state.
In the first term $\not \! q q^\mu / q^2$ is
included for completeness, but does not
contribute for elastic reactions.

The quark current $j_i$ ($i=1,2$) in Eq.~(\ref{eqJi}),
can be decomposed as the sum of operators
acting on quark 3 in SU(3) flavor space
\be
j_i=
\sfrac{1}{6} f_{i+} \lambda_0
+  \sfrac{1}{2}f_{i-} \lambda_3
+ \sfrac{1}{6} f_{i0} \lambda_s,
\ee
where
\ba
&\lambda_0=\left(\begin{array}{ccc} 1&0 &0\cr 0 & 1 & 0 \cr
0 & 0 & 0 \cr
\end{array}\right), \hspace{.3cm}
&\lambda_3=\left(\begin{array}{ccc} 1& 0 &0\cr 0 & -1 & 0 \cr
0 & 0 & 0 \cr
\end{array}\right),
\nonumber \\
&\lambda_s \equiv \left(\begin{array}{ccc} 0&0 &0\cr 0 & 0 & 0 \cr
0 & 0 & -2 \cr
\end{array}
\right),
\label{eqL1L3}
\ea
are the flavor operators.
These operators act on
the quark wave function in flavor space,
$q=  (\begin{array}{c c c} \! u \, d \, s \!\cr
\end{array} )^T$.

The functions $f_{i\pm}(Q^2)$  ($i=1,2$) are normalized
by $f_{1 \, n}(0)=1$ ($n=0,\pm$),
$f_{2\pm}(0)=\kappa_\pm$,
and $f_{20}(0)=\kappa_s$.
The isoscalar ($\kappa_+$) and isovector ($\kappa_-$)
anomalous magnetic moments
are defined in terms of the $u$ and $d$ quark
anomalous magnetic moments, $\kappa_+= 2 \kappa_u -\kappa_d$
and $\kappa_-= \sfrac{2}{3} \kappa_u+ \sfrac{1}{3} \kappa_d$.
In the previous works the
quark anomalous magnetic moments were adjusted
to reproduce the experimental magnetic moments of the
nucleon and the $\Omega^-$~\cite{Nucleon,Omega}.
In this work however, we will readjust
the $u$ and $d$ quark anomalous magnetic moments
as will be explained later.

To see explicitly
the quark flavor contributions
for the electromagnetic current~(\ref{eqJi}),
we sum over the quark flavors following Refs.~\cite{Octet,Omega},
and get the coefficients
\ba
j_i^A&=&
\bra{M_A} j_i \ket{M_A},
\label{eqjiA}\\
j_i^S&=&
\bra{M_S} j_i \ket{M_S},
\label{eqjiS}
\ea
for $i=1,2$.
The results, corresponding to the
states given in Table~\ref{tablePHI},
are presented in Table~\ref{tablePhiB}.

\subsection{Valence quark contributions
for the electromagnetic form factors}
\label{secBareFF}

Using the expressions derived in the previous work for the
nucleon form factors in the S-state approach~\cite{Nucleon},
we obtain the corresponding expressions
for the octet baryons $B$ by replacing the nucleon
coefficients $j_i^A$ and $j_i^S$ ($i=1,2$)
by the respective baryon state
\ba
\tilde e_{0B} &=&
B(Q^2) \times \nonumber \\   
& & \left(\frac{3}{2} j_1^A +
\frac{1}{2}
\frac{3 -\tau}{1+ \tau}
j_1^S - 2 \frac{\tau}{1+\tau} \frac{M_B}{M_N}
j_2^S \right),  \label{eqF1} \\
\tilde \kappa_{0B} &=&
B(Q^2) \times  \nonumber \\
& &
\left[
\left(
\frac{3}{2}
j_2^A
-\frac{1}{2} \frac{1-3\tau}{1+\tau} j_2^S \right) \frac{M_B}{M_N}
-2 \frac{1}{1+\tau} j_1^S
\right] , \nonumber \\
& &
\label{eqF2}
\ea
with $\tau=\sfrac{Q^2}{4M_B^2}$, and
\be
B(Q^2)= \int_k \psi_B(P_+,k) \psi_B(P_-,k),
\ee
the overlap integral between
the initial and final scalar wave functions.
The normalization of the wave function
leads to $B(0)=1$.
The expressions in Eqs.~(\ref{eqF1}) and~(\ref{eqF2})
given for the nucleon~\cite{Nucleon}
are briefly reviewed in the Appendix.  

Another, possible electromagnetic transition
between the octet baryon members is $\gamma^\ast \Lambda \to \Sigma^0$.
This transition is very interesting, and will be studied
in a separate work~\cite{InPreparation},
since it is an inelastic reaction.

The expressions for $\tilde e_{0B}$
and $\tilde \kappa_{0B}$ [Eqs.~(\ref{eqF1}) and~(\ref{eqF2})]
are particularly simplified for $Q^2=0$
\ba
e_{0B} &=&
= \frac{3}{2}
\left( j_1^A + j_1^S \right), \nonumber \\
\kappa_{0B} &=&
\left(
\sfrac{3}{2}  j_2^A
-  \sfrac{1}{2} j_2^S \right)  \sfrac{M_N}{M_B}
- 2 j_1^S.
\label{eqK}
\ea
We require that the bare charge $e_{0B}$
and the dressed charge $e_B$ are the same
\be
e_{0B}= e_B.
\ee

To get the numerical results,
we must specify the functions
$f_{i\,n}(Q^2)$ ($i=1,2$, $n=0,\pm$) and $\psi_B(P,k)$.
The explicit expressions will be given
in Sec.~\ref{secParam}.

\begin{table}[t]
\begin{center}
\begin{tabular}{l c c}
\hline
\hline
$B$   & $j_i^S$  &   $j_i^A$  \\
\hline
$p$     & $\sfrac{1}{6} (f_{i+}-f_{i-}) $ &
        $\sfrac{1}{6} (f_{i+}+3 f_{i-})  $ \\
$n$     & $\sfrac{1}{6} (f_{i+}+  f_{i-}) $ &
        $\sfrac{1}{6} (f_{i+}-  3 f_{i-})$ \\
\hline
$\Lambda^0$ & $\sfrac{1}{6}f_{i+}$ &
 $\sfrac{1}{18} (f_{i+}-  4 f_{i0})$ \\
\hline
$\Sigma^+$  & $\sfrac{1}{18}
(f_{i+} + 3 f_{i-} -4 f_{i0}) $ &
 $\sfrac{1}{6} (f_{i+}+3 f_{i-})  $ \\
$\Sigma^0$ &  $\sfrac{1}{36} (2 f_{i+}-  8 f_{i0})$
& $\sfrac{1}{6}f_{i+}$ \\
$\Sigma^-$ & $\sfrac{1}{18}
(f_{i+} - 3 f_{i-} -4 f_{i0}) $ &
        $\sfrac{1}{6} (f_{i+}-  3 f_{i-})           $ \\
\hline
$\Xi^0$ & $\sfrac{1}{18} (2 f_{i+} + 6  f_{i-} -2 f_{i0}) $ &
$-\sfrac{1}{3}f_{i0}$ \\
$\Xi^-$ & $\sfrac{1}{18} (2 f_{i+} -6  f_{i-} -2 f_{i0}) $ &
$-\sfrac{1}{3}f_{i0}$ \\
\hline
\hline
\end{tabular}
\end{center}
\caption{Mixed symmetric and antisymmetric coefficients for the octet baryons
appearing in Eqs.~(\ref{eqjiA}) and~(\ref{eqjiS}).}
\label{tablePhiB}
\end{table}

The form factors described in this section,
corresponding to Eqs.~(\ref{eqF1}) and~(\ref{eqF2}),
include only the valence quark contributions.
For a realistic estimate we need to include
the pion cloud effects explicitly.

\section{Pion cloud contributions for the
electromagnetic form factors}
\label{secPionCloud}

We discuss here the pion cloud contributions
for the electromagnetic current and form factors,
represented by the diagrams
in Fig.~\ref{figPionCloud}.
Following Ref.~\cite{Octet}, we assume
the pion as the dominant meson excitation
to be included in the octet baryon form factors.
Then, the meson cloud contributions for
the octet baryon electromagnetic form factors can be described
in terms of 6 independent functions of $Q^2$,
related to the pion-baryon Feynman integral,
as will be described next.

\subsection{Pion-baryon couplings in SU(3)}

The meson-baryon interaction vertices between
the baryon octet and the pseudoscalar meson octet,
$\pi,K,\bar K$ and $\eta$,
can be described by the
two independent coefficients $D$ and $F$
based on SU(3) symmetry~\cite{GellMann62,Carruthers}.
The coupling constant of the pion ($\pi$) and baryons ($B$ and $B^\prime$),
$g_{\pi BB^\prime}$, can
be represented in terms of the ratio,
$\alpha=\frac{D}{F+D}$ and a global coupling constant
$g = g_{\pi NN}$, the $\pi NN$ coupling constant.
As a result, the interaction currents and baryon self-energies,
at the one-pion loop level, which depend on
the coupling constants of,
$\pi NN$, $\pi \Lambda \Sigma$, $\pi \Sigma \Sigma$
and $\pi \Xi \Xi$,
can be expressed in terms
of the independent coefficients $\beta_B$
for $B= \Lambda, \Sigma,\Xi$,
\ba
& &
\beta_\Lambda = \frac{4}{3} \alpha^2, \\
& &
\beta_\Sigma= 4(1-\alpha)^2, \\
& &
\beta_\Xi = (1-2 \alpha)^2.
\ea
The factor for the nucleon is $\beta_N=1$.
Absorbing the global coupling constant $g$
in the functions associated with the pion-baryon loops,
we can express the pion cloud contributions
entirely in terms of $\alpha$.

In the exact SU(3) symmetry limit all the octet
baryon masses are the same.
We break the symmetry by the
pion cloud effects on the mass, but only
by the one-pion loop in the self-energies.
Following Ref.~\cite{Octet}, we represent
the mass of the octet baryon $B$ as $M_B=M_0+ \Sigma(M_B)$,
where $M_0$ is a mass parameter and $\Sigma(M_B)$
the self-energy at the physical baryon mass point.
Then, we can write
$\Sigma(M_B)= G_{0B} {\cal B}_0$,
with ${\cal B}_0$ being a scalar integral and $G_{0B}$
a factor that includes the couplings
of the pion with the baryon (see Table 2 in Ref.~\cite{Octet}).
Using the value, $\alpha=0.6$ given by an SU(6) quark model
\cite{Thomas83},
we get $M_0=1.342$ GeV and ${\cal B}_0=-0.127$ GeV,
which can describe the octet baryon masses within
an accuracy of 7\%.

The choice of $\alpha=0.6$ given by SU(6) symmetry,
defines the strength of the pion cloud contributions as
$\beta_\Sigma=\sfrac{16}{25}$,
$\beta_\Lambda=\sfrac{12}{25}$ and
$\beta_\Xi= \sfrac{1}{25}$~\cite{Thomas83,TsushimaCBM}.
Note that, the intermediate baryons
in the one-pion loop diagrams in Fig.~\ref{figPionCloud},
are always those in the same isomultiplet with the external baryons
($N$, $\Sigma$ or $\Xi$), with
the exception of the $\Lambda$-$\Sigma$ mixture
for the $\Sigma$ case,
where the mass difference of them is small.

A comment on the value of $\alpha$ is in order.
Contrary to the previous study
of the spectator quark model~\cite{Octet},
where $\alpha \simeq 0.69$  was obtained by fitting
to the octet baryon masses and octet baryon magnetic moments,
here we use the value obtained by SU(6) symmetry.
This enables us to treat all
the octet and decuplet baryons in a unified manner
based on SU(6) symmetry.

\subsection{Pion cloud dressing}

The pion cloud corrections, namely
the coupling of the photon
to the pion $J_\pi^\mu$, and
the coupling to the intermediate baryons
$J_{\gamma B}^\mu$, can be written~\cite{Octet}
\ba
J_{\pi}^\mu
&=&
\left(
\tilde B_1 \gamma^\mu + \tilde B_2
\frac{i \sigma^{\mu \nu}q_\nu}{2M_{B}}
\right) G_{\pi B}, \label{eqJpi}\\
J_{\gamma B}^\mu &=&
\left(
\tilde C_1 \gamma^\mu + \tilde C_2
\frac{i \sigma^{\mu \nu}q_\nu}{2M_{B}} \right) G_{e B}+ \nonumber \\
& &
\left(
\tilde D_1 \gamma^\mu + \tilde D_2
\frac{i \sigma^{\mu \nu}q_\nu}{2M_{B}}\right)  G_{\kappa B}.
\label{eqJgB}
\ea
In the above, $\tilde B_i$, $\tilde C_i$ and $\tilde D_i$
($i=1,2$) are arbitrary functions of $Q^2$ (as $\tilde e_{0B}$ and
$\tilde \kappa_{0B}$ are), and
$G_{\pi B}$,
$G_{e B}$, $G_{\kappa B}$ are the coefficients
that depend on the baryon flavors ($B=N,\Sigma,\Lambda,\Xi$).
We assume that the functions $\tilde B_i$, $\tilde C_i$
and $\tilde D_i$ are only weakly dependent on the baryon masses,
and the same for all the octet baryons as in Ref.~\cite{Octet}.
That allows a description of the pion cloud dressing
with a reduced number of coefficients.
We write $B_i$, $C_i$, and $D_i$ to represent, respectively,
the functions $\tilde B_i$, $\tilde C_i$ and $\tilde D_i$
at $Q^2=0$.
The coefficients $\tilde B_1$ and $\tilde B_2$
are proportional to the pion electromagnetic
form factor $F_\pi(Q^2)$, but
we absorb it in the
definition of $\tilde B_i$ for simplicity.

Combining Eqs.~(\ref{eqJdecomp}), (\ref{eqJpi}) and (\ref{eqJgB}),
we get
\ba
J_B^\mu&=&
Z_B \left\{ \sfrac{}{} \!
 \tilde e_{0B} + G_{\pi B} \tilde B_1
+ G_{eB} \tilde C_1 + G_{\kappa B} \tilde D_1\right\} \gamma^\mu \nonumber \\
&+&
Z_B
\left\{ \sfrac{}{} \!  \tilde \kappa_{0B} + G_{\pi B} \tilde B_2
+ G_{e B} \tilde C_2 + G_{\kappa B} \tilde D_2\right\}
\frac{i \sigma^{\mu \nu}q_\nu}{2M_{B}}. \nonumber \\
& &
\label{eqJgen2}
\ea

The renormalization constant $Z_B$ is determined by the
relation between the dressed form factor $F_{1B}$
at $Q^2=0$ and the charge $e_B$ [$F_{1B}(0)=e_{B}$],
or by the self-energy. (See Ref.~\cite{Octet} for details.)
The condition for the nucleon leads to
\be
D_1=0, \hspace{1cm} B_1=C_1,
\ee
and gives
\bea
&&Z_N=\Big[1+3B_1\Big]^{-1}.
\eea
Similarly, we can get the renormalization constants
for the other octet baryons~\cite{Octet}
\bea
&&Z_\Lambda=\Big[1+3\beta_\Lambda B_1\Big]^{-1},
\nonumber\\
&&Z_\Sigma=\bigg[1+\Big(2\beta_\Sigma+\beta_\Lambda\Big)B_1\bigg]^{-1},
\nonumber\\
&&Z_\Xi=\Big[1+3\beta_\Xi B_1\Big]^{-1}.
\eea

\begin{table}[t]
\begin{center}
\begin{tabular}{l c c}
\hline
\hline
 $B$ & $G_{\pi B}$ & $G_{ z B}$ \\
\hline
  $N$   & $2 \tau_3$    &   $\sfrac{1}{2}\left(3 z_N^s- z_N^v \right)$       \\
 $\Lambda$ & 0  & $\beta_\Lambda \left( 3 z_\Sigma^0 + z_\Sigma^2 \right)$  \\
 $\Sigma$ & $\qquad\left(\beta_\Sigma + \beta_\Lambda  \right) {\bf J}_3 \qquad$ &
 $\left[\beta_\Sigma \left( 2 z_\Sigma^0 + z_\Sigma^2\right) +
\beta_\Lambda  z_\Lambda \right]{\bf 1}  $
\\
  &  &
$+ \frac{1}{2} \beta_\Sigma  \left(
z_\Sigma^1 {\bf J}_3   - z_\Sigma^2 {\bf J}_3^2\right)$
\\
 $\Xi$  & $2\beta_\Xi \tau_3$ &
º$\beta_\Xi \left( 3 z_\Xi^s - z_\xi^v \tau_3 \right)$ \\
\hline
\hline
\end{tabular}
\end{center}
\caption{Coefficients $G_{\pi B}$ and $G_{zB}$,
where $z=e,\kappa$ for the octet baryons $B=N,\Lambda,\Sigma$ and $\Xi$.
See also Ref.~\cite{Octet} for details.}
\label{tableGB}
\end{table}

\subsection{Coefficients $G_{\pi B}$, $G_{e B}$ and $G_{\kappa B}$}

The coefficients $G_{\pi B}$ and
$G_{zB}$, where $z$ is either $e$ or $\kappa$,
were calculated in Ref.~\cite{Octet}.
To express the results, it is convenient to introduce
a general operator decomposition
for the bare current.
We use $z_B$ to represent $\tilde e_B$ or
$\tilde \kappa_B$.
For the $N$ and $\Xi$ isospin doublets, we use
the standard isoscalar-isovector notation,
\bea
z_B=\sfrac12(z^s_B+z^v_B\tau_3).
\eea
At $Q^2=0$ we have $e^s_N=e^v_N=1$, and $e^s_\Xi=-e^v_\Xi=-1$.
The $\Lambda$ case is given by the scalar functions
$\tilde e_{0 \Lambda}$ and $\tilde \kappa _{0 \Lambda}$,
with $e_{\Lambda}= e_{0 \Lambda}=0$.

For the $\Sigma$ isospin operators,
the decomposition of the three states ($0,\pm$) can be given by
\bea
z_\Sigma=z^0_\Sigma {\bf 1}+\sfrac12\Big(z^1_\Sigma {\bf J}_3+z^{2}_\Sigma {\bf J}^2_3\Big),
\eea
where ${\bf J}_3$ is the third component of the isospin 1 operator.
With this notation we get
\bea
z^0_{\Sigma}&=&z_{_{\Sigma^0}},
\nonumber\\
z^1_{\Sigma}&=&z_{_{\Sigma^+}}-z_{_{\Sigma^-}},
\nonumber\\
z^{2}_{\Sigma}&=&z_{_{\Sigma^+}}+z_{_{\Sigma^-}}-2z_{_{\Sigma^0}}\, ,
\eea
where $e^0_\Sigma=e^2_\Sigma=0$ and $e^1_\Sigma=2$, for $Q^2=0$.
The results for the coefficients are listed in Table~\ref{tableGB}.
(See Ref.~\cite{Octet} for details.)

\subsection{Octet baryon electromagnetic form factors with pion cloud dressing}
\label{secDressedFF}

From the current~(\ref{eqJgen2}),
using the expressions~(\ref{eqF1}) and~(\ref{eqF2}),
and the coefficients in Table~\ref{tableGB},
we can write down the Dirac form factors
$F_{1B}$ for the octet baryons $B$,
\ba
F_{1p} &=& Z_N
\left\{
\tilde e_{0p} + 2 \tilde B_1 + (\tilde e_{0p} + 2 \tilde e_{0n}) \tilde C_1  \sfrac{}{}
\right.
\nonumber \\
& & \left.
+ (\tilde \kappa_{0p} + 2 \tilde \kappa_{0n})\tilde D_1 \sfrac{}{} \!
\right\},  \label{eqF1p} \\
F_{1n} &=&
Z_N
\left\{ \tilde e_{0n} - 2 \tilde B_1 + (2 \tilde e_{0p} + \tilde e_{0n}) \tilde C_1
\sfrac{}{}
\right.
\nonumber \\
& & \left.
+ (2\tilde \kappa_{0p} + \tilde \kappa_{0n})\tilde D_1 \sfrac{}{} \!
\right\}, \label{eqF1n} \\
& & \nonumber \\
F_{1 \Lambda} &=&
Z_\Lambda
\left\{ \tilde e_{0 \Lambda}  +
\beta_\Lambda(\tilde e_{0 \Sigma^+} + \tilde e_{0\Sigma^0}
+ \tilde e_{0 \Sigma^-}) \tilde C_1  \sfrac{}{}
\right.
\nonumber \\
& & \left.
+ \beta_\Lambda (\tilde \kappa_{0 \Sigma^+} + \tilde \kappa_{0 \Sigma^0}
+ \tilde \kappa_{0 \Sigma^-})\tilde D_1 \sfrac{}{} \!
\right\},  \label{eqF1Lambda}\\
& & \nonumber \\
F_{1 \Sigma^+} &=&
Z_\Sigma
\left\{
\tilde e_{0 \Sigma^+}  + (\beta_\Sigma + \beta_\Lambda) \tilde B_1 \frac{}{} \!
\right.
\nonumber \\
& & \left.
+ \left[
\beta_\Sigma(\tilde e_{0 \Sigma^+} + \tilde e_{0\Sigma^0})
+ \beta_\Lambda \tilde e_{0 \Lambda} \right] \tilde C_1  \frac{}{} \!
\right.
\nonumber \\
& & \left. +
\left[
\beta_\Sigma(\tilde \kappa_{0 \Sigma^+} + \tilde \kappa_{0\Sigma^0})
+ \beta_\Lambda \tilde \kappa_{0 \Lambda} \right] \tilde D_1
 \frac{}{} \!
\right\}, \\
F_{1 \Sigma^0} &=&
Z_\Sigma
\left\{
\tilde e_{0 \Sigma^0}   \frac{}{} \!
\right.
\nonumber \\
& & \left.  +
\left[
\beta_\Sigma(\tilde e_{0 \Sigma^+} + \tilde e_{0\Sigma^-})
+ \beta_\Lambda \tilde e_{0 \Lambda} \right] \tilde C_1  \frac{}{} \!
\right.
\nonumber \\
& & \left. +
\left[
\beta_\Sigma(\tilde \kappa_{0 \Sigma^+} + \tilde \kappa_{0\Sigma^-})
+ \beta_\Lambda \tilde \kappa_{0 \Lambda} \right] \tilde D_1
 \frac{}{} \!
\right\}, \\
F_{1 \Sigma^-} &=&
Z_\Sigma
\left\{
\tilde e_{0 \Sigma^-} - (\beta_\Sigma+ \beta_\Lambda) \tilde B_1 \frac{}{} \!
\right.
\nonumber \\
& & \left.  +
\left[
\beta_\Sigma(\tilde e_{0 \Sigma^0} + \tilde e_{0\Sigma^-})
+ \beta_\Lambda \tilde e_{0 \Lambda} \right] \tilde C_1  \frac{}{} \!
\right.
\nonumber \\
& & \left. +
\left[
\beta_\Sigma(\tilde \kappa_{0 \Sigma^0} + \tilde \kappa_{0\Sigma^-})
+ \beta_\Lambda \tilde \kappa_{0 \Lambda} \right] \tilde D_1
 \frac{}{} \!
\right\}, \\
& & \nonumber \\
F_{1 \Xi^0} &=&
Z_\Sigma
\left\{
\tilde e_{0 \Xi^0} + 2 \beta_\Xi \tilde B_1  \frac{}{} \!
\right.
\nonumber \\
& & \left.  +
\beta_\Xi(\tilde e_{0 \Xi^0} + 2 \tilde e_{0\Xi^-}) \tilde C_1 \frac{}{} \!
\right.
\nonumber \\
& & \left. +
\beta_\Xi(\tilde \kappa_{0 \Xi^0} + 2 \tilde \kappa_{0\Xi^-}) \tilde D_1
 \frac{}{} \!
\right\}, \\
F_{1 \Xi^-} &=&
Z_\Sigma
\left\{
\tilde e_{0 \Xi^-} - 2 \beta_\Xi \tilde B_1  \frac{}{} \!
\right.
\nonumber \\
& & \left.  +
\beta_\Xi(2 \tilde e_{0 \Xi^0} +  \tilde e_{0\Xi^-}) \tilde C_1 \frac{}{} \!
\right.
\nonumber \\
& & \left. +
\beta_\Xi(2 \tilde \kappa_{0 \Xi^0} + \tilde \kappa_{0\Xi^-}) \tilde D_1
 \frac{}{} \!
\right\}.
\label{eqF1XiM}
\ea

Similarly, for the Pauli form factors $F_{2B}$, we can write down
\ba
F_{2p}&=&
Z_N
\left\{ \tilde \kappa_{0p} + 2 \tilde B_2  \sfrac{}{}\right. \nonumber \\
& &
\left. +
(\tilde e_{0p} + 2 \tilde e_{0n}) \tilde C_2
+ (\tilde \kappa_{0p} + 2 \tilde \kappa_{0n})\tilde D_2 \sfrac{}{} \!
\right\}, \label{eqF2p} \\
F_{2n}&=&
Z_N
\left\{ \tilde \kappa_{0n} - 2 \tilde B_2  \sfrac{}{} \right. \nonumber \\
& &
\left. +
(2 \tilde e_{0p} +  \tilde e_{0n}) \tilde C_2
+ (2 \tilde \kappa_{0p} +  \tilde \kappa_{0n})\tilde D_2 \sfrac{}{} \!
\right\},  \label{eqF2n} \\
&& \nonumber \\
F_{2 \Lambda} &=&
Z_\Lambda
\left\{ \tilde \kappa_{0 \Lambda}  +
\beta_\Lambda(\tilde e_{0 \Sigma^+} + \tilde e_{0\Sigma^0}
+ \tilde e_{0 \Sigma^-}) \tilde C_2  \sfrac{}{}
\right.
\nonumber \\
& & \left.
+ \beta_\Lambda (\tilde \kappa_{0 \Sigma^+} + \tilde \kappa_{0 \Sigma^0}
+ \tilde \kappa_{0 \Sigma^-})\tilde D_2  \sfrac{}{} \!
\right\}, \label{eqF2Lambda} \\
& & \nonumber \\
F_{2 \Sigma^+} &=&
Z_\Sigma
\left\{ \frac{}{} \!
\tilde \kappa_{0 \Sigma^+}  + (\beta_\Sigma + \beta_\Lambda) \tilde B_2
\right.
\nonumber \\
& & \left.  +
\left[
\beta_\Sigma(\tilde e_{0 \Sigma^+} + \tilde e_{0\Sigma^0})
+ \beta_\Lambda \tilde e_{0 \Lambda} \right] \tilde C_2  \frac{}{} \!
\right.
\nonumber \\
& & \left. +
\left[
\beta_\Sigma(\tilde \kappa_{0 \Sigma^+} + \tilde \kappa_{0\Sigma^0})
+ \beta_\Lambda \tilde \kappa_{0 \Lambda} \right] \tilde D_2
 \frac{}{} \!
\right\}, \\
F_{2 \Sigma^0} &=&
Z_\Sigma
\left\{ \frac{}{} \!
\tilde \kappa_{0 \Sigma^0}
\right.
\nonumber \\
& & \left.  +
\left[
\beta_\Sigma(\tilde e_{0 \Sigma^+} + \tilde e_{0\Sigma^-})
+ \beta_\Lambda \tilde e_{0 \Lambda} \right] \tilde C_2  \frac{}{} \!
\right.
\nonumber \\
& & \left. +
\left[
\beta_\Sigma(\tilde \kappa_{0 \Sigma^+} + \tilde \kappa_{0\Sigma^-})
+ \beta_\Lambda \tilde \kappa_{0 \Lambda} \right] \tilde D_2
 \frac{}{} \!
\right\}, \\
F_{2 \Sigma^-} &=&
Z_\Sigma
\left\{
\tilde \kappa_{0 \Sigma^-}   - (\beta_\Sigma+ \beta_\Lambda) \tilde B_2 \frac{}{} \!
\right.
\nonumber \\
& & \left.  +
\left[
\beta_\Sigma(\tilde e_{0 \Sigma^0} + \tilde e_{0\Sigma^-})
+ \beta_\Lambda \tilde e_{0 \Lambda} \right] \tilde C_2  \frac{}{} \!
\right.
\nonumber \\
& & \left. +
\left[
\beta_\Sigma(\tilde \kappa_{0 \Sigma^0} + \tilde \kappa_{0\Sigma^-})
+ \beta_\Lambda \tilde \kappa_{0 \Lambda} \right] \tilde D_2
 \frac{}{} \!
\right\}, \\
&& \nonumber \\
F_{2 \Xi^0} &=&
Z_\Sigma
\left\{
\tilde \kappa_{0 \Xi^0} + 2 \beta_\Xi \tilde B_2  \frac{}{} \!
\right.
\nonumber \\
& & \left.  +
\beta_\Xi(\tilde e_{0 \Xi^0} + 2 \tilde e_{0\Xi^-}) \tilde C_2 \frac{}{} \!
\right.
\nonumber \\
& & \left. +
\beta_\Xi(\tilde \kappa_{0 \Xi^0} + 2 \tilde \kappa_{0\Xi^-}) \tilde D_2
 \frac{}{} \!
\right\}, \\
F_{2 \Xi^-} &=&
Z_\Sigma
\left\{
\tilde \kappa_{0 \Xi^-} - 2 \beta_\Xi \tilde B_2  \frac{}{} \!
\right.
\nonumber \\
& & \left.  +
\beta_\Xi(2 \tilde e_{0 \Xi^0} +  \tilde e_{0\Xi^-}) \tilde C_2 \frac{}{} \!
\right.
\nonumber \\
& & \left. +
\beta_\Xi(2 \tilde \kappa_{0 \Xi^0} + \tilde \kappa_{0\Xi^-}) \tilde D_2
 \frac{}{} \!
\right\}.
\label{eqF2XiM}
\ea

The magnetic moment of the octet baryon member $B$ is defined in terms
of the magnetic form factor $G_{MB}=F_{1B}+F_{2B}$
at $Q^2=0$, according to $\mu_B=  G_{MB}(0) \sfrac{e}{2M_B}$.
The results for the magnetic moments are
usually expressed in
terms of the nuclear magneton $\hat \mu_N= \sfrac{e}{2M_N}$,
namely, $\mu_B= G_{MB}(0) \frac{M_N}{M_B} \hat \mu_N$.

\section{Parametrizations}
\label{secParam}

In the previous sections we have defined the general structure
of the valence quark part and the
functions for the pion cloud.
We still need to specify
the quark currents
in terms of the functions $f_{i\, n}(Q^2)$ ($i=1,2$, $n=0,\pm$),
scalar wave functions $\psi_B(P,k)$,
and the functions for the pion cloud
effects, $\tilde B_i$, $\tilde C_i$, and $\tilde D_i$ ($i=1,2$).

\subsection{Parametrization of the quark current}

To parametrize the quark current (\ref{eqJi}),
we adopt the structure inspired by the VMD
mechanism as in Refs.~\cite{Nucleon,Omega},
\ba
& &
\hspace{-1cm}
f_{1 \pm} = \lambda_q
+ (1-\lambda_q)
\frac{m_v^2}{m_v^2+Q^2} + c_\pm \frac{M_h^2 Q^2}{(M_h^2+Q^2)^2},
\nonumber \\
& &
\hspace{-1cm}
f_{1 0} = \lambda_q
+ (1-\lambda_q)
\frac{m_\phi^2}{m_\phi^2+Q^2} + c_0 \frac{M_h^2 Q^2}{(M_h^2+Q^2)^2},
\nonumber \\
& &
\hspace{-1cm}
f_{2 \pm} = \kappa_\pm
\left\{
d_\pm  \frac{m_v^2}{m_v^2+Q^2} + (1-d_\pm)
\frac{M_h^2 }{M_h^2+Q^2} \right\}, \nonumber \\
& &
\hspace{-1cm}
f_{2 0} = \kappa_s
\left\{
d_0  \frac{m_\phi^2}{m_\phi^2+Q^2} + (1-d_0)
\frac{M_h^2}{M_h^2+Q^2}  \right\},
\label{eqQff}
\ea
where $m_v,m_\phi$ and $M_h$ are the masses respectively
corresponding to the light vector meson $m_v \simeq m_\rho$,
the $\phi$ meson (associated with an $s \bar s$ state),
and an effective heavy meson
with mass $M_h= 2 M_N$ to represent
the short range phenomenology.
For the isoscalar component it should be
$m_v=m_\omega$, but we neglect the small mass difference
between the $\rho$ and $\omega$ mesons, and use $m_\rho$.
The coefficients $c_0,c_\pm$ and $d_0,d_\pm$ were
determined in the previous studies for nucleon
(model II)~\cite{Nucleon} and $\Omega^-$~\cite{Omega}.
The values are, respectively,
$c_+= 4.160$, $c_-= 1.160$, $d_+=d_-=-0.686$,
$c_0=4.427$ and $d_0=-1.860$~\cite{Omega}.
The constant $\lambda_q=1.22$ represents
the quark density number in deep inelastic
scattering~\cite{Nucleon}.

The isovector $\kappa_+=2 \kappa_u-\kappa_d$
and isoscalar $\kappa_-=\sfrac{1}{3}(2\kappa_u+ \kappa_d)$
nucleon anomalous magnetic moments were adjusted
differently in the past
based on the other studies:
nucleon elastic form factors~\cite{Nucleon}
or octet baryon magnetic moments~\cite{Octet}.
In Ref.~\cite{Nucleon} the $u$ and $d$
quark anomalous magnetic moments were fixed to describe
the nucleon (proton and neutron) magnetic moments
using the models without the
contributions from the pion cloud at $Q^2=0$.
More recently, these parameters were updated
in a model for the octet baryons with the pion cloud
to reproduce the octet baryon magnetic moments~\cite{Octet}.
That model was not constrained by
the finite $Q^2$ data,
but only by the $Q^2=0$ data.
In the present study because we want to
describe also the finite $Q^2$ data,
we need to relax the conditions to
fix $\kappa_u$ and $\kappa_d$.
The quark anomalous magnetic moments
will then be constrained by the
physical and lattice data.
As for $\kappa_s$ (the strange quark
anomalous magnetic moment) it was fixed as
$\kappa_s=1.462$, to reproduce the
$\Omega^-$ magnetic moment~\cite{Omega}.


The quark form factors parametrized by
a VMD mechanism in Eq.~(\ref{eqQff}) are particularly
convenient to extend the model to the lattice QCD regime,
because they are written in terms of the vector meson
and nucleon masses.
Then, the extension can be done
replacing these masses by those of the lattice regime.

\subsection{Scalar wave functions}

Using the expressions (\ref{eqF1}) and (\ref{eqF2}),
we can determine all the electromagnetic form factors of the octet baryons.
To do so, we need to specify the scalar wave functions $\psi_B$.
Generalizing the
form used previously for the nucleon~\cite{Nucleon},
we assume the scalar wave functions for the octet baryons in the form:
\ba
\psi_N(P,k) &=&
\frac{N_N}{m_D(\beta_1 + \chi_{_N})(\beta_2+\chi_{_N})},
\label{eqPsiSN}\\
\psi_\Lambda(P,k) &=&
\frac{N_\Lambda}{m_D(\beta_1 + \chi_{_\Lambda})
(\beta_3+\chi_{_\Lambda})}, \\
\psi_\Sigma(P,k) &=&
\frac{N_\Sigma}{m_D(\beta_1 + \chi_{_\Sigma})
(\beta_3+\chi_{_\Sigma})}, \\
\psi_\Xi(P,k) &=&
\frac{N_\Xi}{m_D(\beta_1 + \chi_{_\Xi})
(\beta_4+\chi_{_\Xi})},
\label{eqPsiSXi}
\ea
where $N_B$
($B=N,\Lambda,\Sigma,\Xi$) are the normalization constants, and
\be
\chi_{_B} = \frac{(M_B-m_D)^2-(P-k)^2}{M_B m_D}.
\ee
Note that, except for the masses, the
$\Lambda$ and $\Sigma$ scalar wave functions are the same.
The normalization constants $N_B$
are determined by
\be
\int_k | \psi_B(\bar P,k)|^2=1,
\ee
where $\bar P=(M_B,0,0,0)$ is the baryon four-momentum at
its rest frame.

In Eqs.~(\ref{eqPsiSN})-(\ref{eqPsiSXi}) the parameters
$\beta_i$ ($i=1,..,4$) define the momentum range
in units of $m_D$.\footnote{In the baryon $B$ rest frame, it reduces to
$$
\beta_i + \chi_{_B} = \left(\beta_i -2\right) + 2 \frac{E_D}{m_D}.
$$}
In the scalar wave functions (\ref{eqPsiSN})-(\ref{eqPsiSXi})
with the assumption $\beta_2,\beta_3,\beta_4 > \beta_1$,
we associate $\beta_1$ with the long range scale
(low momentum range) that is common to all the octet baryon members,
and the remaining $\beta_i$ with the shorter spatial range scale.
The parameters $\beta_1$ and $\beta_2$ can be
determined by the nucleon data only (no strange quarks)
in a model without the pion cloud.
(See e.g., Ref.~\cite{Nucleon}.)
The parameters $\beta_3$ and $\beta_4$ are associated with the strange quark.
While $\beta_3$ is related to the system with one strange quark
and  $\beta_4$ is related to that of the two strange quarks.
Since the strange quark is heavier than the $u$ and $d$
quarks, we can expect $\beta_2 > \beta_3 > \beta_4$.
The parameters $\beta_i$ ($i=1,..,4$)
will be determined later.

\subsection{Extension of the model for the lattice regime}
\label{secLatticeExt}

We now discuss the extension of the
model for the lattice regime.
In this regime the mass of the octet baryon
$B$ is characterized by the pion mass in
lattice $m_\pi^{latt}$ and
denoted by $M_B^{latt}$.
The current $j_q^\mu$ [Eq.~(\ref{eqJi})] is
also characterized by the corresponding pion mass
in terms of the two components
$j_1$ and $j_2$, which are represented
based on the VMD parametrization in Eq.~(\ref{eqQff}).

In the quark current (\ref{eqJi})
we replace the coefficient of the
Pauli form factor $1/(2 M_N)$
by $1/(2 M_N^{latt})$ in the lattice regime.
As for the quark form factors,
we use Eq.~(\ref{eqQff})
with the meson masses replaced by
the respective lattice masses
as in the previous studies~\cite{Omega,Lattice,LatticeD}.
Namely, we replace $m_v$ by the lattice $\rho$ mass $m_\rho^{latt}$,
and the effective heavy meson mass of $M_h = 2 M_N$ by $2 M_N^{latt}$.
On the other hand, the physical mass is used for $m_\phi$,
since presently lattice simulations are
performed using the physical strange quark mass.
To represent the $\rho$ meson mass in the lattice regime,
we use the following expression based on the lattice studies
made in Ref.~\cite{Leinweber01},
\be
m_\rho^{latt}= a_0 + a_2 \left(m_\pi^{latt}\right)^2,
\label{eqMrho}
\ee
where $a_0=0.766$ GeV and $a_2= 0.427$ GeV$^{-1}$.
With this procedure we can define unambiguously the quark current.

As for the wave function $\Psi_B$,
$M_B$ is replaced by $M_B^{latt}$ in
the lattice regime.
This applies for the scalar wave functions
(\ref{eqPsiSN})-(\ref{eqPsiSXi}).
There is no need of modifying
the diquark mass in the lattice regime,
since the electromagnetic form factors are
independent of it.
We assume that the range parameters $\beta_i$ ($i=1,..,4$)
are independent of the baryon masses
and therefore independent of the lattice pion mass $m_\pi^{latt}$.
We can expect this approximation to work
for a certain range of the $m_\pi^{latt}$ values, but it breaks down
for the larger values of $m_\pi^{latt}$~\cite{Lattice}.

Using the model extended to the lattice regime, namely, using
the quark currents and baryon wave functions in the lattice regime,
we can calculate the form factors $F_{1B}$ and $F_{2B}$ in the lattice regime
via Eqs.~(\ref{eqF1}) and~(\ref{eqF2}),
using the lattice regime masses corresponding to $m_\pi^{latt}$.
However, note that the results include only
the valence quark contributions,
but not the pion cloud contributions.

\subsection{Pion cloud factors}
\label{secPC}

To describe the pion cloud contributions
for the octet baryon electromagnetic form factors,
we use the following parametrization:
\ba
& & \tilde B_1= B_1\left(\frac{\Lambda_{11}^2}{\Lambda_{11}^2 + Q^2} \right)^4,
\label{eqB1}
\\
& & \tilde C_1= B_1\left(\frac{\Lambda_{12}^2}{\Lambda_{12}^2 + Q^2} \right)^2,
\\
& & \tilde D_1= D_1^\prime \frac{Q^2\Lambda_{13}^4}{(\Lambda_{13}^2 + Q^2)^3},
\\
& & \tilde B_2= B_2\left(\frac{\Lambda_{21}^2}{\Lambda_{21}^2 + Q^2} \right)^5,
\\
& & \tilde C_2= C_2\left(\frac{\Lambda_{22}^2}{\Lambda_{22}^2 + Q^2} \right)^3,
\\
& & \tilde D_2= D_2\left(\frac{\Lambda_{23}^2}{\Lambda_{23}^2 + Q^2} \right)^3.
\label{eqD2}
\ea
The parametrization is phenomenological
and motivated by the expected falloff
of the quark-antiquark contributions
at very large $Q^2$~\cite{Brodsky75},
as well as the magnitude of the pion cloud
contributions estimated for the
$\gamma N \to \Delta$ reaction~\cite{NDelta,NDeltaD,LatticeD}.
The pion cloud contributions should falloff by a factor $1/Q^4$
faster than the falloff of the valence quark contributions.
In principle, $\tilde B_1$ and $\tilde B_2$
should contain the pion electromagnetic form factor
$F_\pi (Q^2)\approx \left(1+ \sfrac{Q^2}{0.5}\right)^{-1}$,
but we adopt simplified functions
using just one cutoff parameter
with the powers 4 and 5, respectively.

In the above $B_1$ ($=C_1$), $B_2$, $C_2$ and $D_2$
represent the values of the respective functions at $Q^2=0$.
The values were determined previously
by the octet baryon magnetic moments in Ref.~\cite{Octet}.
$D_1^\prime$ is a new constant defined as
$D_1^\prime = \sfrac{1}{\Lambda_{13}^2}\sfrac{d D_1}{d Q^2}(0)$.
However, we do not use the values determined in Ref.~\cite{Octet}
in this study.
Instead, we will determine them directly by the
nucleon form factor data at finite $Q^2$,
and by the $\Lambda$, $\Sigma$ and $\Xi$ physical magnetic moment data,
after determining the contributions from the valence quarks.
The adjustable parameters of our
pion cloud parametrization are then,
the coefficients for $Q^2=0$, and
the cutoffs $\Lambda_{1i}$ and $\Lambda_{2i}$ ($i=1,2,3$).
But for simplicity, we use and vary only two independent cutoff values
for the Dirac and Pauli form factors,
respectively $\Lambda_1$ and $\Lambda_2$ in this study.

\subsection{Separating the pion cloud contributions}

Based on the discussions in the previous sections,
the baryon form factors
$F_{iB}$ ($i=1,2$) may be decomposed into
\be
F_{iB}(Q^2)= Z_B \left[  F_{i0B}(Q^2) + \delta F_{iB} (Q^2)\right],
\ee
where $F_{10B}= \tilde e_{0B}$, $F_{20B} = \tilde \kappa_{0B}$
and corresponding $\delta F_{1,2 B}$ the pion cloud contributions,
which are given in Eqs.~(\ref{eqF1p})-(\ref{eqF2XiM}).
We may regard that $Z_{B} F_{i0B}$ represent
the effects of the valence quarks, and $Z_{B} \delta F_{iB}$
those of the pion cloud.
The same decomposition can be applied for
the electric and magnetic form factors:
\ba
& &
G_{E B}(Q^2)= Z_B \left[ G_{E0B}(Q^2) + \delta G_{E B}(Q^2)\right], \nonumber \\
& &
G_{M B}(Q^2)= Z_B \left[ G_{M0B} (Q^2)+ \delta G_{M B}(Q^2)\right],
\label{eqGX1}
\ea
where $G_{E0B}=  \tilde e_{0B} - \tau \tilde \kappa_{0B}$,
$G_{M0B}=  \tilde e_{0B} + \tilde \kappa_{0B}$,
$\delta G_{E B}= \delta F_{1B} -\tau \delta F_{2B}$, and
$\delta G_{M B}= \delta F_{1B} + \delta F_{2B}$.
In this case $Z_B \delta G_{EB}$, and $Z_B \delta G_{MB}$ reflect
the dressing of the pion cloud.
To estimate the pion cloud contributions,
we will compare the full result,
$G_{EB}$ or $G_{MB}$, with the total contributions
of the valence quark core,
$Z_B G_{E0B}$ or $Z_B G_{M0B}$.
The difference is
the pion cloud contributions, $Z_B \delta G_{EB}$ or $Z_B \delta G_{MB}$.
Hereafter, we will use $G_{XB}$ to express the electric
form factor $G_{EB}$ ($X=E$) or
the magnetic form factor $G_{MB}$ ($X=M$).

Note that we can alternatively define the contributions
from the pion cloud as the difference between
the bare form factor, $G_{E0B}$ or $G_{M0B}$,
and the full form factor,
$G_{EB}$ or $G_{MB}$, instead of
$Z_B G_{E0B}$ or $Z_B G_{M0B}$ described above.
We will refer to these terms as the {\it effective} pion cloud contributions.
By this definition we may have small {\it effective} pion cloud contributions
in the cases, e.g., where
$Z_B \delta G_{EB}$ or $Z_B \delta G_{MB}$
are significant\footnote{The
{\it effective} pion cloud contributions for the form
factor $G_{XB}$ is $Z_B \delta G_{XB}- (1-Z_B) G_{X0B}$.
If $Z_B$ differs substantially from 1,
the pion cloud contributions $Z_B \delta G_{XB}$
will be modified by the term $- (1-Z_B) G_{X0B}$,
and large cancellation can happen.}.
Such an example is the model in Ref.~\cite{Octet}.
There the bare contributions for the nucleon magnetic
form factor at $Q^2=0$, given by $Z_N G_{M 0 N}$,
was about 60\% (namely, the pion cloud contributions were $\approx 40\%$),
although the {\it effective} pion cloud contributions were only 5\%.

\begin{figure*}[t]
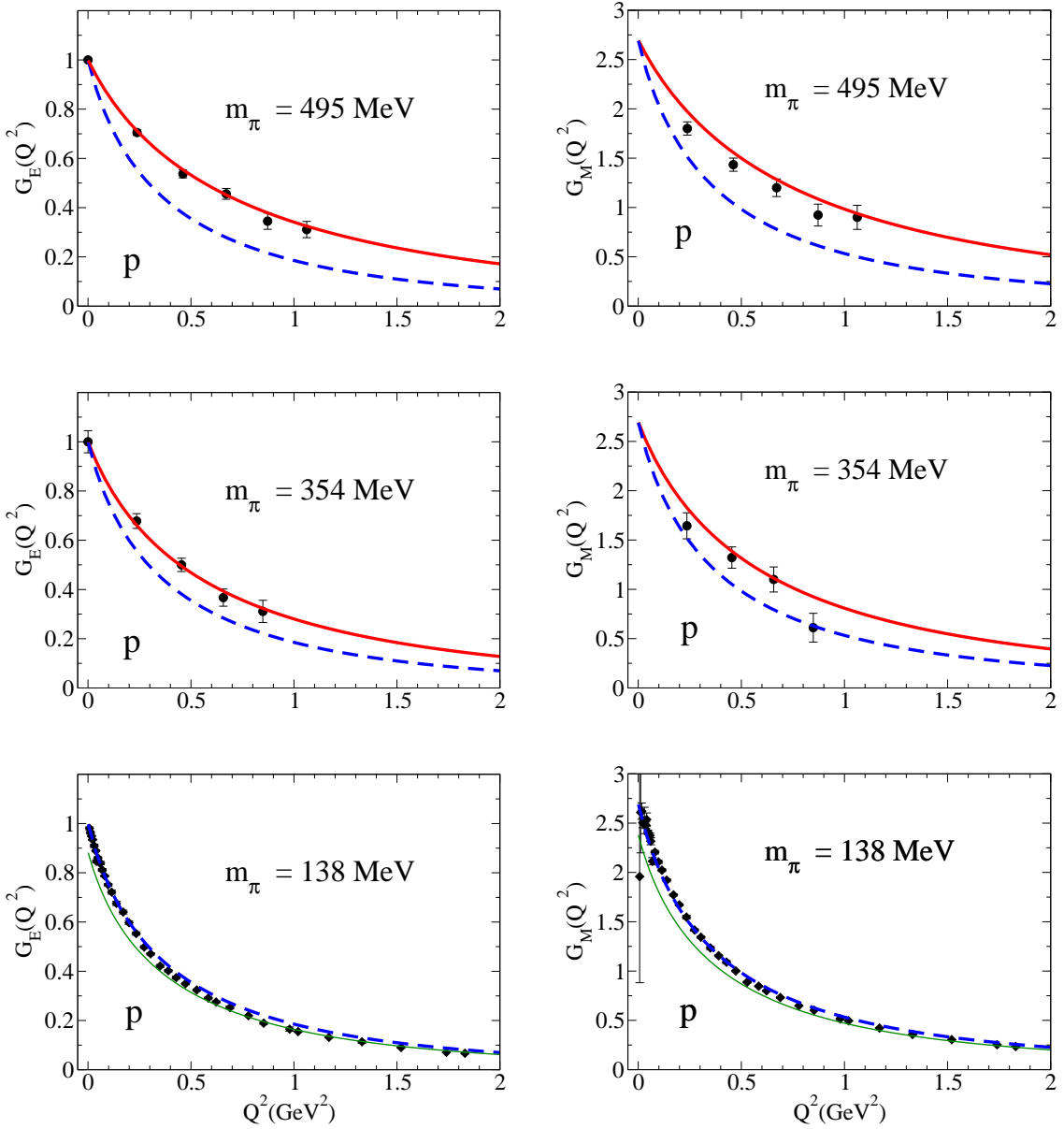

\centerline{
\mbox{
\includegraphics[width=2.8in]{GEp495_5B.eps} \hspace{.5cm}
\includegraphics[width=2.8in]{GMp495_5B.eps}
}}
\vspace{0.75cm}
\centerline{
\mbox{
\includegraphics[width=2.8in]{GEp354_5B.eps} \hspace{.5cm}
\includegraphics[width=2.8in]{GMp354_5B.eps}
}}
\vspace{0.75cm}
\centerline{
\mbox{
\includegraphics[width=2.8in]{GEp138_5C.eps} \hspace{.5cm}
\includegraphics[width=2.8in]{GMp138_5C.eps}
}}
\caption{\footnotesize{
Bare electromagnetic form factors for the proton determined by the global fit,
compared with the lattice data from Ref.~\cite{Lin09}.
The lines are the lattice regime (solid line) and the physical regime (dashed line).
For $m_\pi=138$ MeV, we include the physical data.
The thin solid lines in the bottom panels do not include
the pion cloud contributions.
}}
\label{figProtonLatt}
\end{figure*}

\begin{figure*}[t]
\vspace{0.5cm}
\centerline{
\mbox{
\includegraphics[width=2.8in]{GEn495_5B.eps} \hspace{.5cm}
\includegraphics[width=2.8in]{GMn495_5B.eps}
}}
\vspace{0.75cm}
\centerline{
\mbox{
\includegraphics[width=2.8in]{GEn354_5B.eps} \hspace{.5cm}
\includegraphics[width=2.8in]{GMn354_5B.eps}
}}
\vspace{0.75cm}
\centerline{
\mbox{
\includegraphics[width=2.8in]{GEn138_5C.eps} \hspace{.5cm}
\includegraphics[width=2.8in]{GMn138_5C.eps}
}}
\caption{\footnotesize{
Bare electromagnetic form factors for the neutron determined by the global fit,
compared with the lattice data from Ref.~\cite{Lin09}.
The lines are the lattice regime (solid line)
and the physical regime (dashed line).
For $m_\pi=138$ MeV, we include the physical data.
The thin solid lines in the bottom panels do not include
the pion cloud contributions.
}}
\label{figNeutronLatt}
\end{figure*}

\begin{figure*}[t]
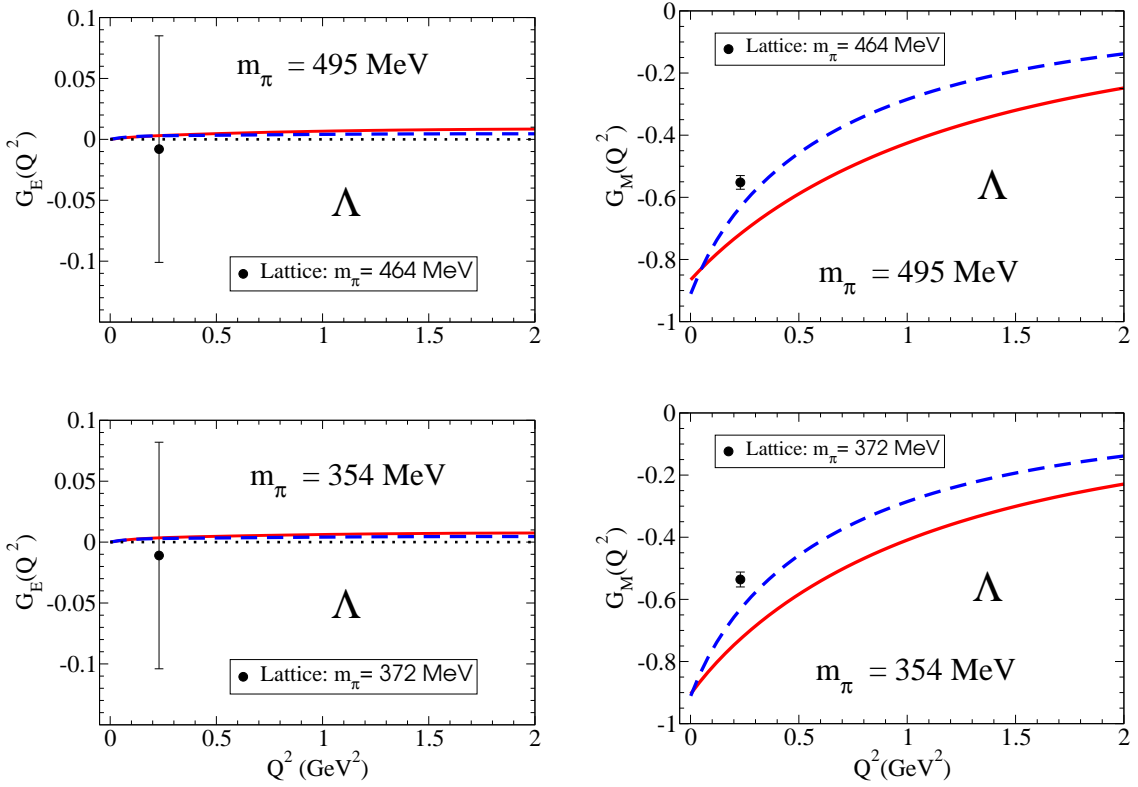

\centerline{
\mbox{
\includegraphics[width=2.8in]{GELamb495_5B.eps} \hspace{.5cm}
\includegraphics[width=2.8in]{GMLamb495_5B.eps}
}}
\vspace{0.75cm}
\centerline{
\mbox{
\includegraphics[width=2.8in]{GELamb354_5B.eps} \hspace{.5cm}
\includegraphics[width=2.8in]{GMLamb354_5B.eps}
}}
\caption{\footnotesize{
$\Lambda$ bare electromagnetic form factors determined by the global fit.
The lines are the lattice regime (solid line) and
the physical regime (dashed line).
For the lattice regime calculations, we use the value
$M_\Lambda= M_\Sigma$, where $M_\Sigma$ is
the $\Sigma$ mass from Table~\ref{tableMasses}.
For the physical point we use the physical $\Lambda$ mass.
The lattice data are from Ref.~\cite{Boinepalli06}
for $m_\pi=372$ and $464$ MeV.}}
\label{figLambdaLatt}
\end{figure*}

\begin{figure*}[t]
\centerline{
\mbox{
\includegraphics[width=2.8in]{GEsigP495_5B.eps}   \hspace{.5cm}
\includegraphics[width=2.8in]{GMsigP495_5B.eps}
}}
\vspace{0.75cm}
\centerline{
\mbox{
\includegraphics[width=2.8in]{GEsigP354_5B.eps}  \hspace{.5cm}
\includegraphics[width=2.8in]{GMsigP354_5B.eps}
}}
\caption{\footnotesize{
$\Sigma^+$ bare electromagnetic form factors determined by the global fit.
The lines are the lattice regime (solid line) and
the physical regime (dashed line).
The lattice data are from Ref.~\cite{Lin09}.
}}
\label{figSigmaPLatt}
\end{figure*}

\begin{figure*}[t]
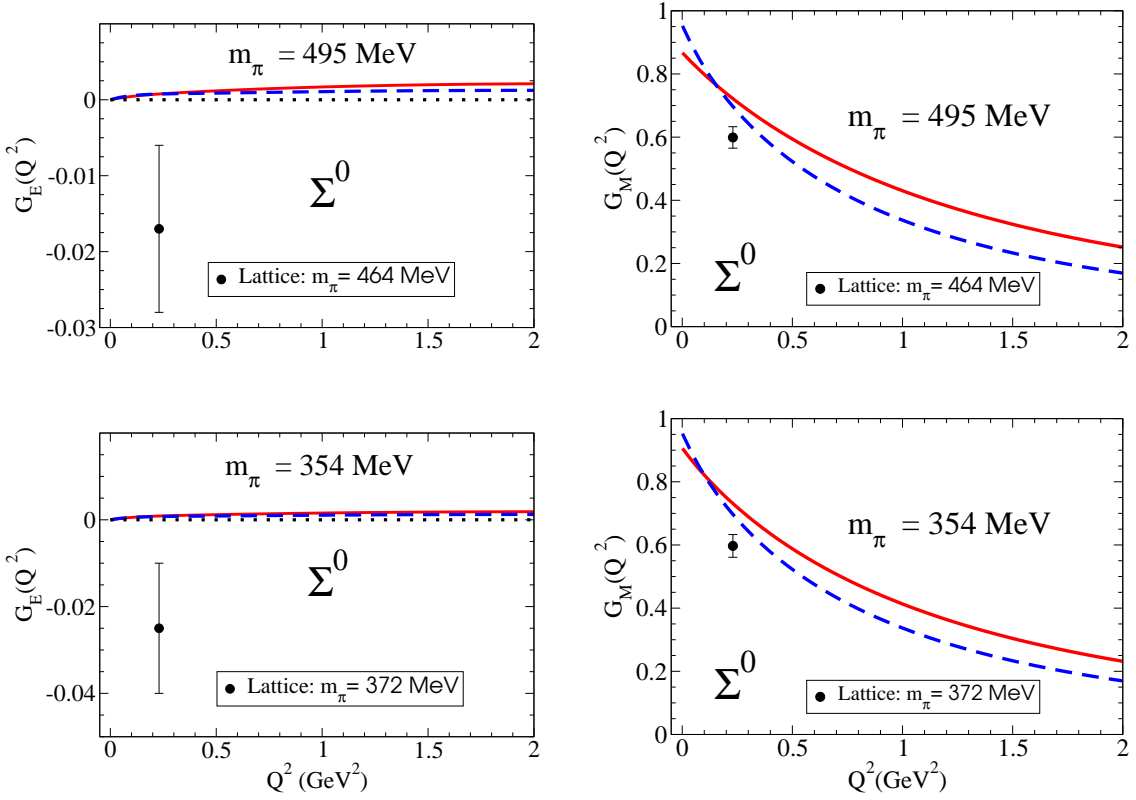

\centerline{
\mbox{
\includegraphics[width=2.8in]{GEsig0495_5B.eps}  \hspace{.5cm}
\includegraphics[width=2.8in]{GMsig0495_5B.eps}
}}
\vspace{0.75cm}
\centerline{
\mbox{
\includegraphics[width=2.8in]{GEsig0354_5B.eps}  \hspace{.5cm}
\includegraphics[width=2.8in]{GMsig0354_5B.eps}
}}
\caption{\footnotesize{
$\Sigma^0$ bare electromagnetic form factors determined by the global fit.
The lines are the lattice regime (solid line) and
the physical regime (dashed line).
The lattice data are from Ref.~\cite{Boinepalli06}
for $m_\pi=372$ and $464$ MeV.
\vspace{0.4cm}
}}
\label{figSigma0Latt}
\end{figure*}

\begin{figure*}[t]
\centerline{
\mbox{
\includegraphics[width=2.8in]{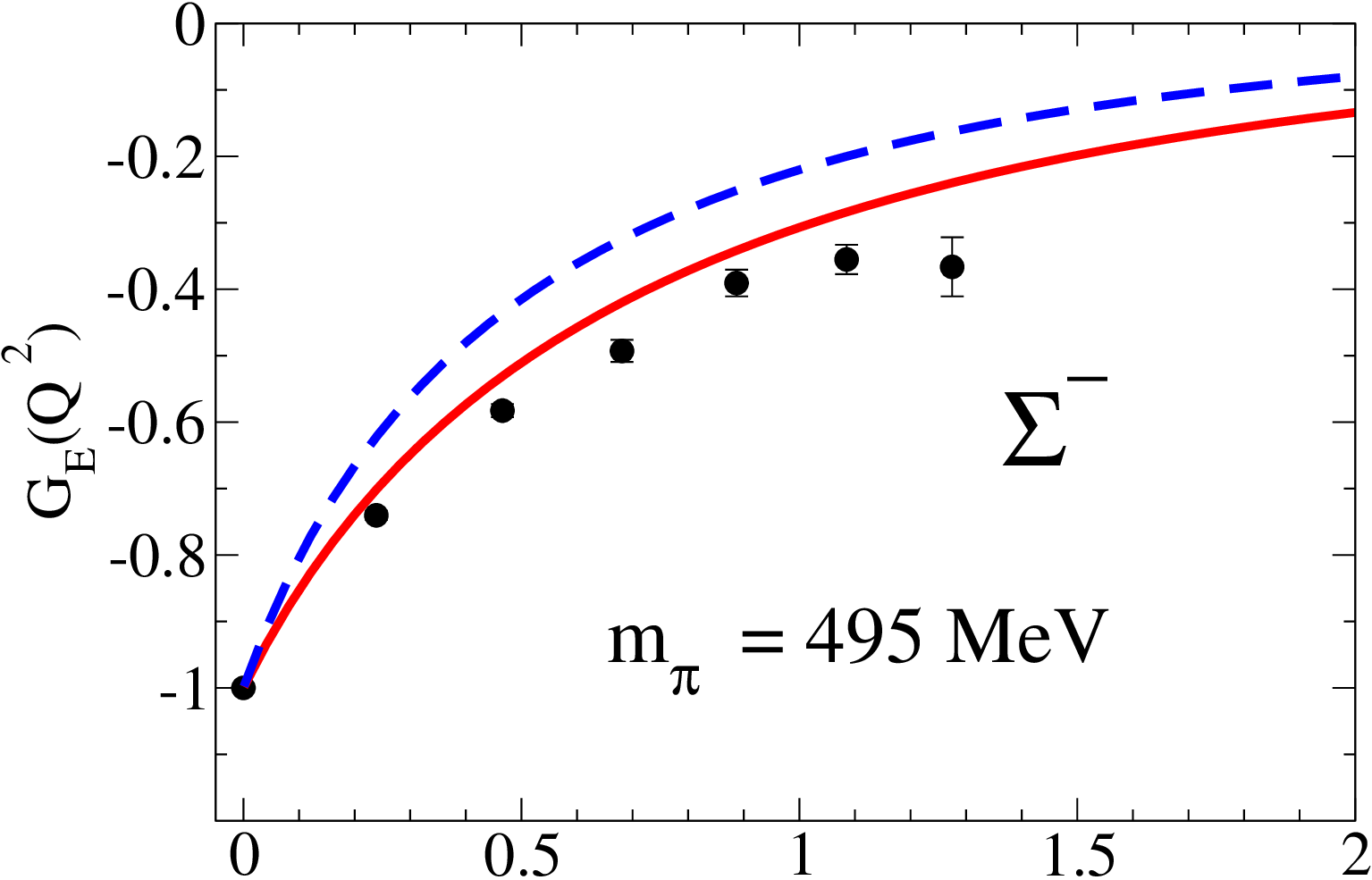} \hspace{.5cm}
\includegraphics[width=2.8in]{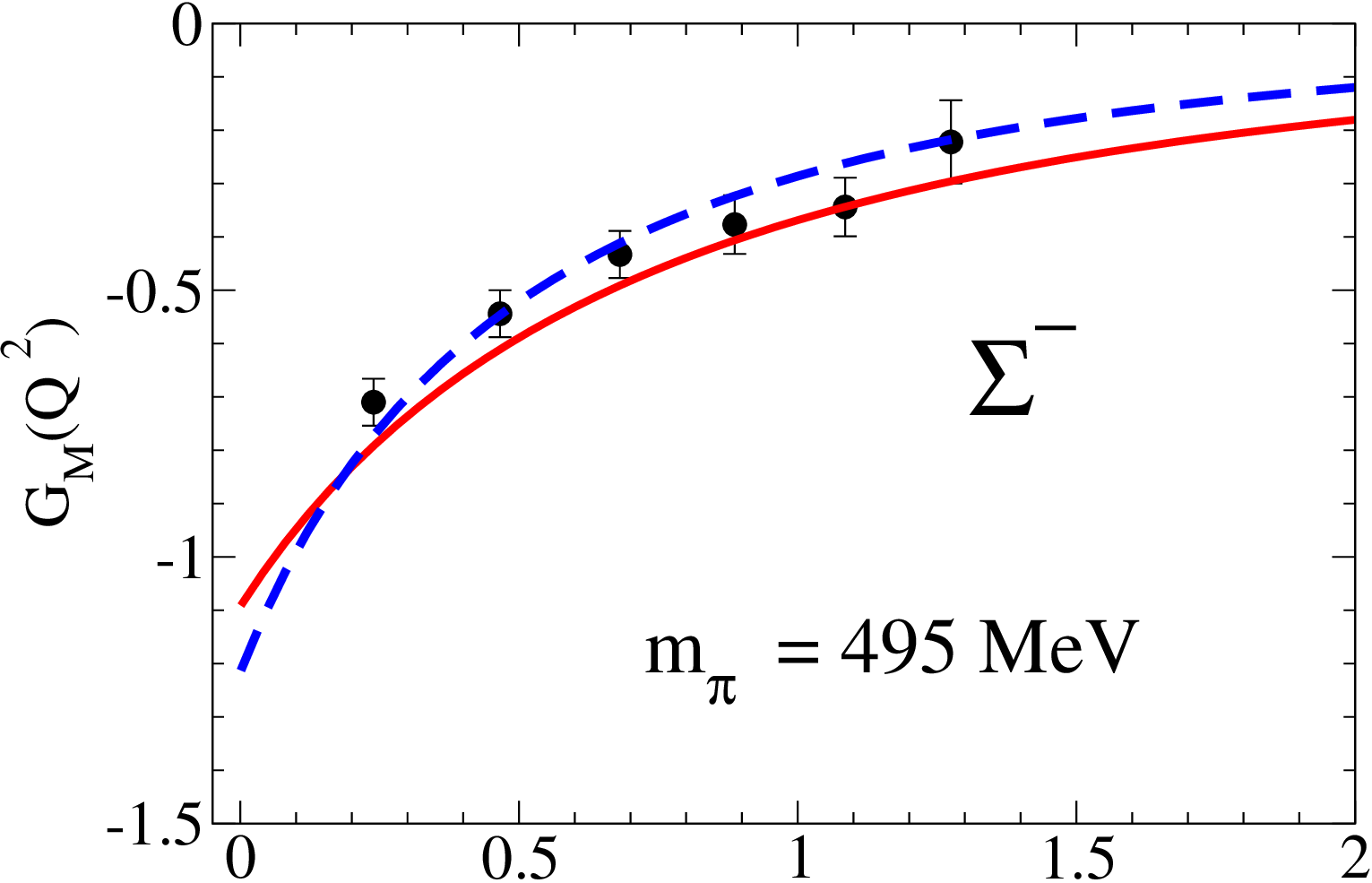}
}}
\vspace{0.75cm}
\centerline{
\mbox{
\includegraphics[width=2.8in]{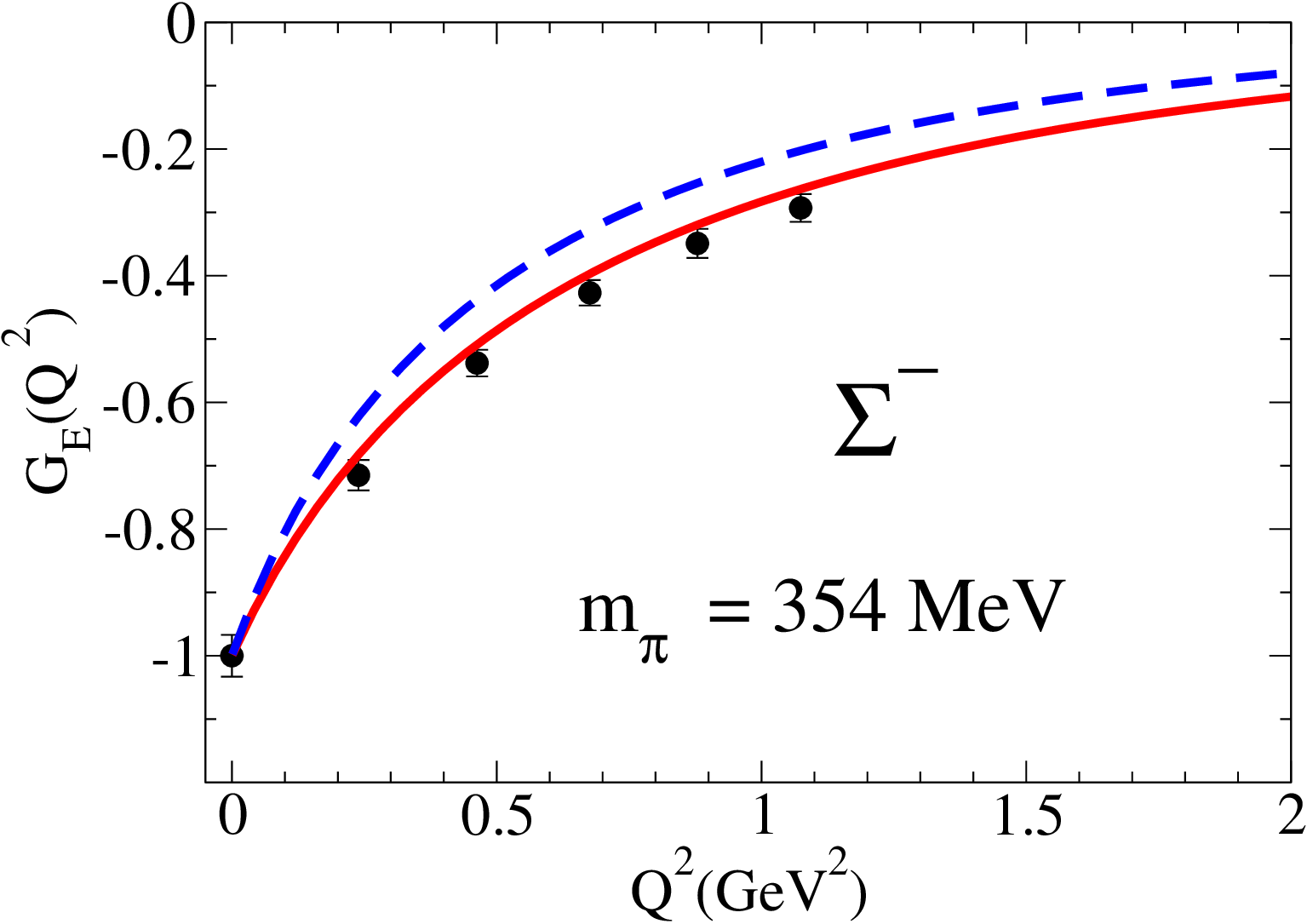} \hspace{.5cm}
\includegraphics[width=2.8in]{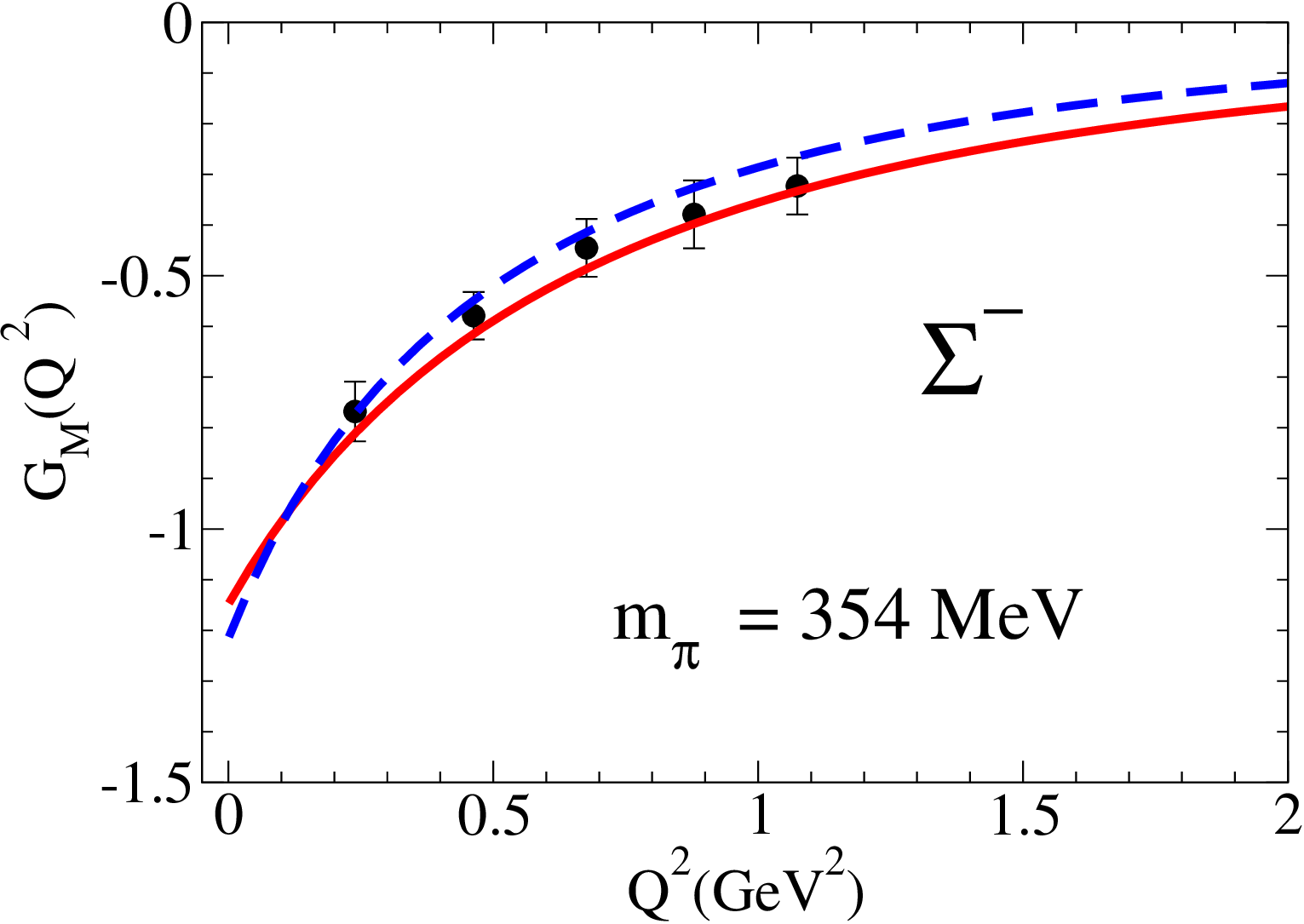}
}}
\caption{\footnotesize{
$\Sigma^-$ bare electromagnetic form factors determined by the global fit.
The lines are the lattice regime (solid line) and
the physical regime (dashed line).
The lattice data are from Ref.~\cite{Lin09}.
\vspace{0.4cm}
}}
\label{figSigmaMLatt}
\end{figure*}

\begin{figure*}[t]
\centerline{
\mbox{
\includegraphics[width=2.8in]{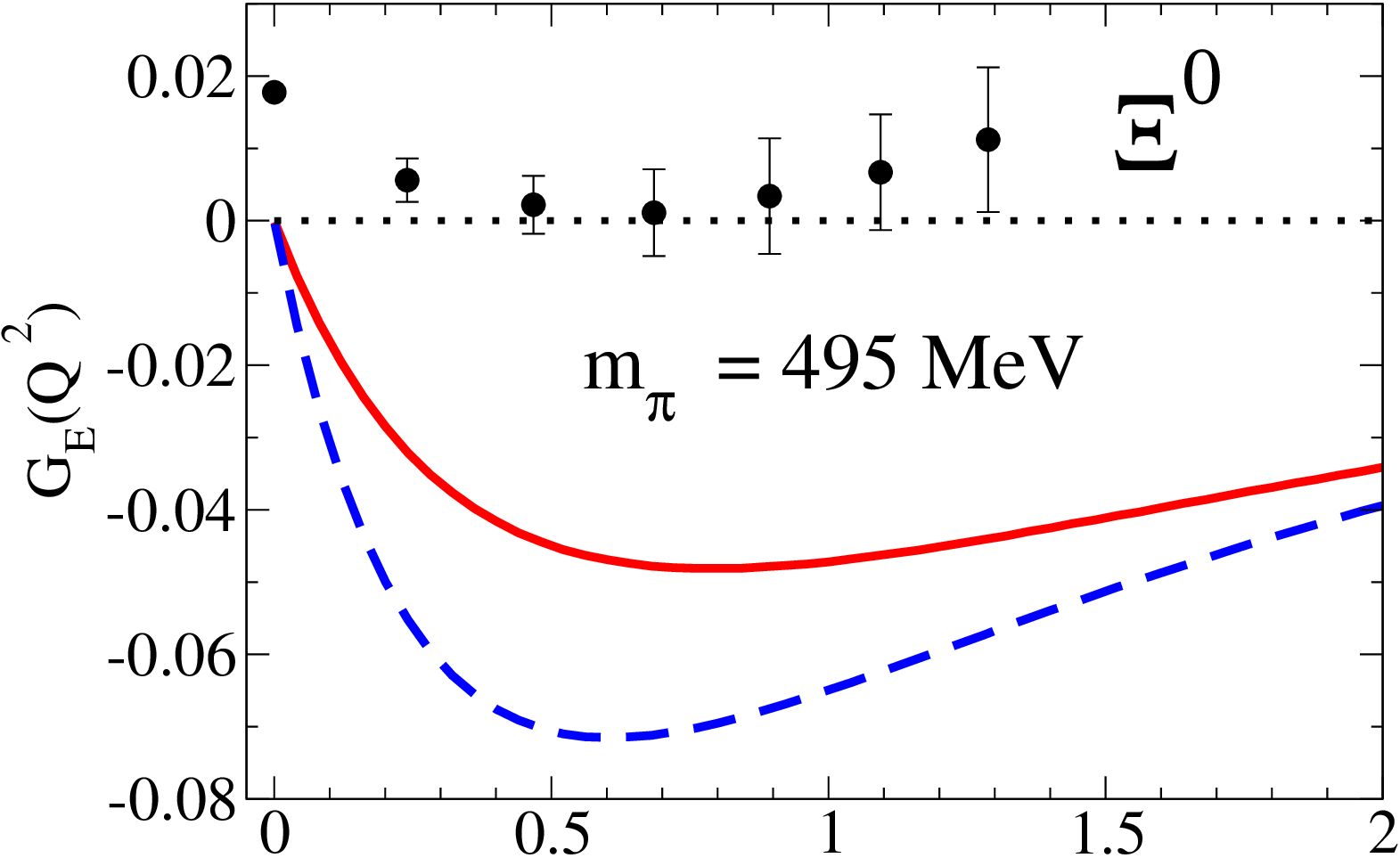} \hspace{.5cm}
\includegraphics[width=2.8in]{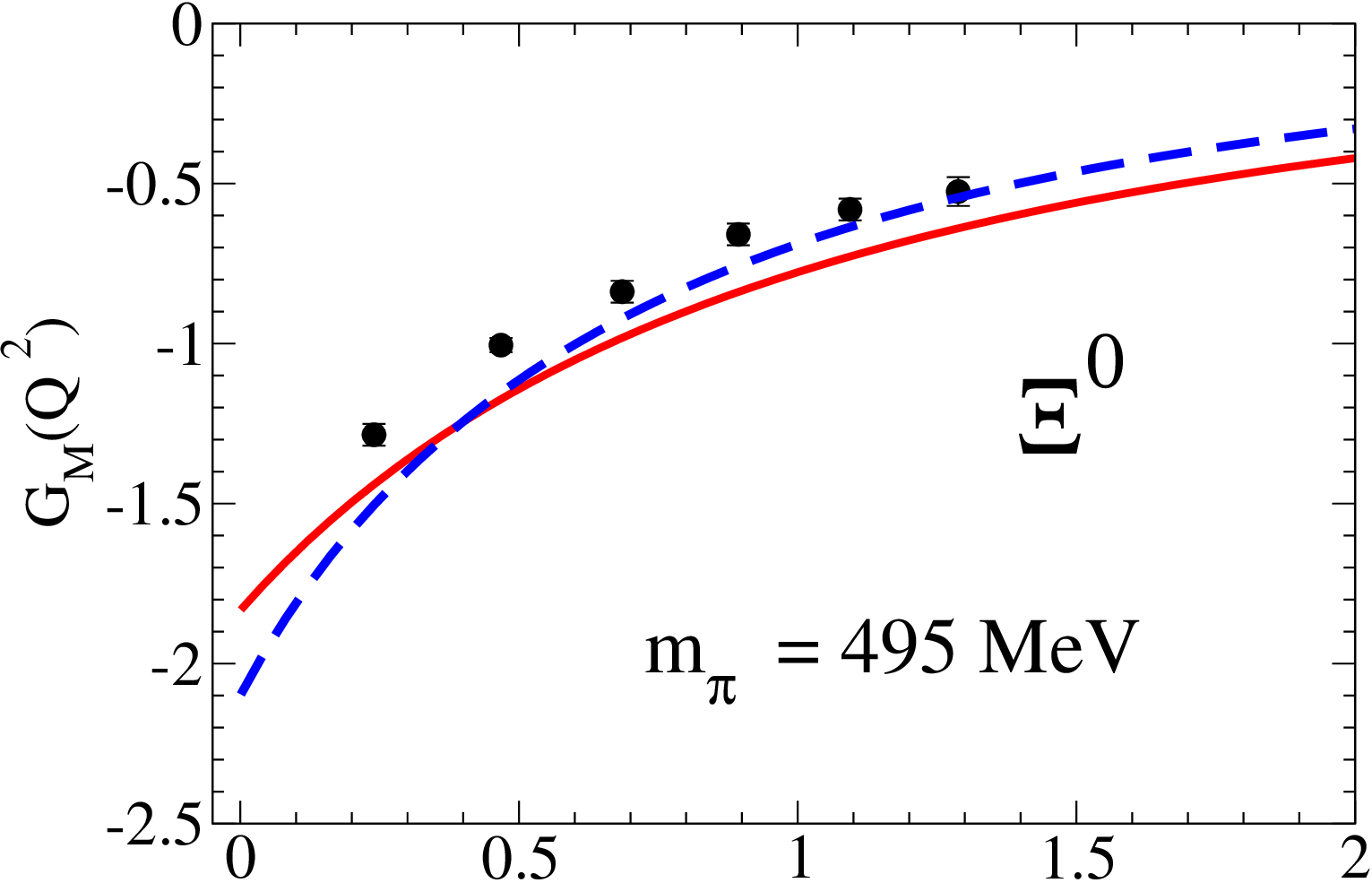}
}}
\vspace{0.75cm}
\centerline{
\mbox{
\includegraphics[width=2.8in]{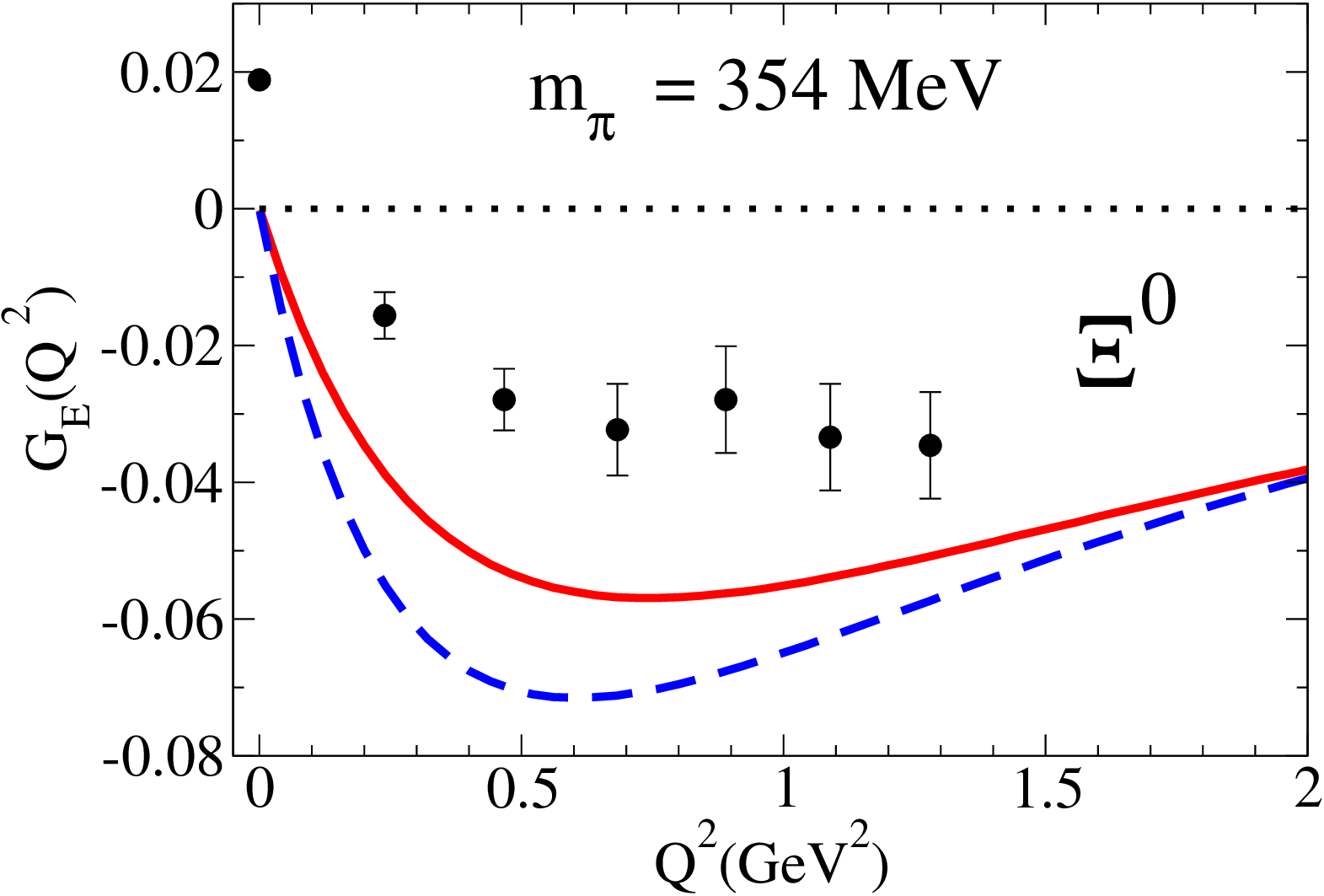} \hspace{.5cm}
\includegraphics[width=2.8in]{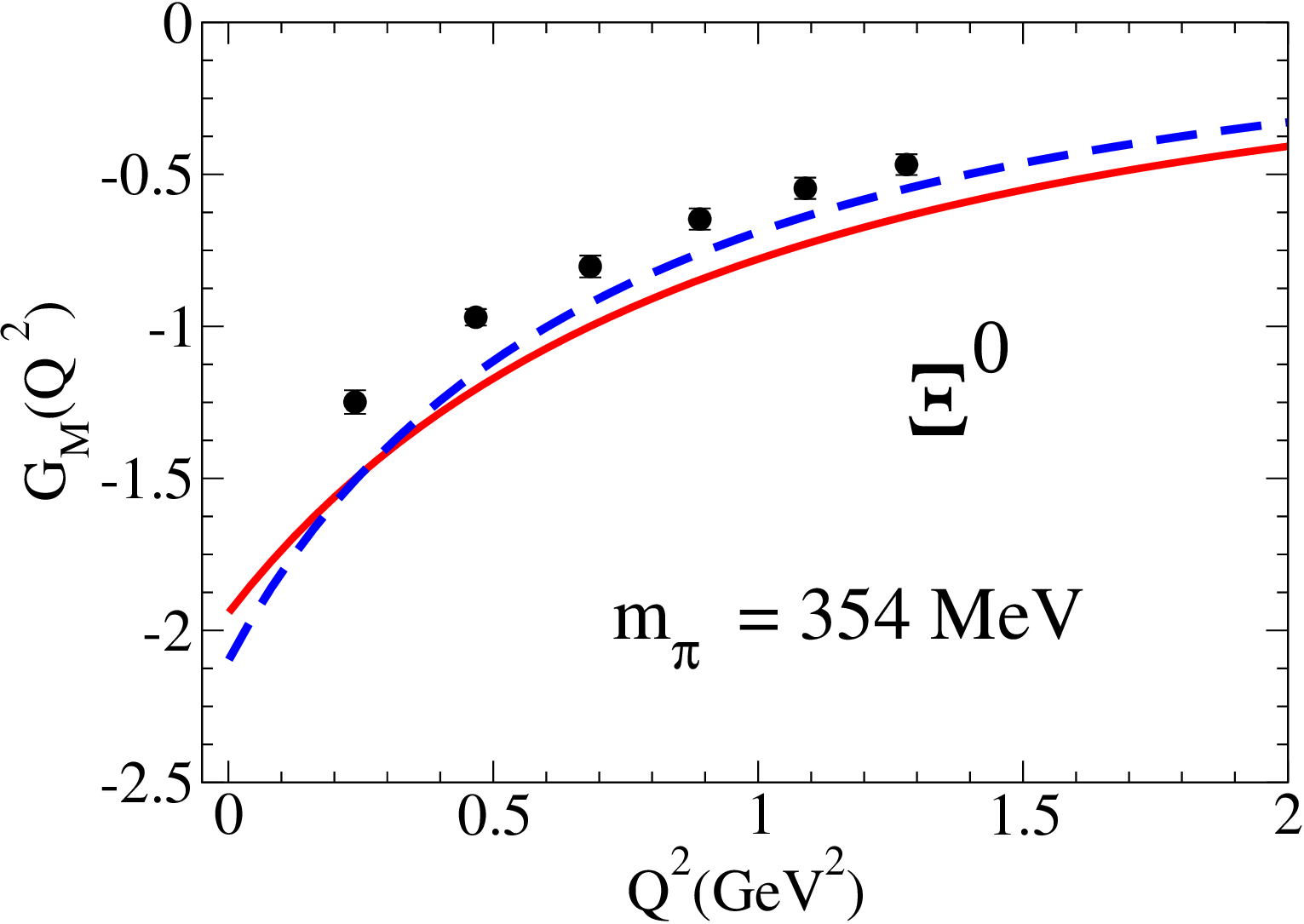}
}}
\caption{\footnotesize{
$\Xi^0$ bare electromagnetic form factors determined by the global fit.
The lines are for the lattice regime (solid line), and
the physical regime (dashed line).
The lattice data are from Ref.~\cite{Lin09}.
\vspace{0.4cm}
}}
\label{figXi0Latt}
\end{figure*}

\begin{figure*}[t]
\centerline{
\mbox{
\includegraphics[width=2.8in]{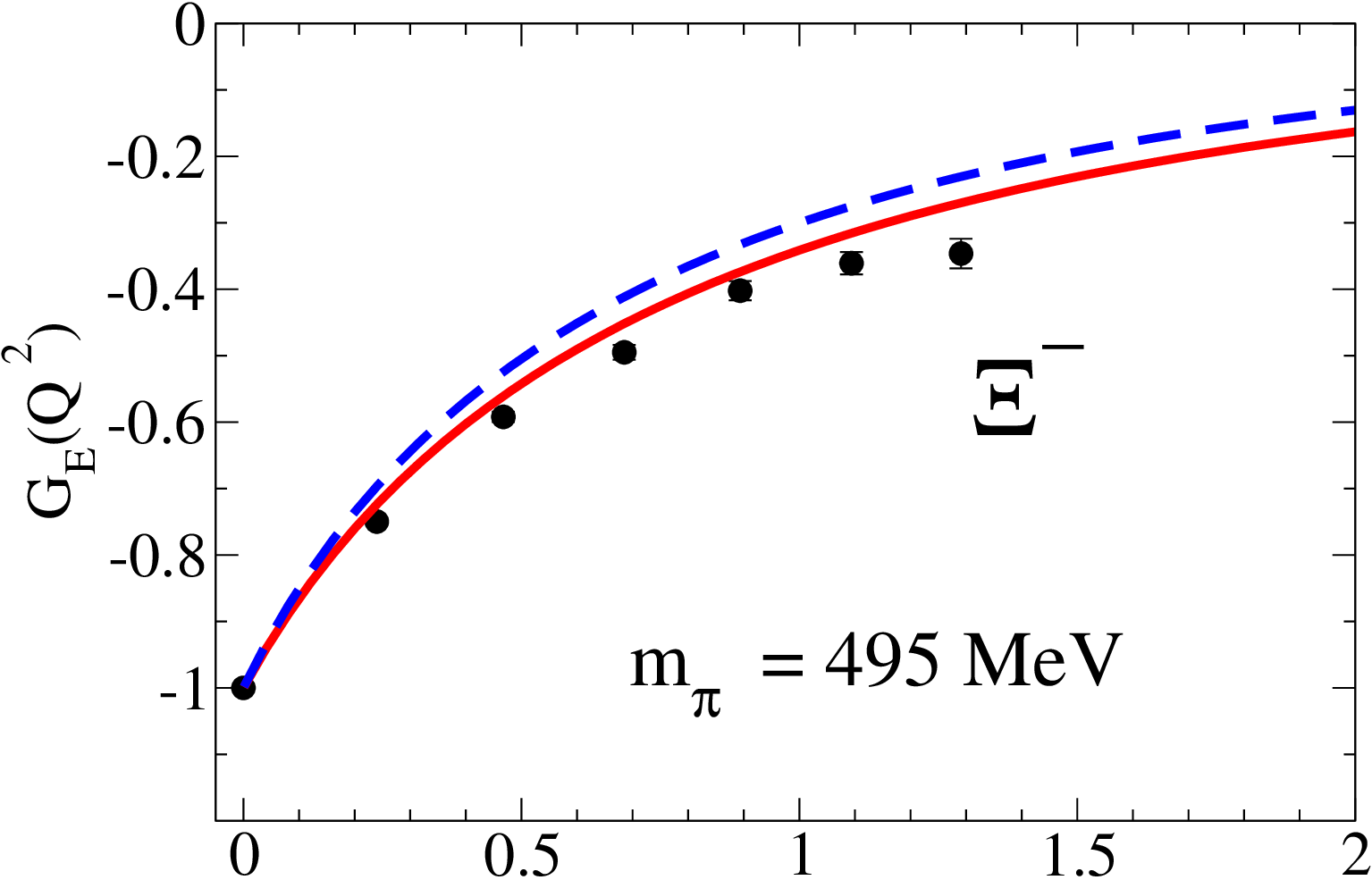} \hspace{.5cm}
\includegraphics[width=2.8in]{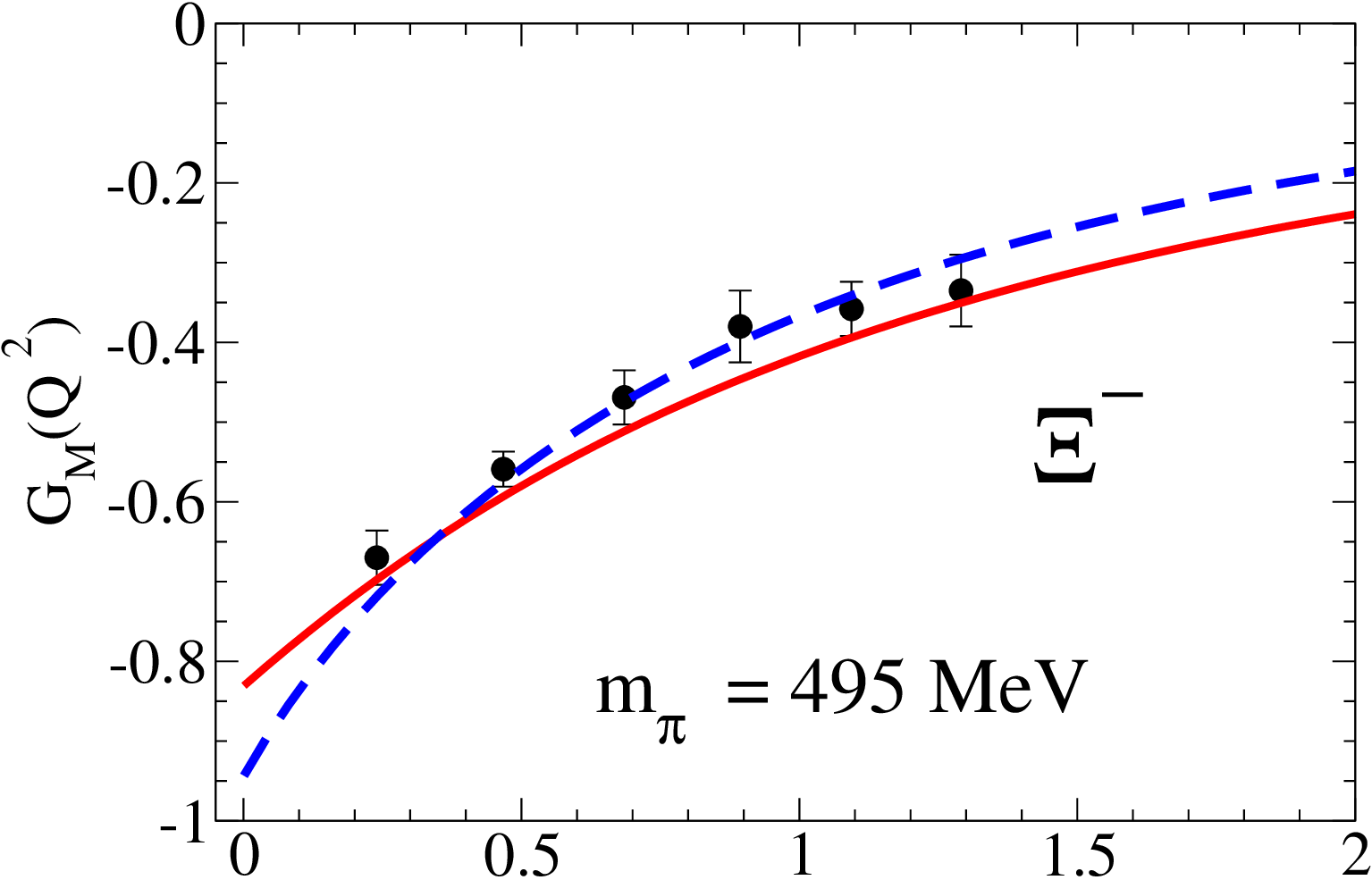}
}}
\vspace{0.75cm}
\centerline{
\mbox{
\includegraphics[width=2.8in]{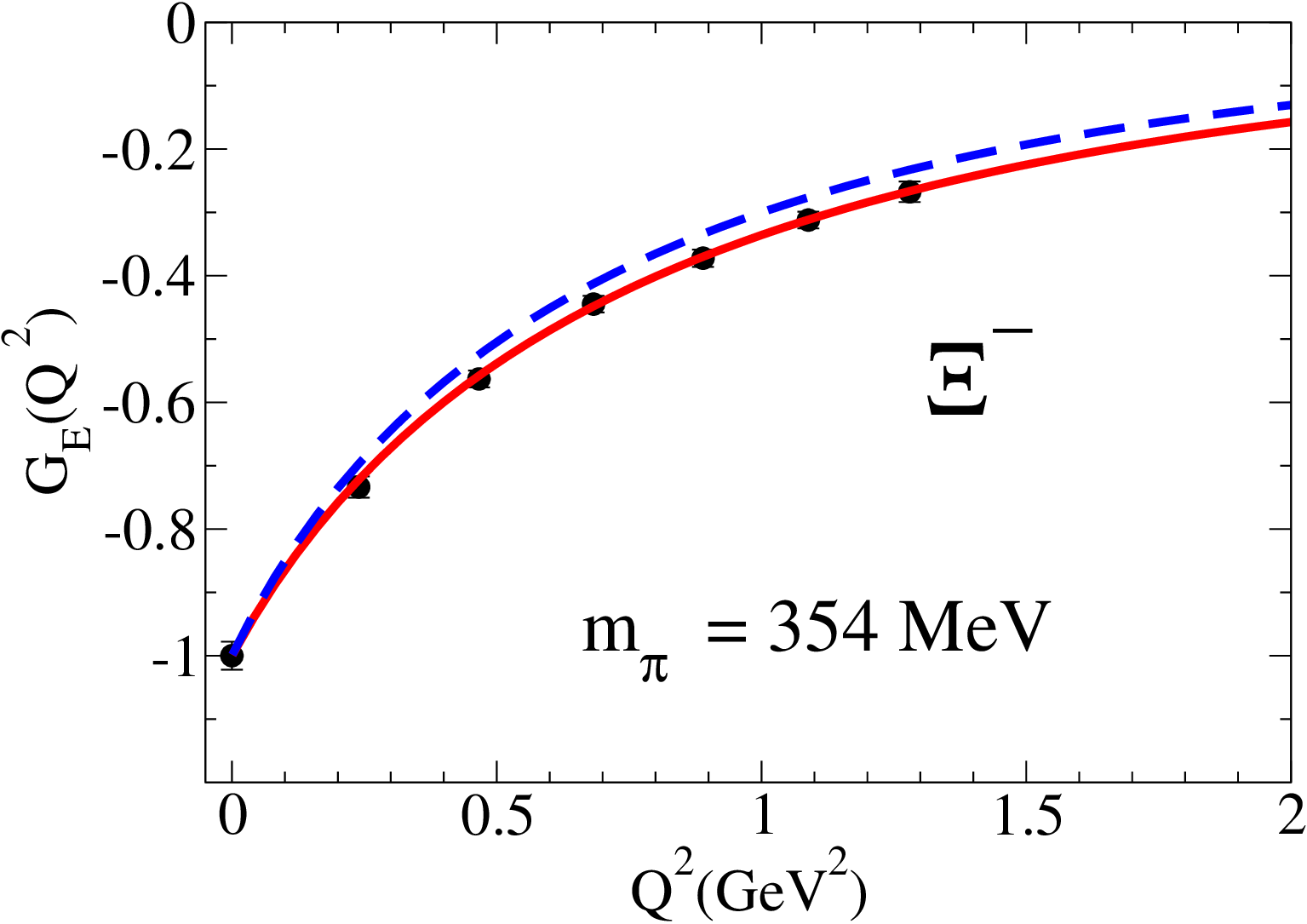} \hspace{.5cm}
\includegraphics[width=2.8in]{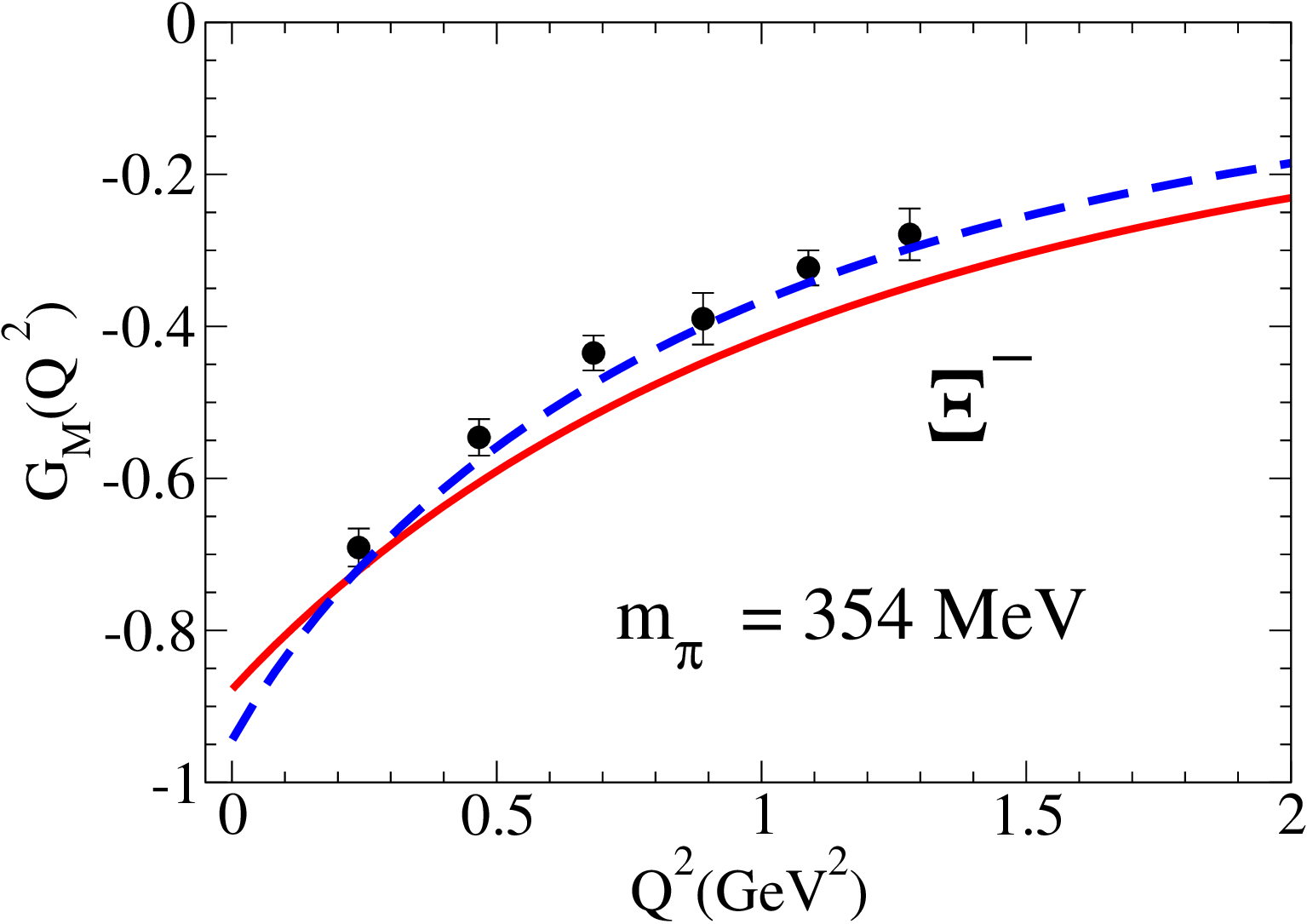}
}}
\caption{\footnotesize{
$\Xi^-$ bare electromagnetic form factors determined by the global fit.
The lines are the lattice regime (solid line), and
the physical regime (dashed line).
The lattice data are from Ref.~\cite{Lin09}.
\vspace{0.4cm}
}}
\label{figXiMLatt}
\end{figure*}


\section{Results}
\label{secResults}

In this section we determine the parameters
of the model, and then present results
for the valence quark contributions
for the octet baryon electromagnetic form factors
in the lattice regime,
as well as those for the physical regime
which includes the pion cloud effects.
While the expressions related to the valence quark contributions
have been presented in Sec.~\ref{secBareFF},
the formalism related to the pion cloud dressing
has been presented in Sec.~\ref{secDressedFF}.

The parameters associated with the
valence quark degrees of freedom and
the ones associated with the pion cloud dressing,
are determined by a global fit
to the octet baryon electromagnetic form factors,
namely, the lattice data, nucleon physical data,
and the octet baryon physical magnetic moments.
Details are described next.
Once we have fixed the relevant parameters, we will discuss
the results for the valence quark contributions and
the effects of the pion cloud dressing.

\subsection{Global fit}

In a regime where the pion cloud effects are small,
such as the lattice QCD simulations with heavy pions,
or the physical regime at high $Q^2$,
the electromagnetic form factors of the octet baryons
should be well described only by the valence
quark degrees of freedom.
Then, we can use the lattice QCD data
with large pion masses
to calibrate our model extended
to the lattice regime based on the formalism described
in Sec.~\ref{secLatticeExt},
without the pion cloud effects.

To describe the physical octet baryon systems,
only the valence quark degrees of freedom are
usually insufficient, and
explicit pion cloud effects are necessary.
Except for the magnetic moments~\cite{Octet},
there are no physical data available for the
octet baryon form factors besides the nucleon system.
Therefore, the nucleon system in the physical regime, is
ideal to study the pion cloud effects
on the form factors at finite $Q^2$.
In this case the nucleon form factors
can be described by the mixture
of the valence quark and pion cloud contributions
given by Eqs.~(\ref{eqF1p}), (\ref{eqF1n}),
(\ref{eqF2p}) and~(\ref{eqF2n}).
Also the
$\Lambda$, $\Sigma^\pm$ and $\Xi^{0,-}$
magnetic moments,
defined by  Eqs.~(\ref{eqF1Lambda})-(\ref{eqF1XiM})
and (\ref{eqF2Lambda})-(\ref{eqF2XiM}) at $Q^2=0$,
can be used to constrain the
effects of the pion cloud at $Q^2=0$.

Summarizing,
to adjust the parameters of our model,
we perform a fit to the lattice data
for the valence quark part
by extending the model to the lattice regime,
and for the pion cloud contributions
we perform a fit to the physical data of
nucleon form factors and octet baryon magnetic moments.
In the latter, the physical regime,
the expressions for the form factors
are given by extrapolating the
bare form factors to the physical case ($m_\pi =138$ MeV), and
the pion cloud contributions are parametrized
according to Eqs.~(\ref{eqB1})-(\ref{eqD2}).
Next, we describe specifically the
lattice QCD data and the nucleon physical data.
As for the magnetic moments,
we use the experimental
data for the
$\Lambda$, $\Sigma^\pm$, and $\Xi^{0,-}$~\cite{PDG}.
We do not include the neutron and proton magnetic moments
in the fit, since the nucleon magnetic form factor data
are also included in our analysis.

\subsubsection{Lattice data}

The lattice data adopted in this work,
are the octet baryon electromagnetic form factor
data from Ref.~\cite{Lin09},
where $n$, $p$, $\Sigma^\pm$, $\Xi^{0,-}$ form factors
were calculated systematically
at finite $Q^2$ for the first time.
The data from Ref.~\cite{Lin09}
are composed of four sets of
unquenched simulations associated with the
pion masses 354, 495, 591 and 680 MeV,
but restricted to the region $Q^2< 1.5$ GeV$^2$.
As a total we have 136 data points for both $G_{EB}$ and $G_{MB}$.
To extend our model (valence quark contributions)
to the lattice regime,
we follow the procedure described in Sec.~\ref{secLatticeExt}.
The relevant variables necessary are
the masses associated with the lattice QCD simulations.
The corresponding values are presented in Table~\ref{tableMasses}
according to the simulations of Ref.~\cite{Lin09},
and the $m_\rho$ values determined through Eq.~(\ref{eqMrho}).

\begin{table}[t]
\begin{center}
\begin{tabular}{c c c c c c}
\hline
\hline
$m_\pi$(GeV) & $m_\rho$(GeV) & $M_N$(GeV) & $M_\Sigma$(GeV) & $M_\Xi$(GeV) \\
\hline
    0.138 &   0.779 &  0.939  &  1.192  & 1.318 \\
\hline
    0.351 &   0.820 &  1.150  &  1.349  & 1.438 \\
    0.495 &   0.871 &  1.290  &  1.410  & 1.475 \\
    0.591 &   0.915 &  1.366  &  1.448  & 1.491 \\
    0.690 &   0.964 &  1.490  &  1.524  & 1.546 \\
\hline
\hline
\end{tabular}
\end{center}
\caption{
Masses of the octet baryons ($M_N$, $M_\Sigma$ and $M_\Xi$)
obtained in lattice QCD in Ref.~\cite{Lin09}.
The $\rho$ meson mass is obtained using the
parametrization (\ref{eqMrho}).
from Ref.~\cite{Leinweber01}.
The first row corresponds to the physical values.
In the physical case it is also
$M_\Lambda=1.116$ GeV.}
\label{tableMasses}
\end{table}

\subsubsection{Nucleon physical data}

For the nucleon we have included the proton
electric and magnetic form factors ($G_{Ep}$ and $G_{Mp}$),
and neutron electric and magnetic form
factors ($G_{En}$ and $G_{Mn}$).
Since one of our goals in this work is
to describe the octet baryon lattice data
for $G_{EB}$ and $G_{MB}$,
we do not adopt the ratio $G_{EB}/G_{MB}$,
although it is considered in many studies of
the proton electromagnetic form factors.

The proton data can be extracted using
the {\it classical} Rosenbluth separation technique
from the cross section data, or the
polarization transfer method developed at
Jefferson Lab to measure the ratio
$G_{Ep}/G_{Mp}$~\cite{JlabR1,Puckett10}.
The results of the two methods show discrepancies
that can possibly be
explained by including the two-photon
exchange corrections in the results
of the Rosenbluth separation
method~\cite{Perdrisat07,Arrington07}.
An analysis that takes into account the two-photon exchange
corrections and uses the cross section information
to determine the values of $G_{Ep}$ and $G_{Mp}$ separately
(not just the ratio), was presented in Ref.~\cite{Arrington07}.
Since our calculations are performed in the
impulse approximation, we compare our
results with the analysis of Arrington \etal~\cite{Arrington07},
where the two-photon exchange contributions were subtracted.
To include the recent, high
$Q^2$ results for $G_{Ep}/G_{Mp}$
from Jefferson Lab~\cite{Puckett10}, we convert
the ratio $G_{Ep}/G_{Mp}$ into $G_{Ep}$ using
the fit to the $G_{Mp}$ presented in Ref.~\cite{Arrington07}.
The fit is accurate for a large $Q^2$-range ($Q^2=0 -10$ GeV$^2$).
Overall, we have 50 data points for $G_{Ep}$
and 56 data points for $G_{Mp}$.

As for the neutron form factors, we collect
the data from different groups.
We prefer to use the data extracted from a deuterium target
rather than those extracted from a $^3$He target,
since the former is expected to have fewer nuclear corrections.
For $G_{En}$ we use the data from Mainz~\cite{MainzR1},
NIKHEF~\cite{Passchier99}, MIT-Bates~\cite{Eden94} and
Jefferson Lab~\cite{JlabR2,Riordan10}. 
The results from Ref.~\cite{Riordan10} obtained
using the $^3$He target, corresponding to the highest $Q^2$ result
for $G_{En}$ ($Q^2=3.4$ GeV$^2$), are also included.
Also the results obtained from the deuteron electric
quadrupole moment are included~\cite{Schiavilla01}.
For $G_{Mn}$ we adopt the data used by Bosted \etal\,~\cite{Bosted95}
in the global fit of the nucleon data,
as well as more recent data from Mainz~\cite{MainzR2}
and Jefferson Lab~\cite{Lachniet09}.
Totally, we have 29 data points for $G_{En}$
and 67 data points for $G_{Mn}$.

The inclusion of the
high $Q^2$ region nucleon data,
where pion cloud effects are expected to be small,
is important for the calibration of our
model in the valence quark sector,
since the information of lattice simulations is restricted to
the low $Q^2$ region.
We can test the parametrization directly by
the nucleon elastic form factor data
that are extended for the proton up to 9 and
31 GeV$^2$ respectively for $G_{Ep}$ and $G_{Mp}$,
and for the neutron up to 3.4 and 10 GeV$^2$,
respectively, for $G_{En}$ and $G_{Mn}$.

\subsubsection{Details of the fit}

The parameters associated
with the valence quark contributions
are adjusted to the lattice data (bare form factors),
the bare part of the nucleon form factors
at the physical point, and octet baryon magnetic moments.
They are the momentum range parameters of the
scalar wave functions, $\beta_i$ ($i=1,..,4$), and
the $u$ and $d$ quark anomalous magnetic moments
$\kappa_u$ and $\kappa_d$, respectively.
The strange quark anomalous magnetic moment $\kappa_s$
is kept unchanged, since it was fixed by the $\Omega^-$
magnetic moment, which is not affected by the pion cloud~\cite{Omega}.

The pion cloud contributions
are adjusted by the nucleon physical data
and the octet baryon magnetic moment data.
We start by calculating the bare form factors
$\tilde e_{0B}$ and $\tilde \kappa_{0B}$
at the physical point ($m_\pi=138$ MeV),
using only the parametrization for the bare form factors.
Next, we add for each form factor,  $F_1$ and $F_2$,
or alternatively  $G_E$ and $G_M$,
the pion cloud contributions given by
Eqs.~(\ref{eqB1})-(\ref{eqD2}).
For this, we need to fix  the parameters
$B_1,B_2,C_1,D_1^\prime,D_2$ and the cutoffs $\Lambda_1$ and $\Lambda_2$.
These parameters are adjusted by the nucleon
form factor data for $G_E$ and $G_M$, and
octet baryon magnetic moments, since
the magnetic moments are proportional to $G_{MB}(0)$.
In the latter we have additional constraints
for the coefficients $B_1, C_1, C_2$ and $D_2$~\cite{Octet}.
(The calibration of $D_1^\prime$,
$\Lambda_1$ and $\Lambda_2$ can be done
only with $Q^2 > 0$ data).
Once we have fixed the parameters associated with the pion cloud,
we can obtain explicit expressions
for the pion cloud contributions
for the octet baryon electromagnetic form factors.

\begin{table*}
\begin{center}
\begin{tabular}{ c r r r r r r r r}
\hline
\hline
$m_\pi$(MeV)
    &   $p$ \sp  &  $n$\sp\sp &  $\Sigma^+$\sp & $\Sigma^-$\sp & $\Xi^0$\sp\sp &
$\Xi^-$ \sp&
$\chi^2$ \sp\\
\hline
 354 &
       0.244  &   3.025  &   4.841  &   1.891  &  19.197  &   0.121 &    1.768\\
    &  1.729  &   4.872  &   7.197  &   0.350  &  38.909  &   6.101 &    4.063\\
    &  0.987  &   3.948  &   6.019  &   1.120  &  29.053  &   3.111 &
      {\bf 2.915}\\
\hline
   495 &
       0.557  &   1.790  &   1.254  &  17.313 &   78.699 &   14.361 &    8.014\\
 &   2.612 &    7.696 &    5.200  &   1.448  &  25.125 &    1.340  &   2.440 \\
 &   1.585 &    4.743 &    3.227  &   9.381  &  51.912  &   7.850  &
       {\bf  5.227}\\
\hline
 591   &
    21.297 &   19.267 &    8.332 &   16.865  & 156.552 &    9.733 &   14.057 \\
  & 13.540 &   34.159 &   24.882 &    5.558  &  37.069 &   12.414 &   14.099 \\
  & 17.418 &   26.713 &   16.607 &   11.212  &  96.810 &   11.073 &   14.078\\
\hline
 680   &
      13.219  &  22.808  &   3.116 &    2.994 &  115.990 &    8.981  &   7.077\\
  &   4.079 &   13.787 &    5.265 &    0.067 &    9.063 &    3.682 &    3.273\\
  &   8.649 &   18.297 &    4.190 &    1.530 &   62.526 &    6.332 &    5.175\\
\hline
\hline
 138  &  1.600  &   1.872    & & &  & &   1.700   \\
      &  1.857  &   2.273    & & &  & &   2.083    \\
      &  1.736  &   2.152    & & &  & &  {\bf 1.933}    \\
\hline
\hline
\end{tabular}
\end{center}
\caption{
$\chi^2$ decomposition of the best fit to the lattice data
and nucleon physical data.
The numbers in the rows corresponding to $m_\pi=138$ MeV (nucleon data)
include the effects of the pion cloud.
See also the discussion in the text.
For the lattice data $\chi^2$ per data point is $5.00$
($2.93$ excluding the $n$ and $\Xi^0$ data).}
\label{tableCHI2}
\end{table*}

For very large pion masses, it is not reasonable
to assume the baryon wave functions can be described
by the same parameters, but some dependence
on the baryon masses (indirect dependence on the quark mass)
may enter.
Also, for small pion masses ($m_\pi < 400$ MeV)
one can expect that lattice simulations may be affected by
the pion cloud effects,
and as a consequence
form factor data should differ from the calculation
based solely on the valence quark degrees of freedom.
The ideal situation would be to use
several sets of lattice data in the region
$400$ MeV $<m_\pi<600$ MeV.
In this way we expect to avoid both the pion cloud
contamination, and the very large $m_\pi$ region
where the model can fail.
In the previous studies, this way of extension
to the lattice regime was very successful~\cite{Lattice,LatticeD}.

A preliminary analysis of the lattice data has revealed
that the set $m_\pi=$ 591 MeV is particularly
difficult to describe in our model
(large $\chi^2$ per data point as presented in Table
\ref{tableCHI2}), although the quality of the fit is
improved for the next set\footnote{Although
the quality of a global fit
is measured by the total $\chi^2$
divided by,
[(the number of data points) - (the number of parameters)]
considered in the fit,
to estimate the quality of the subsets of data,
it is simpler to ignore  the subtraction of
the number of parameters,
since the parameters are not fixed by one subset
of the data.} ($m_\pi =691$ MeV).
Using efficiently
the available lattice QCD data
for the octet baryon electromagnetic form factors,
and at the same time
to keep a reasonable description of the present model,
we perform a fit to
the first two sets of lattice data, $m_\pi =351$ and
495 MeV.
(The risk of the pion cloud contamination in the $m_\pi =351$ MeV
set is compensated by the increase of accuracy in the set.)

Another point to be noted in our analysis
is the lattice data for the neutral particles, $n$ and $\Xi^0$.
The results for the $n$ and $\Xi^0$
electric charge form factors at
$Q^2=0$ (their charges) differ from zero,
and this fact suggests that the lattice
results have some systematic errors.
This may be a consequence of incomplete cancellations
among the contributions from the different quark flavors
for the form factors~\cite{Lin09},
and we have to take this into account.
In order to use the $n$ and $\Xi^0$
data but achieve a reasonable accuracy,
we reduce the respective impact on $\chi^2$
by doubling the respective statistical errors.

As already mentioned,
the pion cloud parametrization is calibrated
using the $\Lambda$, $\Sigma^\pm$, $\Xi^{0,-}$
physical magnetic moment data and the nucleon physical data.
Some of the octet magnetic moment data are
extremely accurate with error bars of less
than 1\% ($\Lambda$ and $\Sigma^-$).
A minimization of $\chi^2$ with
such small error bars of $G_{MB}(0)$
will impose strong contributions on $\chi^2$ in the
$Q^2=0$ region, and reduce the relative
impact of the region $Q^2> 0$, represented
by the nucleon form factor data.
To achieve a good description
(low $\chi^2/n_p$, where $n_p$ is the number of the data points)
for the nucleon data, we reduce the weight
of the octet baryon magnetic moment data,
by doubling the experimental errors.

We define the best model as the model
that minimizes the total $\chi^2$ associated with
the lattice data, the octet magnetic moment data
and nucleon physical data as described before.
Totally, we have 272 lattice data points,
5 experimental magnetic moment data points, and
202 nucleon physical data points.

A fit with no constraints leads to a good
description of the lattice data (small $\chi^2$ per data point),
but a poor description of the nucleon physical data
with $\chi^2$ per data point $> 2$.
That fit generates also a large
contributions from the pion cloud effects,
and has a large extension in $Q^2$
from the pion cloud effects, in particular for $G_M$
due to a large cutoff $\Lambda_2$ compared to 1 GeV.
We interpret this result as a consequence
of the dimension of the lattice database
(272 data points) for 6 baryons including the nucleon,
to be compared with the nucleon physical data (202 data points).
This procedure reduces the relative importance of the nucleon physical data.
Since we want to describe well the nucleon system
but simultaneously to have
a reasonable description of the lattice data,
we reinforce the impact of the
nucleon physical data by doubling the
contributions of their $\chi^2$
in the global evaluation of $\chi^2$.
The fit with this constraint
leads a good qualitative description
of the nucleon physical data ($\chi^2$ per data point $\simeq 1.9$),
and also reasonable values for the cutoffs
$\Lambda_1$ and $\Lambda_2$,
where the values
are smaller than 1 GeV, or closer to 1 GeV,
consistent with the pion cloud effects
restricted to low $Q^2$ region.
The values of the cutoffs will be discussed
in more detail later.
As we will see, the final fit is consistent
with the small pion cloud contributions for
the octet baryon form factors.

The quality of the fit, measured in terms
of $\chi^2$ per data point,
can be understood from Table~\ref{tableCHI2}.
In the table we represent for each set,
$\chi^2(G_{EB})$ in the first row,
$\chi^2(G_{MB})$ in the second row, and
the combined result in the last row.
With bold face ==> In boldface
In boldface
we represent the $\chi^2$ for the
sets considered in the fit.
The rows associated with the physical regime ($m_\pi=138$ MeV)
are the only ones that reflect the effects
of the pion cloud.
Note that the sets $m_\pi= 591$ and $680$ MeV,
are not included in our fit.
In the last column we represent
the partial $\chi^2(G_{E})$ and $\chi^2(G_{M})$, and
the total $\chi^2$,
associated with respective sets.
In this case the contributions from $n$ and $\Xi^0$ are not included.

\subsection{Bare octet form factors}

The results of the fit for the octet baryon electromagnetic form factors
are presented in Figs.~\ref{figProtonLatt}-\ref{figXiMLatt}.
The values of the pion mass in the fit are
$m_\pi= 354$ and $495$ MeV.
For the nucleon [Figs.~\ref{figProtonLatt} and \ref{figNeutronLatt}],
we present also the physical case
($m_\pi= 138$ MeV).
The parameters associated with the results are
\ba
& &
\beta_1= 0.0440, \hspace{.1cm} \beta_2= 0.9077,\nonumber \\
& &
\beta_3= 0.7634, \hspace{.1cm} \beta_4= 0.4993,\nonumber \\
& &
\kappa_u= 1.6690, \hspace{.2cm} \kappa_d= 1.9287.
\label{eqBetas}
\ea
We note that the (bare) quark anomalous magnetic
moments, $\kappa_u$ and $\kappa_d$, are similar to the
previous results obtained in Refs.~\cite{Nucleon,Octet}, within
a 16\% variation.
(In Refs.~\cite{Nucleon,Omega} the values obtained are
$\kappa_u=1.778$ and $\kappa_d=1.915$, while in Ref.~\cite{Octet}
they are $\kappa_u=1.929$ and $\kappa_d =1.919$.)

Furthermore, the values for $\beta_1$ and $\beta_2$ are also similar
to those of the model in Ref.~\cite{Nucleon} ($\beta_1=0.049$ and
$\beta_2= 0.717$).
Interpreting $\beta_1$ as the spatial long range parameter
common to all the systems associated with one light quark
($N,\Sigma,\Lambda$ and $\Xi$), $\beta_2,\beta_3$ and $\beta_4$
are interpreted as spatial short range
parameters associated with spatial extension
of the $q_l q_l$, $q_l s$ and $ss$ quark pairs respectively,
where $q_l$ stands for a light quark.
In this respect we should expect $\beta_2 > \beta_3 > \beta_4$,
consistently with Eq.~(68),
a decreasing of the spatial extension
of the systems, gradually from the nucleon
followed by the $\Lambda,\Sigma$ and
finally by the $\Xi$ system.
The systems with one strange quark are
more compact than the ones with no strange quarks,
and the systems with two strange quarks are
even more compact than the ones with only one
strange quark.
We will return to this issue later
when we discuss the charge and magnetic
squared radii.

In Figs.~\ref{figProtonLatt}-\ref{figXiMLatt}
we compare the results of our fit for the
$n$, $p$, $\Lambda$, $\Sigma^{0,\pm}$, and $\Xi^{0,-}$
with the lattice data for
$m_\pi=354$ and $495$ MeV.
The lattice data~\cite{Lin09} used in the fit are shown
in Figs.~\ref{figProtonLatt}, \ref{figNeutronLatt},
\ref{figSigmaPLatt}, \ref{figSigmaMLatt},
\ref{figXi0Latt}, and~\ref{figXiMLatt},
respectively, for the $p$, $n$, $\Sigma^+$, $\Sigma^-$,
$\Xi^0$, and $\Xi^-$.
For the proton and neutron we include
also the physical case ($m_\pi= 138$ MeV)
and compare the results with the experimental data
(details will be discussed later).
We have no data for the $\Lambda$ and $\Sigma^0$
from Ref.~\cite{Lin09}, respectively,
shown in Figs.~\ref{figLambdaLatt} and~\ref{figSigma0Latt},
but we compare our results with the quenched lattice QCD simulation
results from Ref.~\cite{Boinepalli06} with the
closest pion mass values, respectively $m_\pi= 372$
and $464$ MeV, for the single point at $Q^2=0.23$ GeV$^2$
calculated in that work.

Overall, we have a very good description
of the data for $p$, $\Sigma^\pm$ and even $\Xi^-$.
As mentioned already, the results for
$n$ and $\Xi^0$ should be taken with caution,
since the lattice simulations have systematic deviations
from the expected results, particularly for
the electric charge form factors.
Nevertheless, we have a good
global description of the
nucleon and $\Sigma$ systems.
As for the $\Xi$,
the lattice results are closer
to the extrapolation to the
physical limit (dashed line)
than the calculation
in the lattice regime (solid line).
This can be caused by the poor quality
of the $G_E$ data for $\Xi^0$ as discussed before,
or a limitation of our simplified approach.
Further lattice QCD simulation data, consistent
with the $n$ and $\Xi^0$ charges, are necessary to clarify this point.
As for the other neutral
particles, $\Lambda$ and $\Sigma^0$, our predictions are
consistent with the simulations of Boinepalli \etal~\cite{Boinepalli06},
within about 2 standard deviations,
for the closest pion mass values used.
A final remark is on the difficulty of
the present approach
in describing the $n$ and $\Xi^0$ data.
Since we try to describe well $F_1$ and $F_2$
(or $G_E$ and $G_M$) simultaneously,
inaccuracy in one set (say $G_E$)
will affect the description of the other set (say $G_M$).

For the nucleon system we compare also
the results extrapolated to the physical limit of
the bare form factors with the physical data
(see the bottom panels with $m_\pi=138$ MeV
in Figs.~\ref{figProtonLatt} and~\ref{figNeutronLatt}).
Note that these results should not
be compared directly, since we have not
yet included the pion cloud effects in the calculation.
From the figures,
our aim is to see whether or not the pion cloud effects are
indeed important, and if reasonable description of the data can
be achieved without the pion cloud effects.
We plot then the results of two different calculations.
The first one is the extrapolation to the physical limit
of the model, represented by the dashed line
(the same as in the upper panels).
The result is obtained setting $Z_N=1$,
in the expression for the
nucleon form factors
[Eqs.~(\ref{eqF1p}), (\ref{eqF1n}), (\ref{eqF2p}), and~(\ref{eqF2n})],
and removing the pion cloud contributions.
Plotted in the same figures (thin solid line),
is the same calculation but for the case $Z_N=$0.885,
given by the fit, the contributions exclusively
from the valence quark core.
From the results shown in the figures, we conclude that
the data are well described
by a model with no pion cloud effects, although
in the region of high $Q^2$ (say $Q^2 > 1$ GeV$^2$),
the model with $Z_N=0.885$ given by
the thin solid lines (solely from the valence quark contributions),
are closer to the data than those of $Z_N=1$.

\begin{figure*}[t]
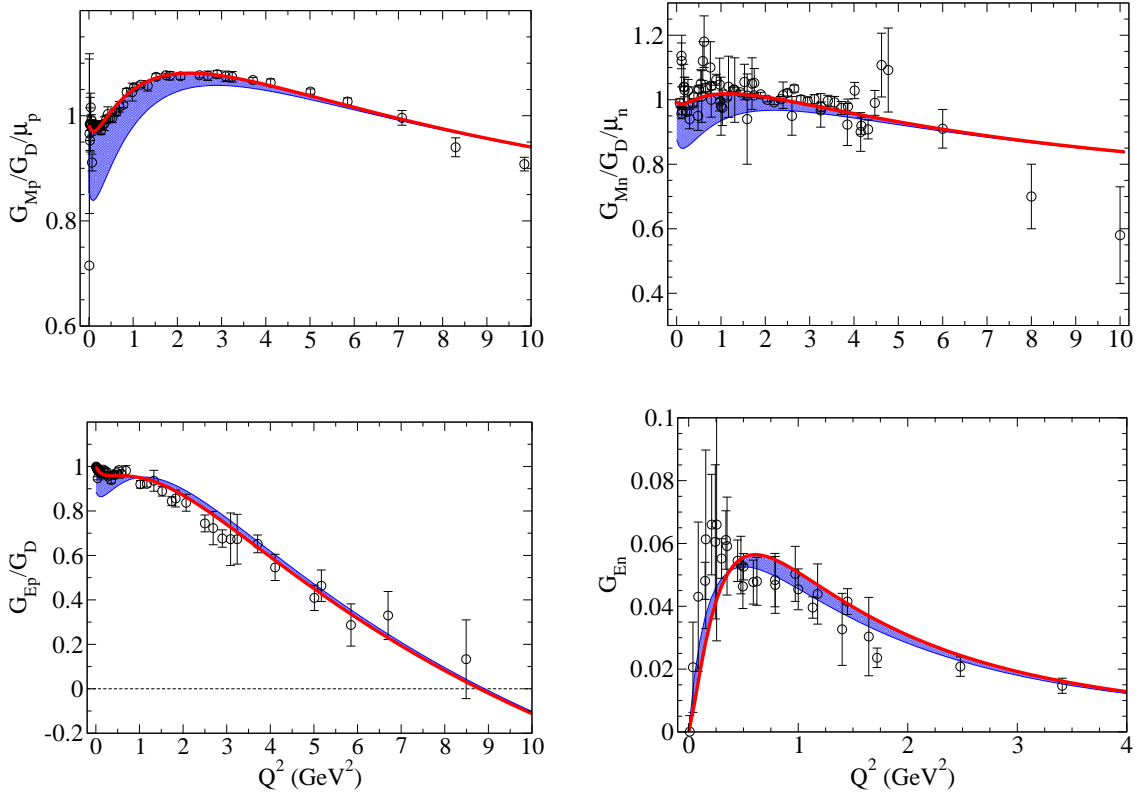

\centerline{
\mbox{
\includegraphics[width=2.8in]{GMpGD612.eps} \hspace{.5cm}
\includegraphics[width=2.8in]{GMnGD612.eps}
}}
\vspace{0.75cm}
\centerline{
\mbox{
\includegraphics[width=2.8in]{GEpGD612.eps} \hspace{.5cm}
\includegraphics[width=2.8in]{GEn612.eps}
}}
\caption{\footnotesize{
Nucleon electromagnetic form factors including the sum
of the bare and the pion cloud (bands).
Data are from Refs.~\cite{Arrington07,JlabR1,Puckett10,MainzR1,Passchier99,Eden94,JlabR2,Riordan10,Schiavilla01,Bosted95,MainzR2,Lachniet09}.
See text for details.}}
\label{figNucleon1}
\end{figure*}

\begin{figure*}[ht]
\vspace{0.3cm}
\centerline{
\mbox{
\includegraphics[width=2.8in]{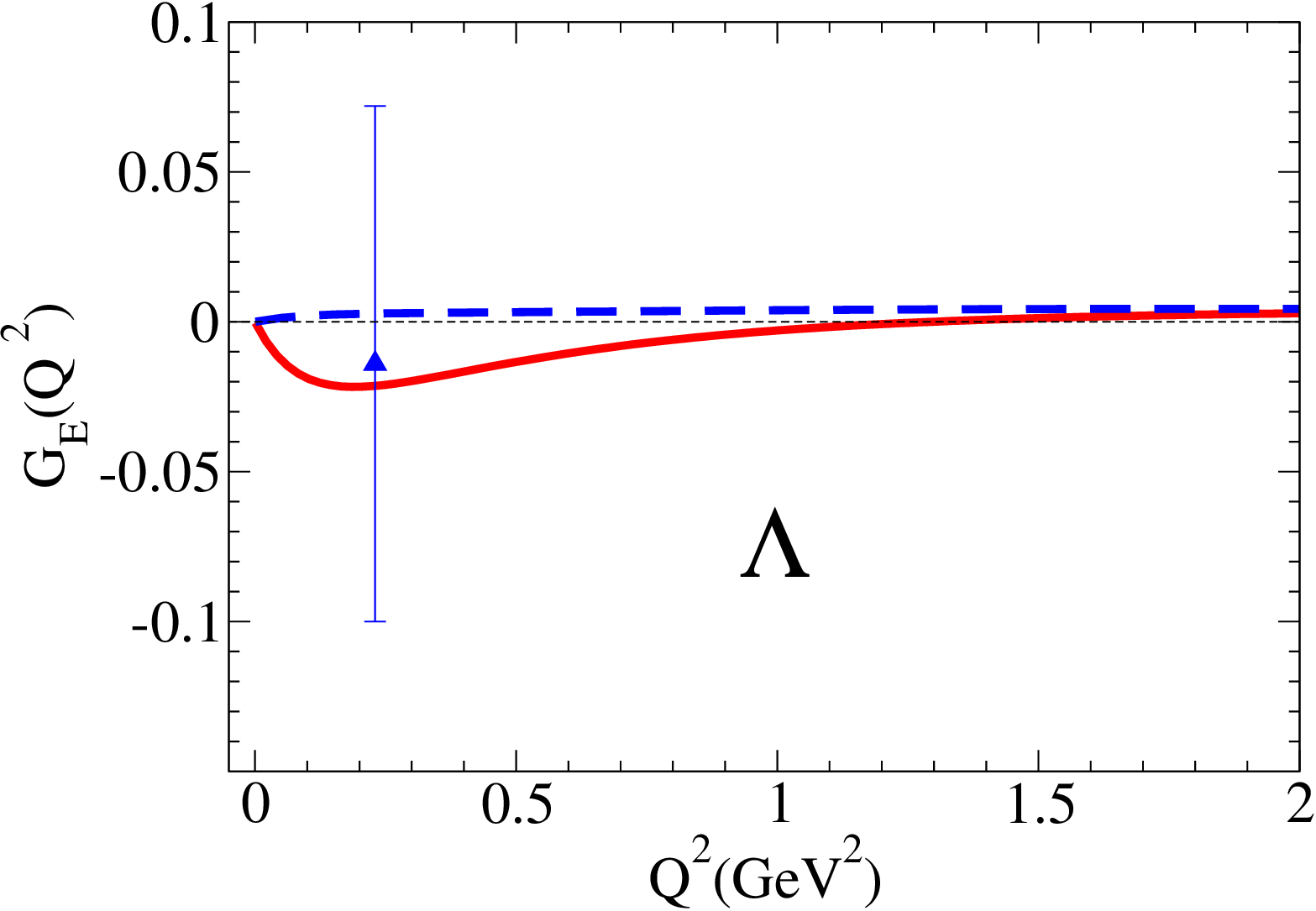} \hspace{.5cm}
\includegraphics[width=2.8in]{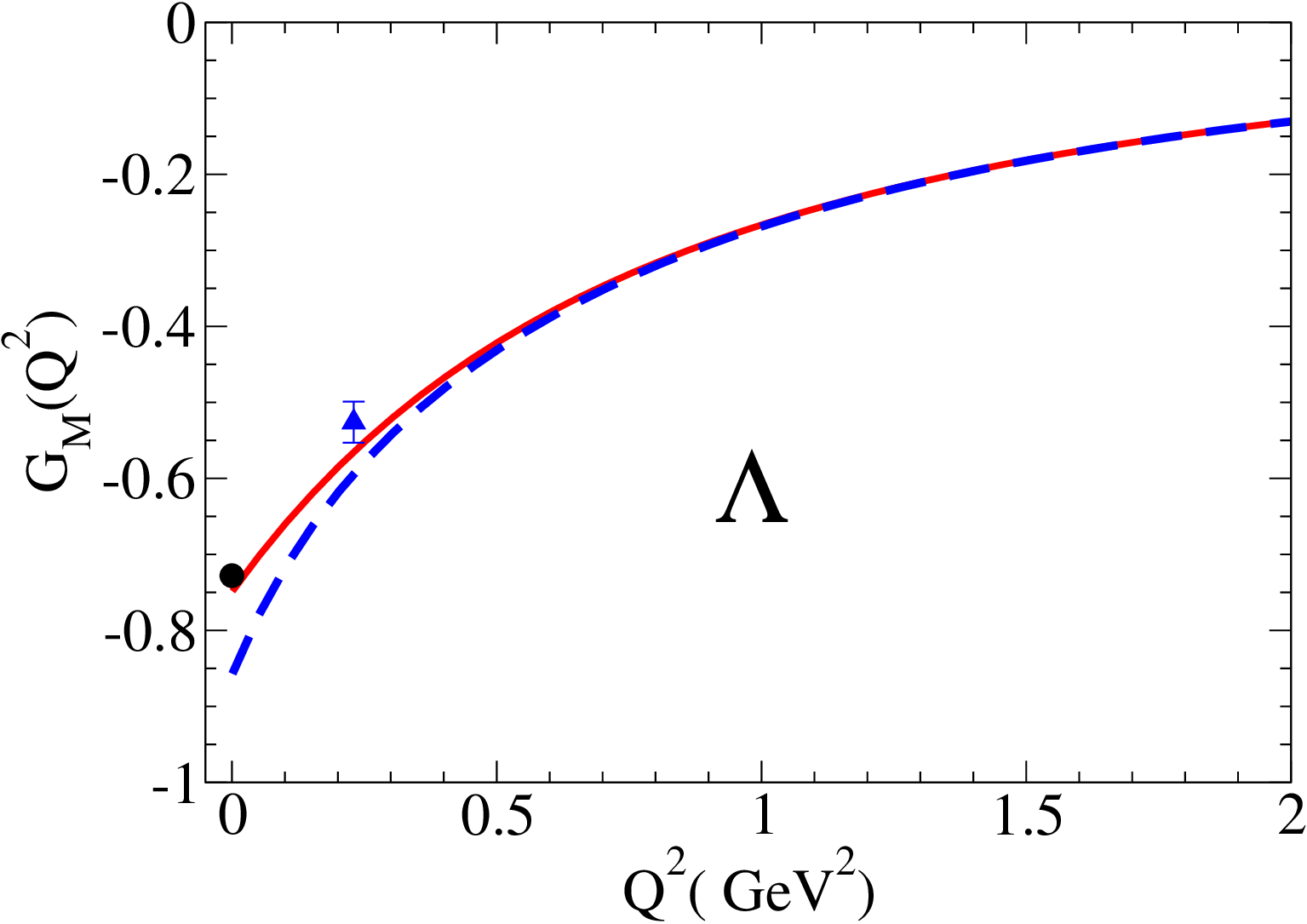} }}
\caption{\footnotesize{
$\Lambda$ electromagnetic form factors for
the total result (solid line) and
the bare result (dashed line).
The data point for $Q^2=0$ is the result of
the magnetic moment~\cite{PDG}.
The lattice data point (filled triangle) for $Q^2=0.23$ GeV$^2$
is from Ref.~\cite{Boinepalli06} ($m_\pi=$ 306 MeV).
}}
\label{figLambda}
\end{figure*}

\begin{figure*}[t]
\centerline{
\mbox{
\includegraphics[width=2.8in]{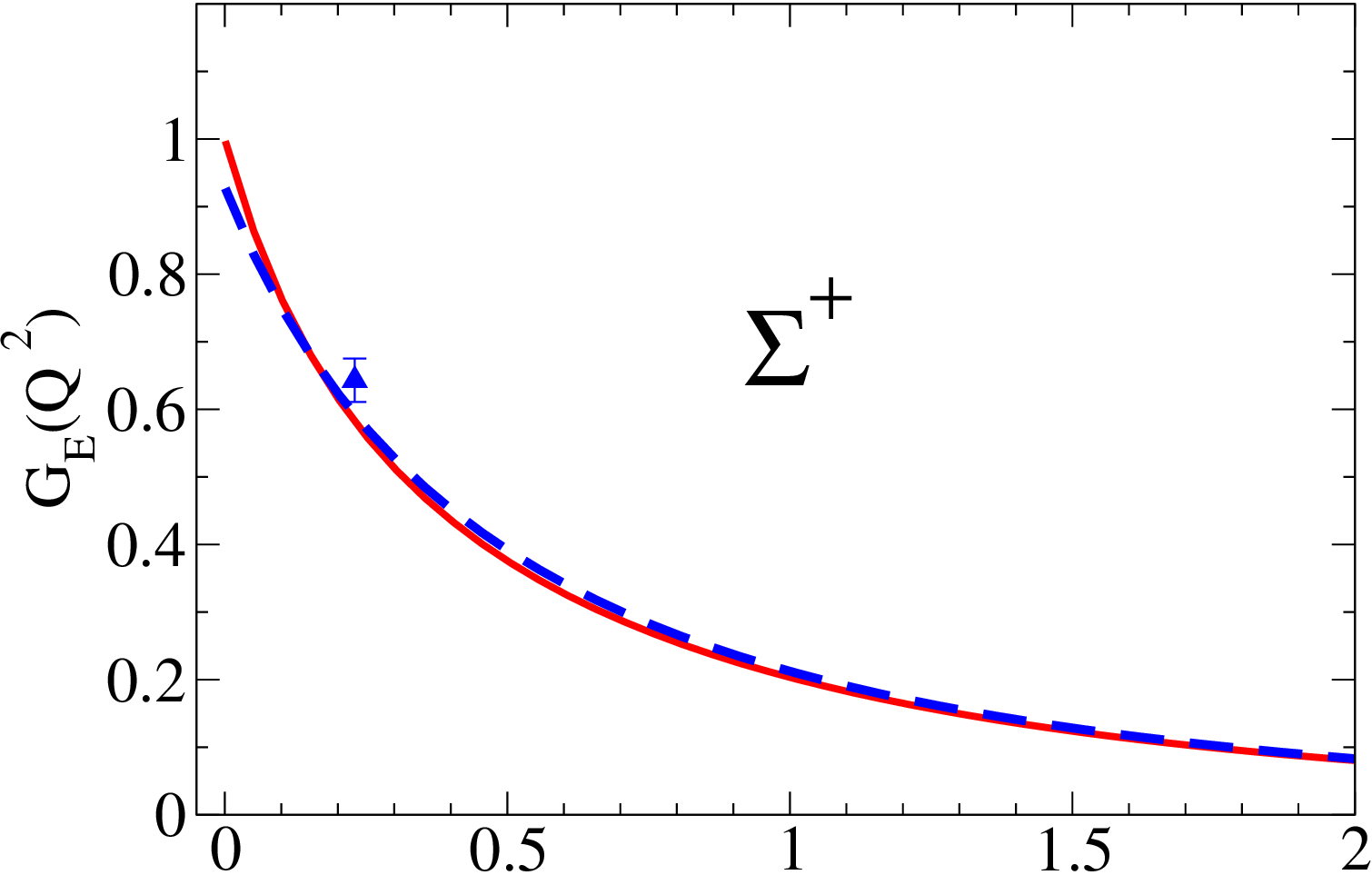} \hspace{.5cm}
\includegraphics[width=2.8in]{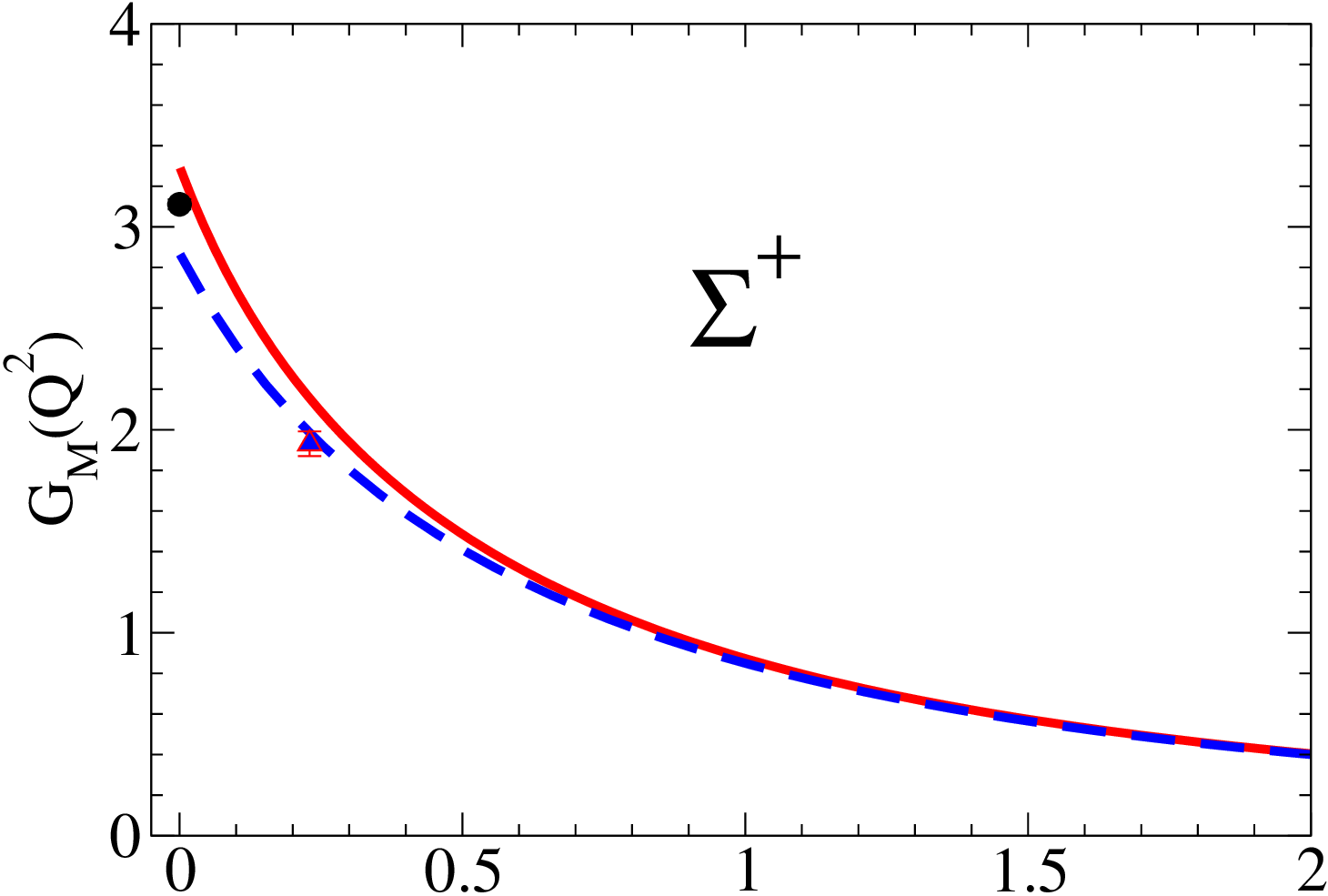} }}
\vspace{0.75cm}
\centerline{
\mbox{
\includegraphics[width=2.8in]{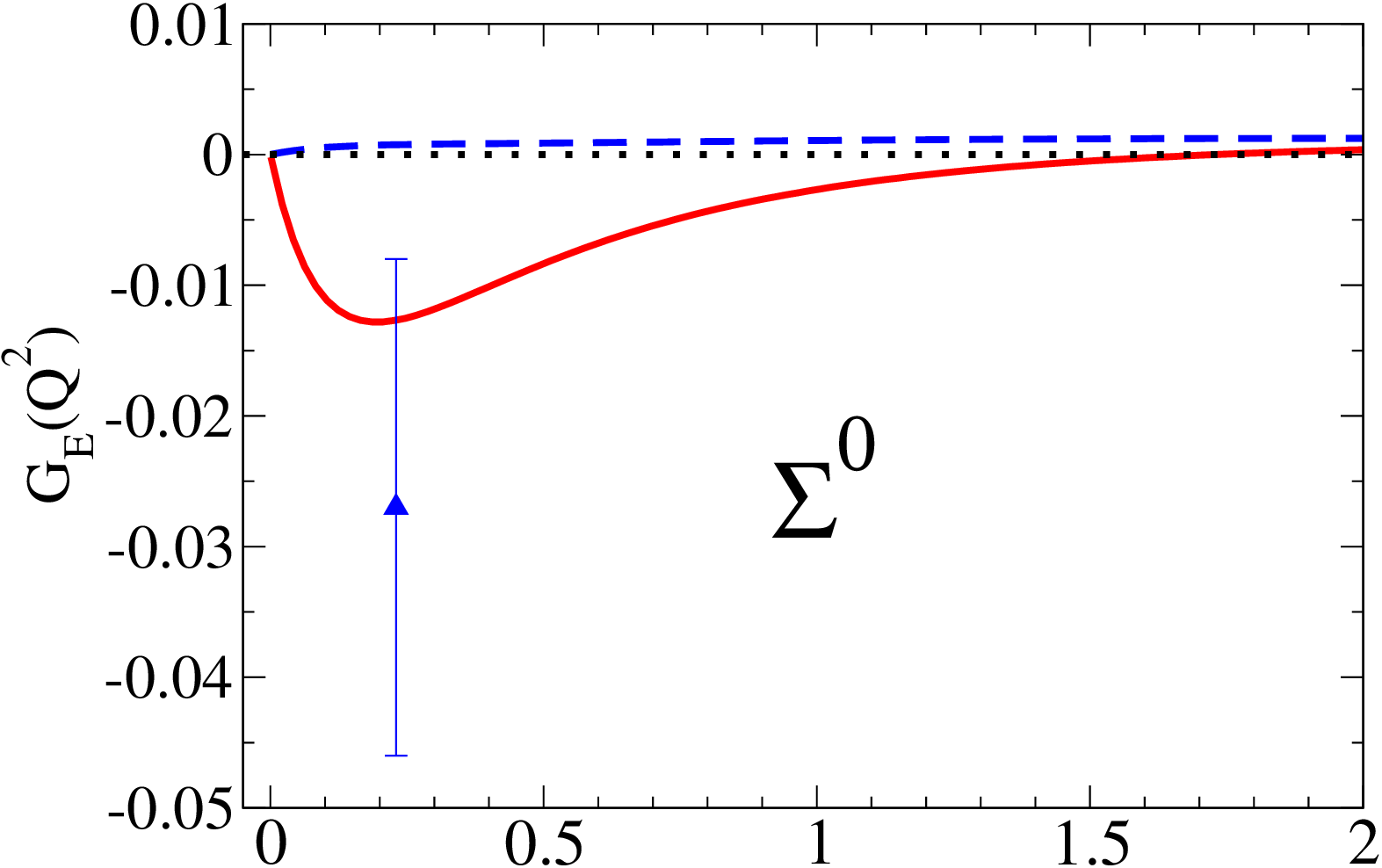} \hspace{.5cm}
\includegraphics[width=2.8in]{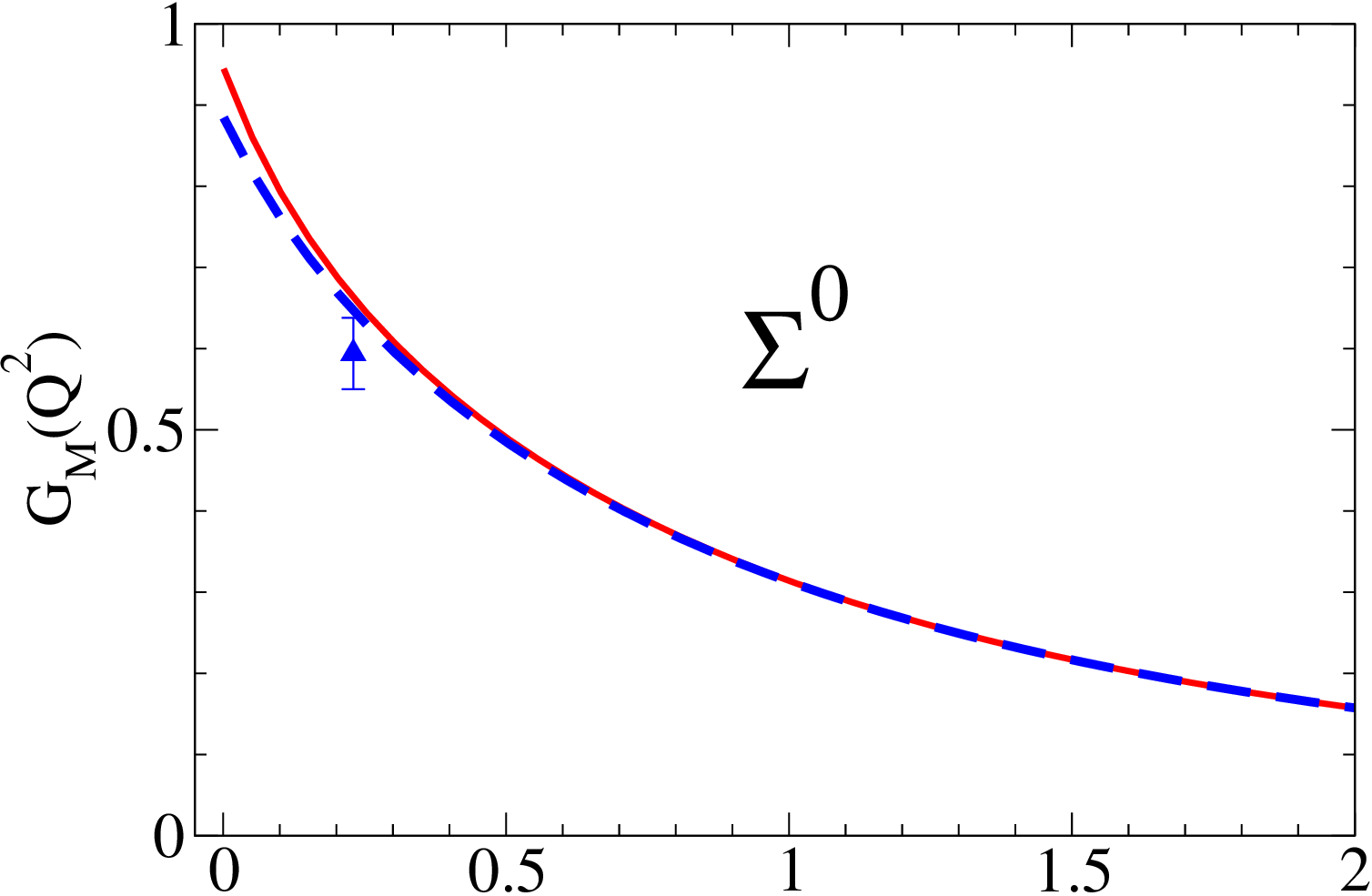} }}
\vspace{0.75cm}
\centerline{
\mbox{
\includegraphics[width=2.8in]{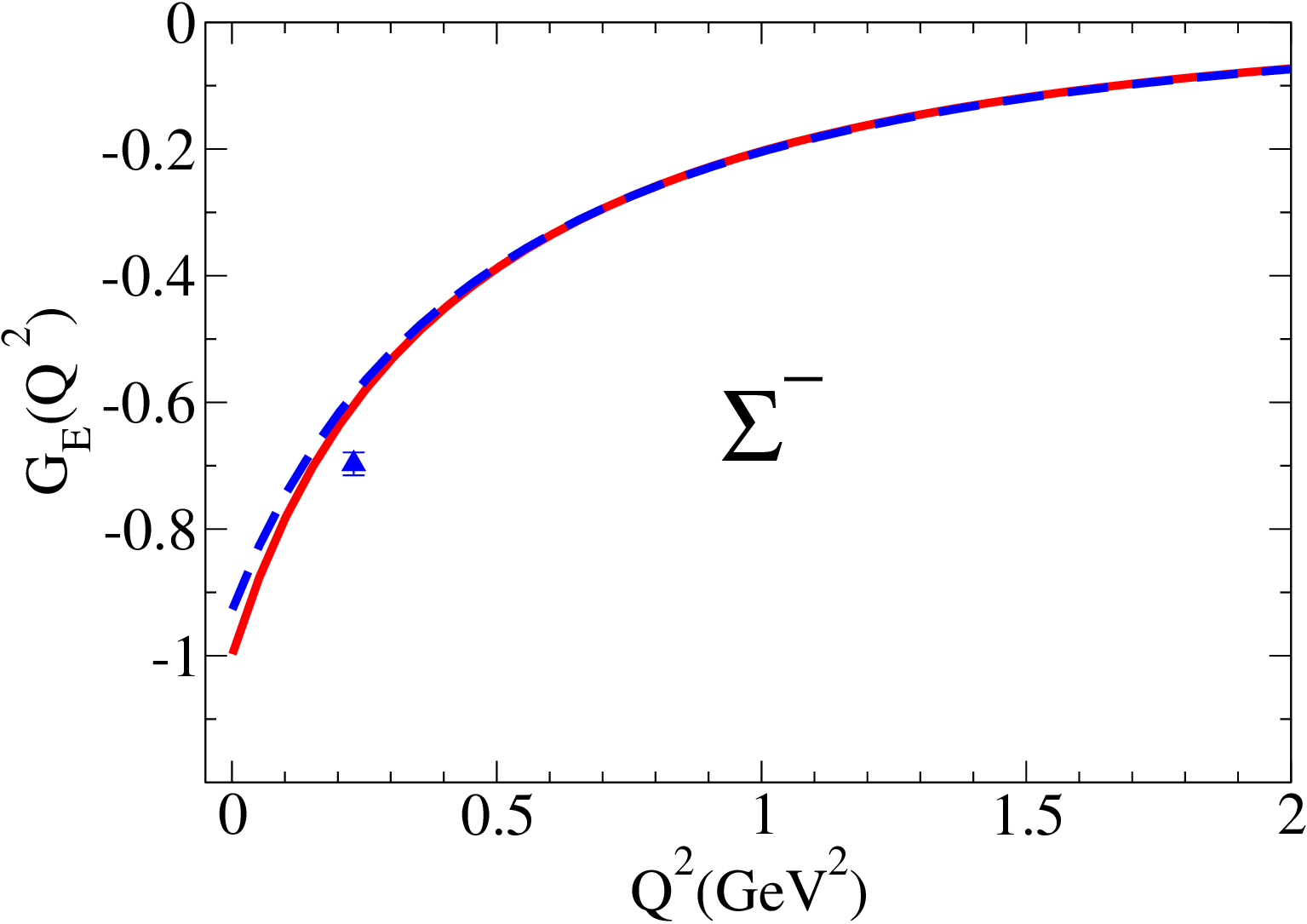} \hspace{.5cm}
\includegraphics[width=2.8in]{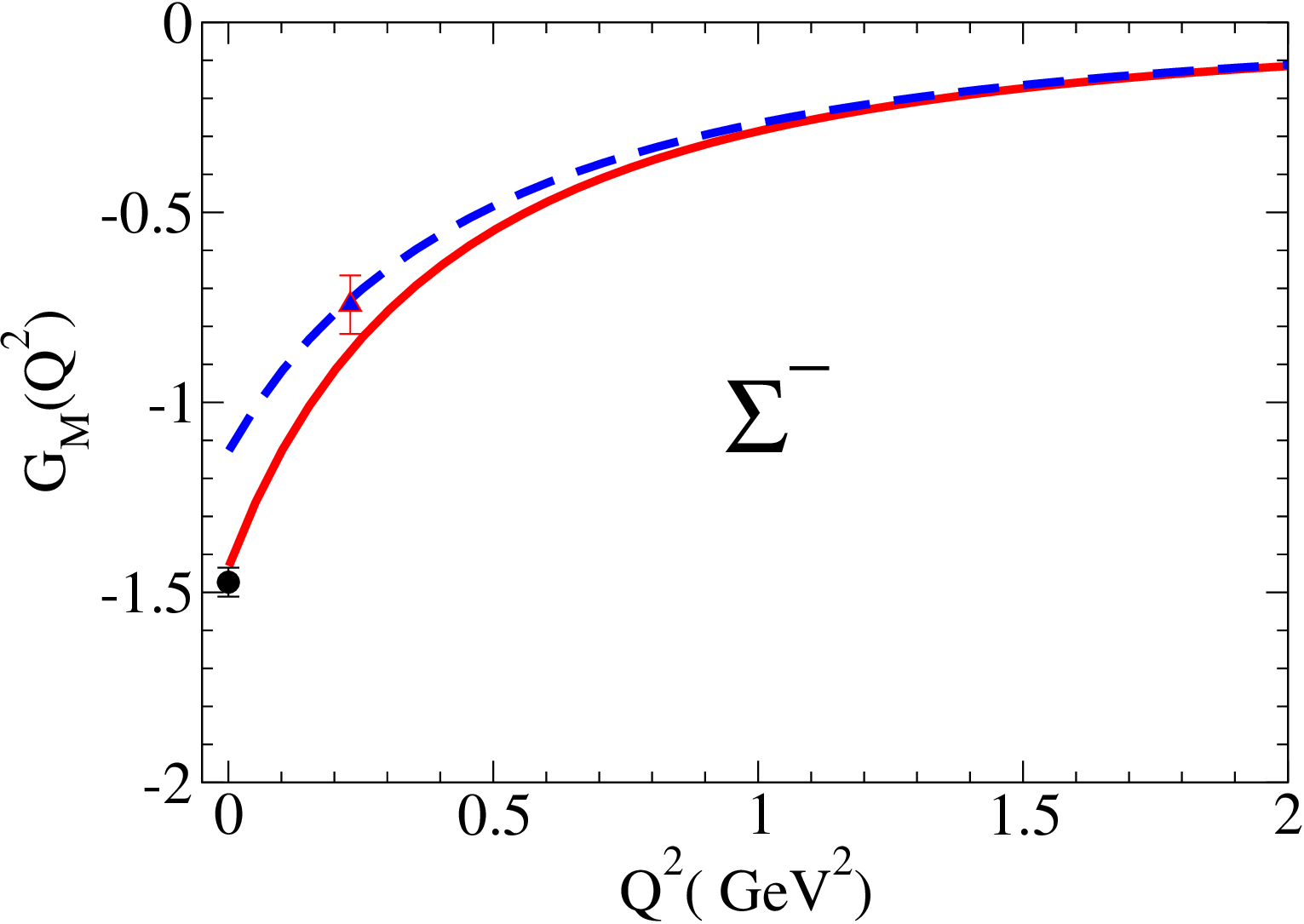} }}
\caption{\footnotesize{
$\Sigma$ electromagnetic form factors for
the total (solid line) and
the bare result (dashed line).
The data point for $Q^2=0$ is the result of
the magnetic moment~\cite{PDG}.
The lattice data point (filled triangle) for $Q^2=0.23$ GeV$^2$
is from Ref.~\cite{Boinepalli06} ($m_\pi=$ 306 MeV).
}}
\label{figSigma}
\end{figure*}

\begin{figure*}[t]
\centerline{
\mbox{
\includegraphics[width=2.8in]{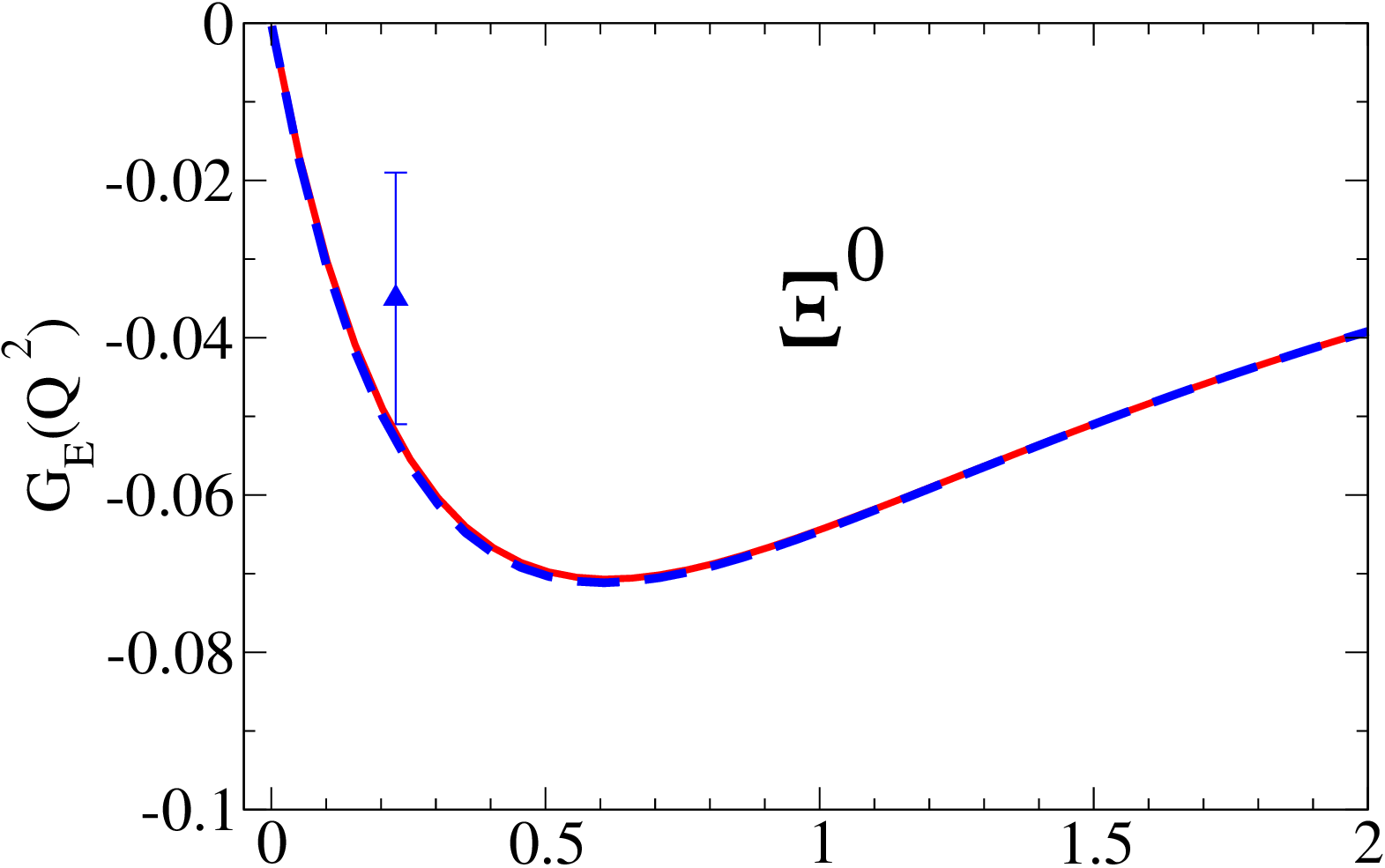} \hspace{.5cm}
\includegraphics[width=2.8in]{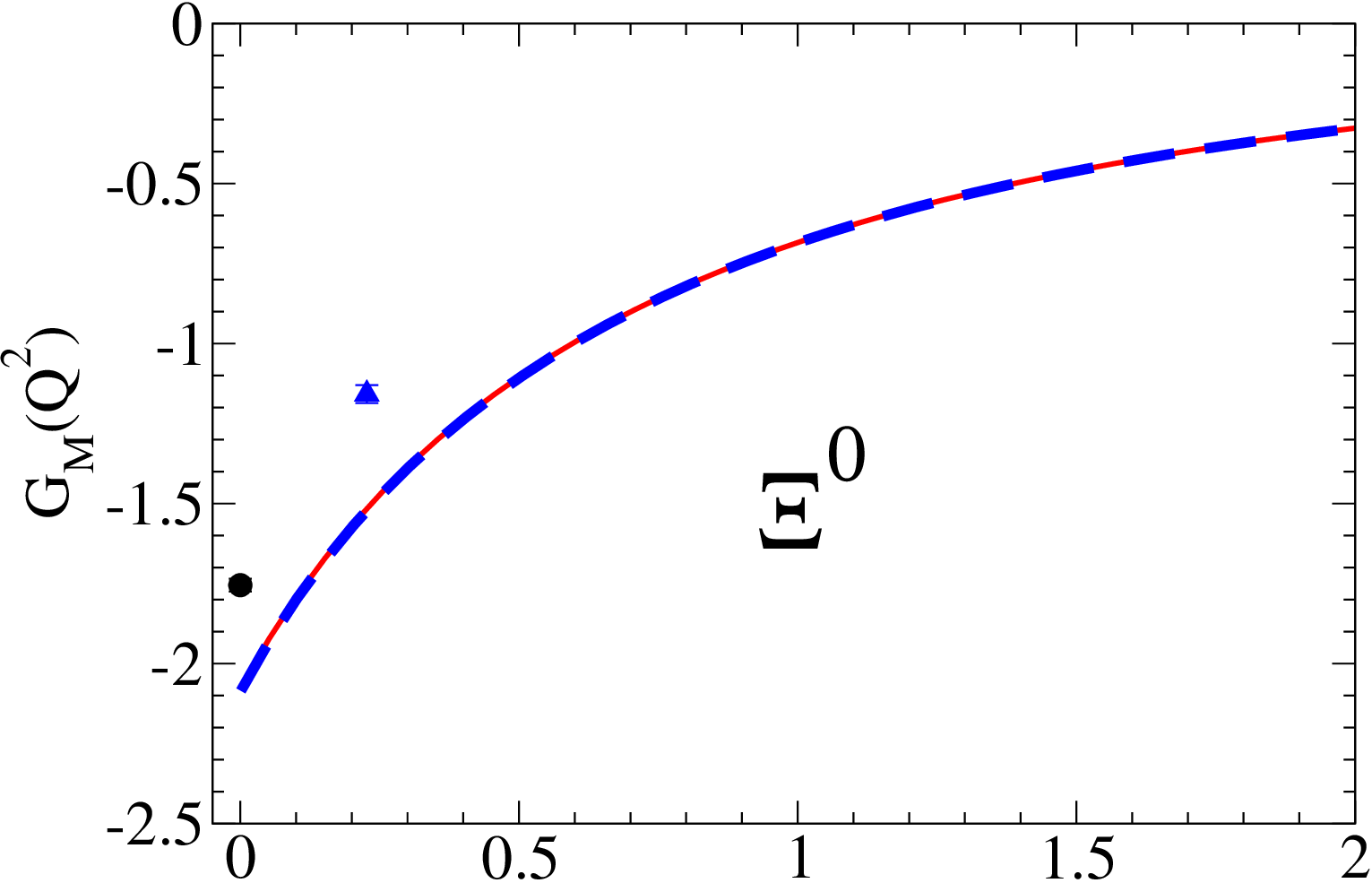} }}
\vspace{0.75cm}
\centerline{
\mbox{
\includegraphics[width=2.8in]{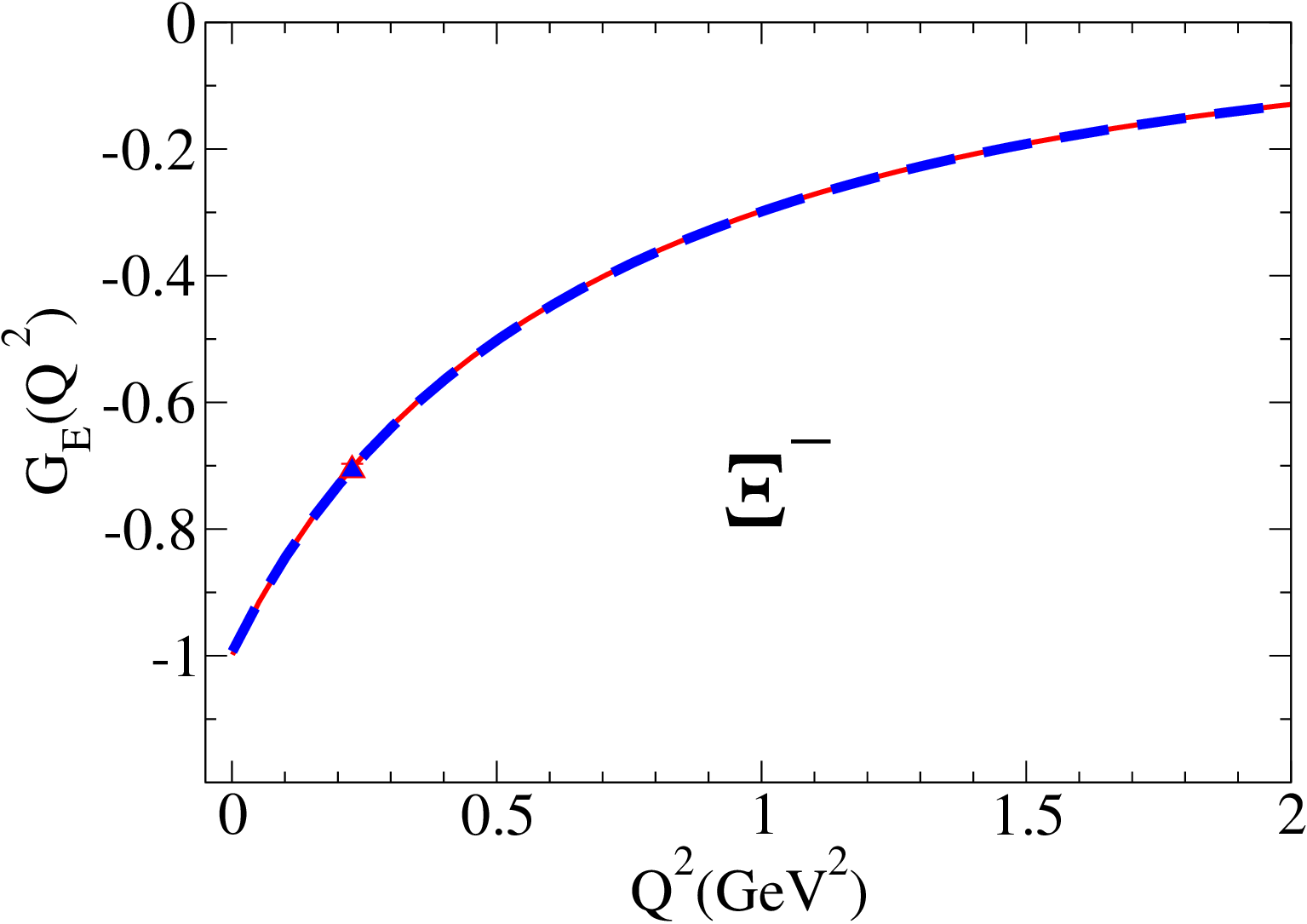} \hspace{.5cm}
\includegraphics[width=2.8in]{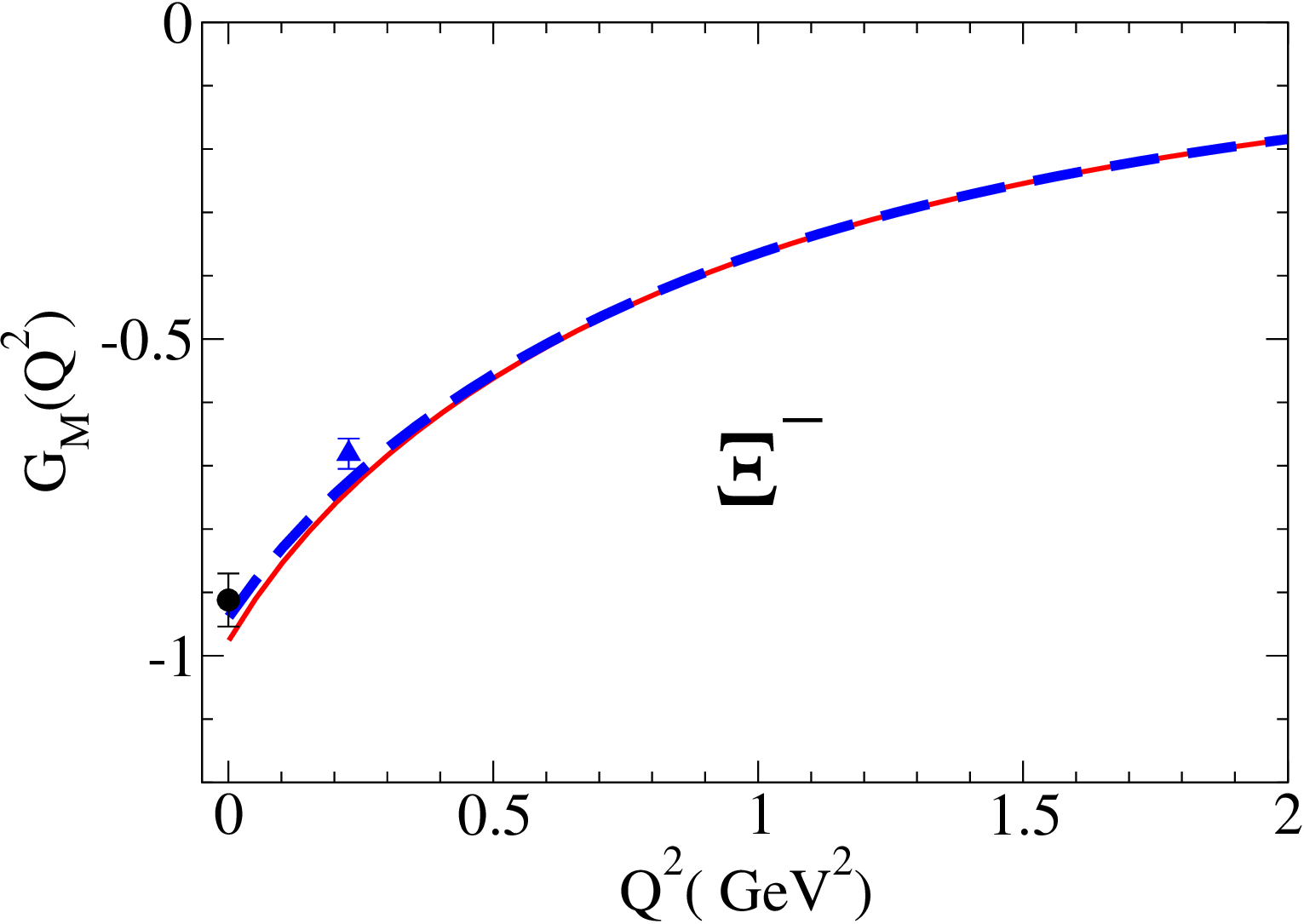} }}
\caption{\footnotesize{
$\Xi$ electromagnetic form factors for
the total result (solid line) and
the bare result (dashed line).
The data point for $Q^2=0$ is the result of
the magnetic moment~\cite{PDG}.
The lattice data point (filled triangle) for $Q^2=0.23$ GeV$^2$
is from Ref.~\cite{Boinepalli06} ($m_\pi=$ 306 MeV).
}}
\label{figXi}
\end{figure*}

\subsection{Pion cloud contributions}

We discuss now the calibration of the
pion cloud effects.
The values in the parametrization associated with
the pion cloud effects are
\ba
& &
B_1=0.04343, \hspace{.2cm} B_2= 0.21477,   \hspace{.2cm}
C_2= 0.02266, \nonumber \\
& &
D_1^\prime = -0.17637, \hspace{.2cm}
D_2= 0.08551,
\ea
with the cutoff values,
\ba
\Lambda_1= 0.7732 \;\mbox{GeV}, \hspace{.2cm}
\Lambda_2 = 1.2455 \;\mbox{GeV}.
\ea

The quality of the fit associated with the
pion cloud effects is measured by the partial $\chi^2$ values
for the nucleon system at the physical point,
given by the last column in Table \ref{tableCHI2}
($\chi^2$ per data point $=1.93$).
Then, we can conclude that the nucleon data are described
better than the lattice data
($\chi^2$ per data point of 2.9 and 5.2
for the sets $m_\pi=354$ and $495$ MeV, respectively).

Since we cannot isolate the pion cloud contributions
from the valence quark contributions in the experimental data,
we analyze the pion cloud effects by comparing
the individual components of the nucleon form factors
with the full result.
The results for the nucleon are presented in
Fig.~\ref{figNucleon1}, where the form factors
are renormalized by the dipole form factor
$G_D= \left( 1+ \sfrac{Q^2}{0.71}\right)^{-2}$.
The exception is the neutron electric form factor.
In the figure, the contributions of the
pion cloud are represented by the bands
that fill the difference between
the valence quark contributions ($Z_B G_{X0B}$) and
the full result ($G_{XB}$, solid line).

Observing the pion cloud contributions for the
nucleon electromagnetic form factors in Fig.~\ref{figNucleon1}, we conclude
that the contributions are similar for both
the proton and neutron magnetic form factors.
In both cases contributions amount to
10--14\% in the region of $Q^2=0-0.5$ GeV$^2$,
and fall to less than 5\% around
$Q^2=2$ GeV$^2$, and even become less than 1\% for $Q^2 > 5$ GeV$^2$.

The analysis for the electric form factors is more delicate.
For the proton there are $\approx 12\%$ contributions
from the pion cloud near $Q^2=0$,
and they fall to 1\% near $Q^2=1$ GeV$^2$,
and stabilize to 5\% negative contributions
for $Q^2 \approx$ 5 GeV$^2$.
In the larger $Q^2$ region one must be careful, since $G_E$ approaches
zero and the ratio is not meaningful.
For $Q^2=10$ GeV$^2$ the valence quark
contributions are larger than 90\%.
As for the neutron near  $Q^2=0$,
where $G_{En}(0)=0$, the pion cloud contributions dominate.
Near $Q^2=1$ GeV$^2$, the pion cloud effects are about 10\%
and drop to less than 4\% for $Q^2=4$ GeV$^2$,
and even smaller for larger $Q^2$.
For $Q^2=10$ GeV$^2$ the valence quark
contributions dominate to give more than 98\%.

The slow falloff of the pion cloud contributions for
the electric form factors
compared with those for the magnetic ones,
is due to the enhancement
of the $F_2$ contributions for $G_E$
by the pre-factor $Q^2$, and the function
form for the pion cloud contributions.
Since the pion cloud contributions are regulated
by the cutoff $\Lambda_2= 1.24$ GeV
($\Lambda_2^2= 1.55$ GeV$^2$)
which is larger than $\Lambda_1 \simeq 0.77$ GeV
($\Lambda_1^2 = 0.59$ GeV$^2$),
the electric form factor is extended to the
higher $Q^2$ region due to
the range of the pion cloud contributions in $F_2$.

We point out that the qualitative differences
of the pion cloud contributions
between the present results and those
of the previous study in
the octet baryon magnetic moments~\cite{Octet}.
In the previous study the $Q^2$ dependence
was not taken into account (no use of lattice data,
neither the nucleon physical form factor data),
and the pion cloud contributions were significantly larger.
Thus, we can conclude that the $Q^2$ dependence of
the pion cloud contributions is very important to constrain
the pion cloud contributions,
in particular for the nucleon system.

\subsection{Dressed form factors}

Taking into account the results for the bare form
factors extracted from the lattice data,
and the pion cloud parametrization from the previous section,
we use the expressions in Sec.~\ref{secDressedFF}
to predict the dressed, physical form factors for the $\Lambda$,
$\Sigma$, and $\Xi$ systems in the physical regime ($m_\pi=138$ MeV).
The predicted results are presented in Figs.~\ref{figLambda},
\ref{figSigma} and \ref{figXi}.
In the figures we show the full results (solid line)
and bare results (dashed line).
Included in the figures are also the
magnetic moments when known ($\Lambda$, $\Sigma^\pm$, and $\Xi^{0,-}$).
To have an idea for the dynamical behavior
($Q^2$ dependence)
of the form factors, we include also the
lattice QCD simulation data corresponding
to the lowest pion mass from Ref.~\cite{Boinepalli06} ($m_\pi=306$ MeV).
In principle, the lattice data should be
compared with the bare form factors (dashed line),
unless strong quenched effects are expected.

An important conclusion from the figures
is that the magnitude of the pion cloud
contributions is small.
Based on the magnetic moments, we have more significant
contributions from the pion cloud for
$\Sigma^-$, $\Sigma^+$, and $\Lambda$,
respectively about 21\%, 14\% and 14\%,
and less than 10\% for all other cases.
With the exception of the neutral particles $\Lambda$, $\Sigma^0$,
and $\Xi^0$, where pion cloud effects can be dominant,
the pion cloud contributions for the electric
form factors are smaller than those for the magnetic form factors.

Overall, our predictions for the octet baryon
electromagnetic form factors, as functions of $Q^2$,
are consistent with the results for the
magnetic moments (at $Q^2=0$),
and close to the lattice QCD simulations.
The major exception is the results for $\Xi^0$,
where we observe the clear deviation from
the experimental result for $G_{M}(0)$, and
also from the lattice data.
We recal again that this can be a consequence
of the difficulty in describing the $\Xi$ lattice data.
[See Figs.~\ref{figXi0Latt} and~\ref{figXiMLatt}.]

\subsection{Electric charge and magnetic dipole radii}

The electric charge squared radius for a charged particle
is usually defined as\footnote{
Some authors~\cite{Boinepalli06} exclude the factor
$G_{EB}(0)$ from the $\left< r_E^2\right>$ definition.}
\ba
\left< r_E^2\right>=
- \frac{6}{G_{EB}(0)}
\left. \frac{d G_{EB}}{d Q^2}\right|_{Q^2=0}.
\label{eqRE2}
\ea
For a neutral particle the same expression
can be used but setting $G_{EB}(0) \to 1$.
The definition~(\ref{eqRE2}) has advantages
for comparing the radii of particles with different
charges such as $p$ and $\Sigma^-$,
and one can relate the corresponding
baryon electric charge radii.
As for the magnetic dipole squared radius,
the most common definition\footnote{
Some authors~\cite{Boinepalli06}
 define $\left< r_M^2\right>$
without the factor $G_{MB}(0)$, but use
$\frac{\left< r_M^2\right>}{G_{MB}(0)}$
to compare the values of different baryons.} is
\ba
\left< r_M^2\right>=
- \frac{6}{G_{MB}(0)}
\left. \frac{d G_{MB}}{d Q^2}\right|_{Q^2=0}.
\label{eqRM2}
\ea
We assume in this case that $G_{MB}(0)$ is not zero,
neither very small.
The results for the electric charge squared radii and
the magnetic squared radii are respectively presented
in Tables~\ref{tableRE2} and~\ref{tableRM2}
(see columns $\left< r_E^2\right>$ and $\left< r_M^2\right>$).
Experimental
values~\cite{PDG,Zhan11,Eschrich01,Mohr08,Bernauer10,Pohl10,Simon80}
are also included in Table~\ref{tableRE2} for $\left< r_E^2\right>$,
and in the caption of Table~\ref{tableRM2} for $\left< r_M^2\right>$.

Since in the present approach we can identify
the valence quark (bare) contributions
and the pion cloud contributions
in the form factor $G_{XB}$ ($X=E,M$),
we follow Eq.~(\ref{eqGX1}) and
decompose $G_{XB}$ into
\ba
G_{XB}(Q^2)= G_{XB}^b(Q^2) + G_{XB}^\pi(Q^2),
\label{eqGX2}
\ea
where $G_{XB}^b(Q^2)= Z_B G_{X0B}(Q^2)$
and $G_{XB}^\pi(Q^2)= Z_B \delta G_{XB}(Q^2)$,
are respectively the bare and pion cloud contributions.
Based on the decomposition~(\ref{eqGX2})
and the definitions of radii (\ref{eqRE2}) and~(\ref{eqRM2}),
we can write
 \ba
& &
\left< r_E^2\right> = \left< r_E^2\right>_b + \left< r_E^2\right>_\pi,
\label{eqREdecomp}\\
& &
\left< r_M^2\right> = \left< r_M^2\right>_b + \left< r_M^2\right>_\pi,
\label{eqRMdecomp}
\ea
where
\ba
& &
\left< r_E^2\right>_b = - Z_B \frac{6}{G_{EB}(0)}
\left. \frac{d G_{E0B}}{d Q^2}\right|_{Q^2=0}, \\
& &
\left< r_M^2\right>_b = - Z_B \frac{6}{G_{MB}(0)}
\left. \frac{d G_{M0B}}{d Q^2}\right|_{Q^2=0},
\ea
and $\left< r_E^2\right>_\pi$ and $\left< r_M^2\right>_\pi$
can be defined in a similar manner,
but using $\delta G_{EB}(Q^2)$ and  $\delta G_{MB}(Q^2)$,
or from Eqs.~(\ref{eqREdecomp}) and~(\ref{eqRMdecomp}),
by subtracting the bare components from the total.

Since our form factors are determined numerically,
we calculate the octet baryon electric charge
squared radii
and magnetic dipole squared radii,
as well as the respective
components, using numerical derivatives.
The contribution from each component is
also presented in Tables~\ref{tableRE2} and~\ref{tableRM2}.

From the global results for
$\left< r_E^2\right>$ and $\left< r_M^2\right>$,
we can conclude that the nucleon system
has larger spatial charge and magnetization distributions
than that of the $\Sigma$ system, and that the $\Sigma$ system
has the larger spatial charge and magnetization distributions
than those of the $\Xi$ system.
We will leave
the neutral particles ($n$, $\Lambda$, $\Sigma^0$ and $\Xi^0$)
out of the discussion for $\left< r_E^2\right>$.
Our results suggest that the
electric charge squared radii of 0.7-0.8  fm$^2$
for proton ($p$), 0.6-0.7  fm$^2$
for $\Sigma^{\pm}$, and $0.4$  fm$^2$ for $\Xi^-$.
These results may support the general idea that
the systems with two strange quarks ($\Xi$)
are more compact than the systems with only one strange quark
(like $\Sigma$), and the latter systems are more compact
than the nucleon in charge density distributions.
The statement becomes more clear when
we look at the magnetic squared radii,
where we have now 0.7 fm$^2$ for $N$,
0.5-0.6 fm$^2$ for $\Sigma$, and
0.3-0.4 fm$^2$ for $\Xi$.
Note that, for $\left< r_M^2\right>$,
the magnitudes hold also for the neutral particles.

Our results for $\left< r_E^2\right>$ are compared
in Table~\ref{tableRE2} with
the experimental results for $p$, $n$ and $\Sigma^-$.
With the exception for the neutron, to be
discussed later, our results are consistent with the
experimental values.
We compare also our results with
the estimates from Ref.~\cite{Wang09}.
In that work the octet baryon electric charge radii
were extrapolated from the quenched
lattice QCD data~\cite{Boinepalli06} to the physical point
using chiral perturbation theory,
including corrections for both
the finite volume and quenched effects.
Aside from $\Lambda$ and $\Sigma^0$
which were not estimated (and $n$
to be discussed latter), the
results of Ref.~\cite{Wang09} are in agreement
with our results.
We recall that $\Lambda$ (like $\Sigma^0$)
has not been included in the calibration
of our model, since the properties of $\Lambda$ were
not calculated
in the lattice simulations in
Ref.~\cite{Lin09}, on which our parametrization is based.
Note however, our result for the $\Lambda$
electric charge squared radius is also
very small (0.08 fm$^2$), although
larger than the result of Ref.~\cite{Boinepalli06},
0.003--0.017 fm$^2$.
We recall again that the electric charge form factors
of neutral particles are difficult to simulate in lattice QCD,
due to the cancellation among
the contributions from the different flavor quarks,
which should cancel out exactly at $Q^2=0$.
Our results can also be compared with other
estimates presented in the
literature~\cite{Kubis01,Cheedket04,Thomas83,Kunz90,Gobbi92,Panda94,Wagner95,Puglia00,Buchmann03,Buchmann07,Dahiya10}.

Using the decompositions of Eqs.~(\ref{eqREdecomp}) and~(\ref{eqRMdecomp}),
we separate also the effects of the valence quarks and
those of the pion cloud.
The results decomposed for $\left< r_E^2\right>$ are
presented in the second and third columns in Table~\ref{tableRE2}.
Focusing first on the charged baryons,
the valence quarks give dominant contributions,
although the pion cloud effects can be important and
as large as 33\% for the $\Sigma^+$.
Since the effective contributions of the
valence quarks were also estimated
in Ref.~\cite{Wang09},
we compare $\left< r_E^2\right>_b$
directly with those results (third column)
in Table~\ref{tableRE2b}.
Although the meson cloud contributions in Ref.~\cite{Wang09}
are not the same as ours, and also kaon cloud contributions
are included (but amounted to small contributions),
it can be illustrative to
compare our results of $\left< r_E^2\right>_b$
with the estimates of Wang \etal~\cite{Wang09}.
Our results are surprisingly close to
those of Ref.~\cite{Wang09},
with a notable exception for $\Sigma^+$.
In this case our estimate
for $\left< r_E^2\right>_\pi$ is 0.2 fm$^2$,
to be compared with that of Ref.~\cite{Wang09},
$-0.061\pm0.045$ fm$^2$, with opposite sign.
In general, we conclude that our
estimates for the valence quark
contributions are
consistent with those of Ref.~\cite{Wang09}.

In Table~\ref{tableRM2} we present also
the results for the bare and pion cloud contributions
for the magnetic squared radii.
Although the physical radii can differ
appreciably from the lattice extracted radii
without chiral extrapolations
(see figures from \ref{figProtonLatt}-\ref{figXiMLatt}),
lattice results can give us
an idea on the magnitude of the valence quark contributions.
For this, we compare our results with those of the lattice
in Ref.~\cite{Boinepalli06} for the lowest pion mass,
$m_\pi=$ 306 MeV, and Ref.~\cite{Lin09} for $m_\pi=$ 354 MeV.
It is still interesting to notice that
$\left< r_M^2 \right>_b$ approach the
lattice extracted values when the number of the valence
strange quarks increases.
The result for $\Sigma$ is closer to that of the lattice
than that of the nucleon, and the results for $\Xi$
is nearly equal to that of the lattice, when
the standard deviation is taken into account.
This is not unexpected, since lattice QCD calculations
use nearly the physical strange quark mass value.
These facts also give us some confidence
on our model to estimate the valence quark contributions
both in the physical and lattice regimes.

\begin{figure}[t]
\centerline{
\mbox{
\includegraphics[width=3.2in]{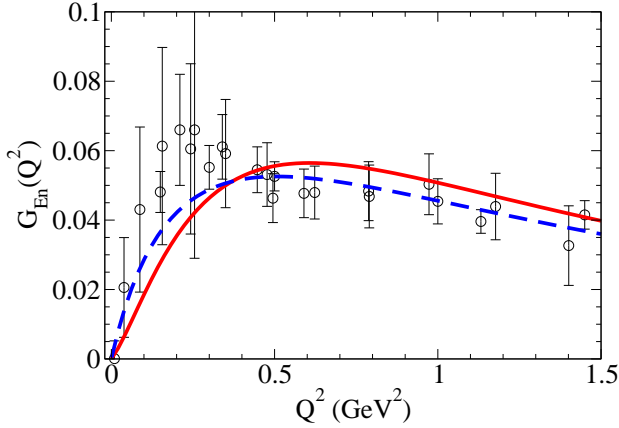}
}}
\caption{\footnotesize{
Neutron electric form factor in the low $Q^2$ region
(magnified from Fig.~\ref{figNucleon1}).
The lines are for the total result (solid line) and
the bare result (dashed line).
}}
\label{figNucleon2}
\end{figure}

We now comment on the neutron electric charge squared radius.
Our result, $-0.029$ fm$^2$, differs appreciably from
the experimental value $-0.12$ fm$^2$.
This deviation is a consequence of our global fit,
and the absence of high accuracy data until recently
for $G_{En}$ in the low $Q^2$ region~\cite{Geis08}.
The result of our fit is presented in Fig.~\ref{figNucleon1},
and magnified in Fig.~\ref{figNucleon2}
for the region $Q^2< 1.5$ GeV$^2$.
As already explained, our fit includes the neutron
physical data, but also includes the octet baryon lattice data.
In particular, the fit also includes the lattice data for
the neutron form factors $G_{En}$ and $G_{Mn}$.
However, since we have reduced the impact
of the lattice data for the neutral particles
to compensate the inaccuracy
of the data, the lattice constraints to
the valence quark contributions are not very strong.
As a consequence, the strongest
constraint for the neutron electric form factor
comes from the very accurate,
high $Q^2$ physical neutron data.
As one can observe in Fig.~\ref{figNucleon2},
our model result deviates from the data in
the region $Q^2< 0.5 $ GeV$^2$.
In the figure the result
associated with the valence quark contributions
is closer to the data than
the full result (dressed).
We then conclude that the deviation of
our result from the physical data
is a consequence of our estimate for the
pion cloud contributions, which are poorly constrained
by the low $Q^2$ neutron physical data.
In short, the poor description of $G_{En}$
is a consequence of our ambition
to describe simultaneously the lattice
and physical data.
It would be improved if
we concentrated only on the physical data in the low $Q^2$ region,
as done in the other studies~\cite{Friedrich03}.
The new generation of experiments for $G_{En}$
with high accuracy~\cite{Geis08} will be important
to clarify the impact of the different effects
in the neutron data, and can constrain quark
models such as the one used in the present work.
Nevertheless, our analysis shows the importance
and sensitiveness of the pion cloud effects
on the neutron and the neutral baryons in general.
The importance of the pion cloud effects is also manifested
in the results of the electric form factors
of $\Sigma^0$, $\Lambda$ and $\Xi^0$.

\begin{table}[t]
\begin{center}
\begin{tabular}{l r r r c c}
\hline
\hline
   & \sp\sp$\left< r_E^2\right>_b$ & \sp\sp $\left< r_E^2\right>_\pi$\sp &
\sp$\left< r_E^2\right>$\sp & \sp \sp Ref~\cite{Wang09}\sp \sp& \sp Exp. \sp\\
\hline
   $p$      &  0.704   & 0.066 &  0.770 &  0.685(66) & 0.769(8) \cite{Zhan11}\\
   $n$      & -0.098   & 0.070 & -0.029 & -0.158(33) & -0.1161(22) \cite{PDG}\\
   $\Lambda$ & -0.0081 & 0.091 &  0.082  &  0.010(9) &   \\
   $\Sigma^+$ & 0.503  & 0.214 &  0.717 & 0.749(72) &  \\
   $\Sigma^0$ &  -0.0020 & 0.049  & 0.048 &          &  \\
   $\Sigma^-$ & 0.519    & 0.113  & 0.632  & 0.657(58)   & 0.61(15)
\cite{Eschrich01}\\
   $\Xi^0$  & 0.090   & -0.0036 & 0.087  & 0.082(29)   &  \\
   $\Xi^-$  & 0.404   &  0.019  & 0.423  & 0.502(47) & \\
\hline
\hline
\end{tabular}
\end{center}
\caption{
Electric charge squared radii of the octet baryons.
All values are in fm$^2$.
The subindexes indicate respectively,
bare ($b$) and pion cloud $(\pi)$ contributions,
as defined in the text.
The experimental value presented for the proton
is an average of Refs.~\cite{Mohr08,Bernauer10,Zhan11}.
An estimate of the proton electric charge squared radius
using muonic hydrogen~\cite{Pohl10}, gives 8\% less than the result
presented in the table.}
\label{tableRE2}
\end{table}

\begin{table}[t]
\begin{center}
\begin{tabular}{l r r r c c c}
\hline
\hline
   & \sp\sp$\left< r_M^2\right>_b$ & \sp\sp \sp $\left< r_M^2\right>_\pi$\sp &
\sp \sp$\left< r_M^2\right>$\sp
&  \sp \sp Ref.~\cite{Boinepalli06} \sp \sp  &
\sp \sp Ref.~\cite{Lin09} \sp \sp
\\
\hline
   $p$      &  0.679   & 0.059 &  0.738 &  0.470(48) & 0.40(8) \\
   $n$      &  0.714   & 0.0034 & 0.718 &  0.478(50) & 0.46(11)\\
   $\Lambda$ & 0.544 & -0.242 & 0.302  & 0.347(24)   & \\
   $\Sigma^+$ & 0.383  & 0.146 &  0.530 &  0.466(42) & 0.36(8)  \\
   $\Sigma^0$ & 0.365  & 0.103  & 0.468 &  0.423(38) & \\
   $\Sigma^-$ & 0.407   & 0.199  & 0.606  & 0.483(49) & 0.37(8)\\
   $\Xi^0$  & 0.370   & -0.0010 & 0.368  & 0.384(22) & 0.32(2) \\
   $\Xi^-$  & 0.295   &  0.045  & 0.340  & 0.336(18) & 0.29(4)\\
\hline
\hline
\end{tabular}
\end{center}
\caption{
Magnetic dipole squared radii of the octet baryons.
All values are in fm$^2$.
The experimental results are
$\left< r_M^2\right>= 0.733\pm0.096$ fm$^2$ for the proton,
and $\left< r_M^2\right>= 0.767\pm0.123$ fm$^2$
for the neutron~\cite{Simon80}.
More recent results for the proton are
$\left< r_M^2\right>= 0.604\pm0.026$ fm$^2$ \cite{Bernauer10},
and $\left< r_M^2\right>= 0.752\pm0.035$ fm$^2$ \cite{Zhan11}.
While the lattice results from Ref.~\cite{Boinepalli06}
correspond to the pion mass $m_\pi= 306$ MeV,
those from Ref.~\cite{Lin09} correspond
to $m_\pi= 354$ MeV.}
\label{tableRM2}
\end{table}

\begin{table}[t]
\begin{center}
\begin{tabular}{l r c c c }
\hline
\hline
   & \sp\sp$\left< r_E^2\right>_b$ &  \sp \sp Ref~\cite{Wang09} \sp \sp &
\sp \sp Ref.~\cite{Boinepalli06}  \sp &
Ref.~\cite{Lin09}

\\
\hline
   $p$       &  0.704  &  0.746(69) & 0.452(53) &  0.41(4)   \\
   $n$       & -0.098  & -0.097(31) & 0.029(41) &         \\
   $\Lambda$ & -0.0081 &  0.026(9)  & 0.025(7)  &            \\
   $\Sigma^+$&  0.503  &  0.820(75) & 0.503(54) &  0.44(3)   \\
   $\Sigma^0$& -0.0020 &            & 0.046(15) &            \\
   $\Sigma^-$& 0.519   &  0.586(56) & 0.410(37) &  0.357(16)  \\
   $\Xi^0$   & 0.090   &  0.113(27) & 0.062(22) &         \\
   $\Xi^-$   & 0.404   &  0.471(46) & 0.389(23) &  0.326(6)  \\
\hline
\hline
\end{tabular}
\end{center}
\caption{
Valence quark contributions for the octet
baryon electric charge squared radii,
compared with the valence quark contributions from
Ref.~\cite{Wang09}, and lattice
QCD simulations~\cite{Lin09,Boinepalli06}.
While the lattice results from Ref.~\cite{Boinepalli06}
correspond to the the pion mass $m_\pi= 306$ MeV,
those from Ref.~\cite{Lin09} correspond
to $m_\pi= 354$ MeV.}
\label{tableRE2b}
\end{table}

\section{Conclusions}
\label{secConclusions}

In the present work we have applied a covariant spectator quark model
to study the valence quark structure of the octet baryons.
Combining the contributions of the valence quarks (covariant model)
with a phenomenological, covariant parametrization
for the pion cloud contributions,
we have described the electromagnetic form factors of
the octet baryons in the lattice and physical regimes.
The octet baryon systems are described in
a simplified quark model with the S-state configuration
for the quark-diquark relative motion.
The simplicity of the wave functions
allows us to calibrate the momentum
dependence of the octet baryon
form factors using four independent
range parameters.
The model has been calibrated using lattice QCD data
for the octet baryons (272 data points),
the physical nucleon form factor data (202 data points),
and the physical octet baryon magnetic moment data (5 data points).
Overall, we have used 6 parameters for
the valence quark structure, and 7 parameters
for the pion cloud effects.
We have derived a parametrization
for the octet baryon wave functions
depending on the baryon flavors.
The parametrized wave functions are very useful
for future studies of the electromagnetic reactions.

Our results suggest a dominance of the
valence quark effects,
but the inclusion of the pion cloud effects improves the global
description of the octet baryon electromagnetic form factor data.
The pion cloud effects are of the order of 15\% for
the nucleon in the low $Q^2$ region, but decrease
in the higher $Q^2$ region.
Surprisingly, the pion cloud effects are still important
for the nucleon electric form factors in the region
of 5-10 GeV$^2$ with the contributions of 5-8\%.
This feature
is a consequence of the value of
$\Lambda_2$ ($\Lambda_2^2 = 1.54$ GeV$^2$)
for the pion cloud parametrization,
which gives a slow falloff for the Pauli
form factor $F_2$.

We have a good description of the nucleon data
($\chi^2$ per data point $\simeq$1.9), but
for the description of the neutron it is slightly worse
($\chi^2$ per data point of 1.9 and 2.3 for
the electric and magnetic form factors, respectively).
This is a consequence of the high accuracy of the recent
$G_{Mn}$ data~\cite{Lachniet09}, and also
our ambition to describe simultaneously
the lattice and physical data .
The large number of the lattice data points
reduces the impact of the nucleon physical data.
Also the inaccuracy of the neutron lattice data (especially
differs from the neutron charge zero at $Q^2=0$) makes it
difficult to constrain those parameters of the model directly
related to the nucleon system.
To improve the description of the nucleon
system, it will be useful
to perform a more systematic and detailed
study for the nucleon form factors.
This can be done using a more complete lattice QCD database
including the results from
Refs.~\cite{Gockeler05,Alexandrou06,Alexandrou08,Syritsyn10,Lin10,Collins11},
and also the results of the new generation
of nucleon form factor data in the low $Q^2$ region~\cite{Geis08,Ron11,Gilad11},
which determine $G_E$ and $G_M$ simultaneously
instead of the ratio\footnote{
That will require the combination of the
polarization transfer measurements and
cross section measurements (Rosenbluth method).}
$G_{E}/G_{M}$.
In this work we have restricted to use the lattice data
from Ref.~\cite{Lin09}, since one of our goals
is the overall description of the octet baryon electromagnetic
structure, and we prefer to avoid possible inconsistencies
arising from different lattice QCD simulations.

Our results for the neutron form factors
require some discussions.
Our global fit has turned up to give a poor
description for the neutron electric form factor
for $Q^2< 0.5$ GeV$^2$
(electric charge squared radius of $-0.029$ fm$^2$).
Although $G_{En}$ is poorly constrained by the lattice data,
it is complemented by the physical data.
Our result for the neutron electric squared radius
differs appreciably from the experimental
value, although our estimate of the bare valence quark core
is closer to the experimental value, and also to that of
the valence quark contribution from Ref.~\cite{Wang09}.
We interpret this as a consequence of the impact
of the high $Q^2$ region data (very precise in general),
which reduce the impact of
the low $Q^2$ region data (less accurate) in the global fit.
As already discussed, a more precise calibration
of the model for the neutron will require
a more detailed analysis of the nucleon system,
and this will be possible when more precise data
as in Ref.~\cite{Geis08} become available.
More accurate lattice QCD data for the neutron
can also be very useful.
Nevertheless, our results show that the pion cloud
effects are indeed important and influential for the neutron data.

Overall, we have a very good global description
of both the lattice QCD and physical data
for the charged particles $p$, $\Sigma^\pm$ and $\Xi^-$.
For the analysis of neutral particles, one must be careful,
because they have been less constrained by
the lattice data ($n$ and $\Xi^0$),
or not at all constrained ($\Lambda$ and $\Sigma^0$).
But we can conclude that the pion cloud effects
are very important for the electromagnetic
form factors $G_{EB}$ of neutral particles.

Concerning the size of the octet baryons,
we conclude that, as expected, the systems with two strange
quarks are more compact than those of the one strange quark,
and the latter is more compact than the nucleon system
with no strange quark.
The estimates of the octet baryon sizes
(electric charge squared and magnetic dipole squared radii)
due to the valence quarks (bare core)
are consistent with this conclusion, and
are also in agreement with the independent
estimates~\cite{Wang09}.
The inclusion of the pion cloud effects
in the electric charge squared radii, in general,
makes the final results approach
to the experimental results,
and also to the values estimated
based on the lattice results extrapolated using
chiral perturbation theory~\cite{Wang09}.

The agreement between our estimates
with the octet baryon electromagnetic form
factor data, and also those of the radii,
gives us some confidence about
the applicability of our model to
estimate the valence quark effects
both in the physical and lattice regimes.
An application of the present model
for the octet to decuplet electromagnetic transitions
is in progress by generalizing the study made for the
$\gamma N \to \Delta$ transition~\cite{NDelta,NDeltaD,LatticeD}.
For this purpose, we can use the octet baryon wave functions obtained
in this work, and those of the decuplet obtained in
Ref.~\cite{Omega} within the same formalism~\cite{InPreparation2}.

\vspace{0.2cm}
\noindent
{\bf Acknowledgments:}

The authors would like to thank Jerry Gilfoyle
for providing the data from Ref.~\cite{Lachniet09}.
This work was supported in part by the European Union
(HadronPhysics2 project ``Study of strongly interacting matter'').
This work was also supported by the University of Adelaide and
the Australian Research Council through the Grant No.~FL0992247 (AWT).
G.~R.\ was supported by the Funda\c{c}\~ao para
a Ci\^encia e a Tecnologia under the Grant
No.~SFRH/BPD/26886/2006.

\appendix

\section{Brief review of nucleon electromagnetic form factors}
\label{apNucleon}

Here we briefly review the results of Ref.~\cite{Nucleon}
for the nucleon electromagnetic form factors.
We consider the wave function from Eq.~(\ref{eqPsiB}),
corresponding to the nucleon case
(see Ref.~\cite{Nucleon} for details).
Below, $M_N$ represents the nucleon mass.

In the following we use
\ba
\Delta^{\alpha \beta} = \sum_{\lambda}
\varepsilon_{P_+}^\alpha(\lambda)
\varepsilon_{P_-}^{\beta \ast}(\lambda),
\ea
where $\varepsilon_{P_\pm}$ is the diquark polarization vectors
of the final ($P_+$) and initial ($P_-$) states
in the fixed-axis representation
\cite{FixedAxis}.
The direct calculation gives \cite{Nucleon,NDelta,FixedAxis}:
\ba
\Delta^{\alpha \beta}&=&
-\left(
g^{\alpha \beta} - \frac{P_-^\alpha P_+^\beta}{P_+ \cdot P_-}
\right)  \label{eqDelta} \\
&+&
a \left(
P_-^\alpha -\frac{P_+ \cdot P_-}{M_+^2}P_+^\alpha
\right)
\left(
P_+^\beta -\frac{P_+ \cdot P_-}{M_-^2}P_-^\beta
\right),
\nonumber
\ea
with
\be
a= - \frac{M_+ M_-}{P_+\cdot P_- (M_+ M_-+ P_+ \cdot P_-)},
\ee
where $M_-$ and $M_+$ are the
masses of the initial and final states.
In the present case $M_+=M_-=M_N$.

Equation (\ref{eqJB0}) leads to
the transition current~\cite{Nucleon},
\ba
J_N^\mu &=&
\frac{3}{2} B(Q^2) \times
\bar u(P_+) \left\{
\left[
j_1 \gamma^\mu + j_2 \frac{i \sigma^{\mu \sigma}q_\sigma}{2M_N}
\right] \right. \\
& &
\left.
-\frac{1}{3}
\gamma_\alpha \gamma_5
 \left[
j_3 \gamma^\mu + j_4 \frac{i \sigma^{\mu \sigma}q_\sigma}{2M_N}
\right]
\gamma_5 \gamma_\beta
 \Delta^{\alpha \beta} \frac{}{}
\right\} u(P_-),
\nonumber
\ea
where $\Delta^{\alpha \beta}$ is given by
Eq.~(\ref{eqDelta}), and
\be
B(Q^2)=
\int_k
\psi_N(P_+,k) \psi_N(P_-,k).
\label{eqB}
\ee
The coefficients, $j_1$ and $j_2$,
can be written in the SU(2) sector,
\be
j_i = \frac{1}{6}f_{1+}(Q^2) + \frac{1}{2} f_{1-} (Q^2) \tau_3,
\label{eqJiN}
\ee
and
\be
j_{(i+2)}= \frac{1}{3} \tau_j j_i \tau_j =
\frac{1}{6}f_{1+} - \frac{1}{6} f_{1-} \tau_3.
\label{eqJiN2}
\ee

Note that the expressions (\ref{eqJiN}) and~(\ref{eqJiN2}),
when applied to the nucleon isospin state
$\chi_+= \left(
\begin{array}{c} 1 \cr 0 \cr
\end{array}
\right)$ and
$\chi_-= \left(
\begin{array}{c} 0 \cr 1 \cr
\end{array}
\right)$,
are equivalent to Eqs.~(\ref{eqjiA}) and~(\ref{eqjiS}).

With some spin algebra
we obtain:
\ba
& &F_1(Q^2)=
\frac{3}{2} B(Q^2)   \nonumber \\
& & \times \left\{ j_1
+ \frac{1}{3} \frac{1}{4M_N^2+Q^2}
\left[(12 M_N^2 -Q^2) j_3 -4 Q^2 j_4 \right]
\right\}, \nonumber \\
& &\\
& &F_2(Q^2)=
\frac{3}{2} B(Q^2)  \nonumber \\
& & \times \left\{j_2
-\frac{1}{3} \frac{1}{4M_N^2+Q^2}
\left[16 M_N^2 j_3 +(4M_N^2-3Q^2)  j_4 \right]
\right\}. \nonumber \\
& &
\ea
To extend these results to the
octet baryon $B$, we use $M_B$ for its mass,
and make the replacements:
\ba
& &j_1 \to j_1^A, \\
& &j_2 \to \frac{M_B}{M_N}j_2^A, \\
& &j_3 \to j_1^S, \\
& &j_4 \to \frac{M_B}{M_N}j_2^S.
\ea
The expressions associated with $j_1$
and $j_3$ are determined directly from the
respective definitions.
For $j_2$ and $j_4$, we need
to take into account that
the quark current Eq.~(\ref{eqJi}) is written
in terms of the nucleon mass, which leads to
\be
\frac{i \sigma^{\mu \nu} q_\mu}{2M_N}=
\frac{M_B}{M_N}
\frac{i\sigma^{\mu \nu} q_\mu}{2M_B}.
\ee
That is why the coefficients $j_2^{A,S}$
are modified by the factor $\sfrac{M_B}{M_N}$.
Finally, in Eq.~(\ref{eqB}) we replace $\psi_N$ by $\psi_B$.



\end{document}